\documentclass[twocolumn,showpacs,showkeys,amssymb,nobibnotes,superscriptaddress,aps,prx,bibnotes]{revtex4-2}

\usepackage{graphicx}
\usepackage{dcolumn}
\usepackage{bm}
\usepackage[T1]{fontenc}
\usepackage{mathptmx}
\usepackage{mathrsfs}
\usepackage{physics}
\usepackage{simplewick}
\usepackage{amsthm,amsmath,amssymb}
\usepackage[bottom]{footmisc}

\usepackage{amssymb}
\usepackage{amsmath}
\usepackage{amsfonts}
\usepackage{bm}
\usepackage{graphicx}
\usepackage{subfigure}
\usepackage[dvipsnames]{xcolor}
\usepackage{calc}
\usepackage{epsfig}
\usepackage{color}
\usepackage[colorlinks,citecolor=blue]{hyperref}
\begin{document}
\bibliographystyle{apsrev4-1}
\newcommand{\be}{\begin{equation}}
\newcommand{\ee}{\end{equation}}
\newcommand{\bs}{\begin{split}}
\newcommand{\es}{\end{split}}
\newcommand{\R}[1]{\textcolor{red}{#1}}
\newcommand{\B}[1]{\textcolor{blue}{#1}}

\newcommand{\suppmaterial}{%
    \clearpage
    \begin{widetext}
    \begin{center}
        \bfseries\large Supplementary Material for ``Testing the quantum nature of gravity through interferometry''\\[1em]
        
\normalfont\normalsize
        Yubao Liu\textsuperscript{1}, 
        Yanbei Chen\textsuperscript{2}, 
        Kentaro Somiya\textsuperscript{3}, 
        Yiqiu Ma\textsuperscript{1,4*}\\[0.5em]
        
             \normalfont\small
        \textsuperscript{1} \emph{National Gravitation Laboratory, Hubei Key Laboratory of Gravitation and Quantum Physics, School of Physics, Huazhong University of Science and Technology, Wuhan, 430074, China}\\
        \textsuperscript{2}  \emph{Burke Institute of Theoretical Physics, California Institute of Technology, Pasadena, CA, 91125, USA} \\
        \textsuperscript{3}  \emph{Department of Physics, Tokyo Institute of Technology,
2-12-1 Ookayama, Meguro-ku, Tokyo 152-8551, Japan} \\
       \textsuperscript{4}  \emph{Department of Astronomy, School of Physics, Huazhong University of Science and Technology, Wuhan, 430074, China} \\ [0.5em]

*Contact: \url{myqphy@hust.edu.cn}
    \end{center}
    \vspace{1em}
    
    \setcounter{equation}{0}
    \renewcommand{\theequation}{S.\arabic{equation}}
}

\newenvironment{suppabstract}{%
  \quad
}{%
    \normalsize
}

\title{Testing the quantum nature of gravity through interferometry}
\author{Yubao Liu}
\affiliation{National Gravitation Laboratory, Hubei Key Laboratory of Gravitation and Quantum Physics, School of Physics, Huazhong University of Science and Technology, Wuhan, 430074, China}
\author{Yanbei Chen}
\affiliation{Burke Institute of Theoretical Physics, California Institute of Technology, Pasadena, CA, 91125, USA}
\author{Kentaro Somiya}
\affiliation{Department of Physics, Tokyo Institute of Technology,
2-12-1 Ookayama, Meguro-ku, Tokyo 152-8551, Japan}
\author{Yiqiu Ma}
\email{myqphy@hust.edu.cn}
\affiliation{National Gravitation Laboratory, Hubei Key Laboratory of Gravitation and Quantum Physics, School of Physics, Huazhong University of Science and Technology, Wuhan, 430074, China}
\affiliation{Department of Astronomy, School of Physics, Huazhong University of Science and Technology, Wuhan, 430074, China}

\begin{abstract}
We propose a Michelson-type interferometric protocol for testing the quantum nature of gravity through testing the phenomenology of semi-classical gravity theory, which predicts a state-dependent Schrödinger-Newton (SN) evolution of the test mass. The protocol's feature lies in utilizing the asymmetry of two interferometric arms induced by SN self-gravity to create cross-talk between the common and differential motion of the test masses. This cross-talk is imprinted as a clean binary signature in the correlation measurements of the interferometer's output light fields. 
Our results demonstrate that, when assisted by 10\,dB squeezed input states, 3\,hours of aggregated measurement data can provide sufficient signal-to-noise ratio to conclusively test the SN theory in $\sim1$\,Kelvin environment. This shows the strong feasibility of using such interferometric protocols to test if gravity operates quantum-mechanically.
\end{abstract}

\maketitle
\emph{Introduction and summary ---} 
The quantum nature of gravity remains an experimental frontier: in order to distinguish between quantum gravity\,(QG) and semi-classical gravity—where classical gravity is sourced by the quantum expectation of the matter density distribution\,\cite{Rosenfeld1963,Mueller1962,Yang2013,Bahrami_2014,Bose2025}—experimental validation is imperative. Recent advancements in quantum optomechanics\,\cite{Chen_2013,Aspelmyer2012,Aspelmyer2014,Schnabel2015,Matsumoto2020,Yu2020,Hoang2016,Ando2010,ubhi2022active,Mason2019,Rossi2018,Aggarwal2020} present a unique opportunity for the experimental scrutiny of macroscopic quantum objects, offering two complementary approaches: testing either (i) quantum gravity through gravity-induced entanglement (GIE) \cite{Bose2017,Marletto2017Gravitationally,Miao2020,Christodoulou2023,krisnanda2020,Carney2021Using,Fujita_2023} or (ii) semi-classical gravity through self-gravity effects like the Schr\"odinger-Newton\,(SN) potential below \cite{Yang2013,Helou2017,Gan2016Optomechanical,Grossardt2016,Datta_2021}.

If gravity is classical, a macroscopic test mass with center-of-mass (CoM) wavefunction $\psi(x)$ must create a classical gravitational field from the convolution of $|\psi(x)|^2$ and the object's internal matter distribution. This self-gravity field provides an additional confinement potential for the quantum evolution of $\psi(x)$~\cite{Yang2013,giulini2011,giulini2014} when CoM displacement uncertainty is much smaller than the ion's internal motion $x_{\rm int}$ in the test mass's crystal field. For an object with mass $M$ inside a harmonic potential with frequency $\omega_m$, the the Schr\"odinger-Newton (SN) Hamiltonian for its CoM is now given by\,\cite{Yang2013,giulini2011,giulini2014}:
\begin{equation}\label{eq:Hamiltonian_self_sn}
\hat{H}_{\rm SN}=\frac{\hat{p}^2}{2M}+\frac{1}{2}M\omega_m^2\hat{x}^2+\frac{1}{2}M\omega^2_{\rm SN}(\hat{x}-\langle\psi|\hat{x}|\psi\rangle)^2.
\end{equation}
Here the confinement potential, centered at the quantum expectation value of the displacement $\langle\psi|\hat{x}|\psi\rangle$, has the SN frequency $\omega_{\rm SN}=\sqrt{Gm/(6\sqrt{\pi}x_{\rm int}^3})$ with ion mass $m$. The value of SN-frequency typically satisfies $\omega_{\rm SN} \gg \omega_g\sim\sqrt{G\rho}$ (with $\rho$ the object's mean mass density), where $\omega_g$ sets the scale for probing quantum nature of gravity via GIE\,\cite{Bose2017,Miao2020}. Crucially, the condition $\hbar\omega_{\rm SN} Q > k_B T$ for observing SN effects is  far less demanding than the GIE requirement $\hbar\omega_g Q > k_B T$ \cite{Miao2020}, where $Q$ is the quality factor of the mechanical oscillators involved. This establishes an experimental hierarchy: the absence of SN effects would challenge semi-classical gravity while supporting QG, making self-gravity tests a natural precursor to GIE experiments.

The state-dependence of Eq.\,\eqref{eq:Hamiltonian_self_sn} introduces  a nonlinearity that had brought considerable conceptual trouble when dealing with the measurement process \cite{Helou2017}. The Causal-Conditional Schrödinger-Newton (CCSN) theory resolved this by replacing expectation values with conditional expectations from measurement outcomes~\cite{Helou2017,scully2022}; CCSN also unifies with stochastic completion of SN theory\,\cite{Nimmrichter2015,Kafri_2014,Diosi1989,Diosi1998} and the post-quantum gravity\,\cite{kryhin2025,oppenheim2023,Oppenheim2023A,Tilloy2016}. In all such theories, gravity remains classical and driven by classical information -- either extracted through observers's measurement or built-in ``auxiliary observers'' (akin to collapse models)~\cite{Miki2025,Tilloy2024}; they map precisely to Quantum Control Theory: classical gravity corresponds to measurement-based feedback, while quantum gravity represents coherent feedback\,\cite{Doherty1999}.

From Hamiltonian~\eqref{eq:Hamiltonian_self_sn}, one had naively expected that as the oscillator is monitored by light, classical thermal noise will peak at $\omega_m$, while quantum fluctuations will peak at $\omega_q \equiv \sqrt{\omega_m^2+\omega_{\rm SN}^2}$, a very observable signature~\cite{Yang2013,Helou2017}. As the CCSN theory appropriately treats the information flow into the classical gravity field, the correctly predicted experimental signatures are often weaker than naively expected. For example: the quantum trajectory of a test mass evolving under a continuous Gaussian measurement has conditional expectations $\langle x \rangle_c$ and $\langle p\rangle_c$ follow the same evolutions as standard quantum mechanics (resonant at $\omega_m$), while the conditional covariance matrix $\mathbf{V}^c$ would have $\omega_m$ replaced by $\omega_q$~ \cite{Liu2023}.  The fact that conditional expectations have resonance at $\omega_m$ makes the double peaks disappear, replaced by more subtle distortions~\cite{Liu2023,Liu2024,Diosi2025}.  The CCSN also predicts false gravity-induced entanglement signatures in steady-state tests that mimic quantum gravity\,\cite{Liu2024}, rendering these approaches inconclusive. Consequently, clean SN signatures now require more technically demanding non-stationary or delayed measurement protocols~\cite{Miki2025}.

\begin{figure}[h]
\centering
\includegraphics[width=0.45\textwidth]{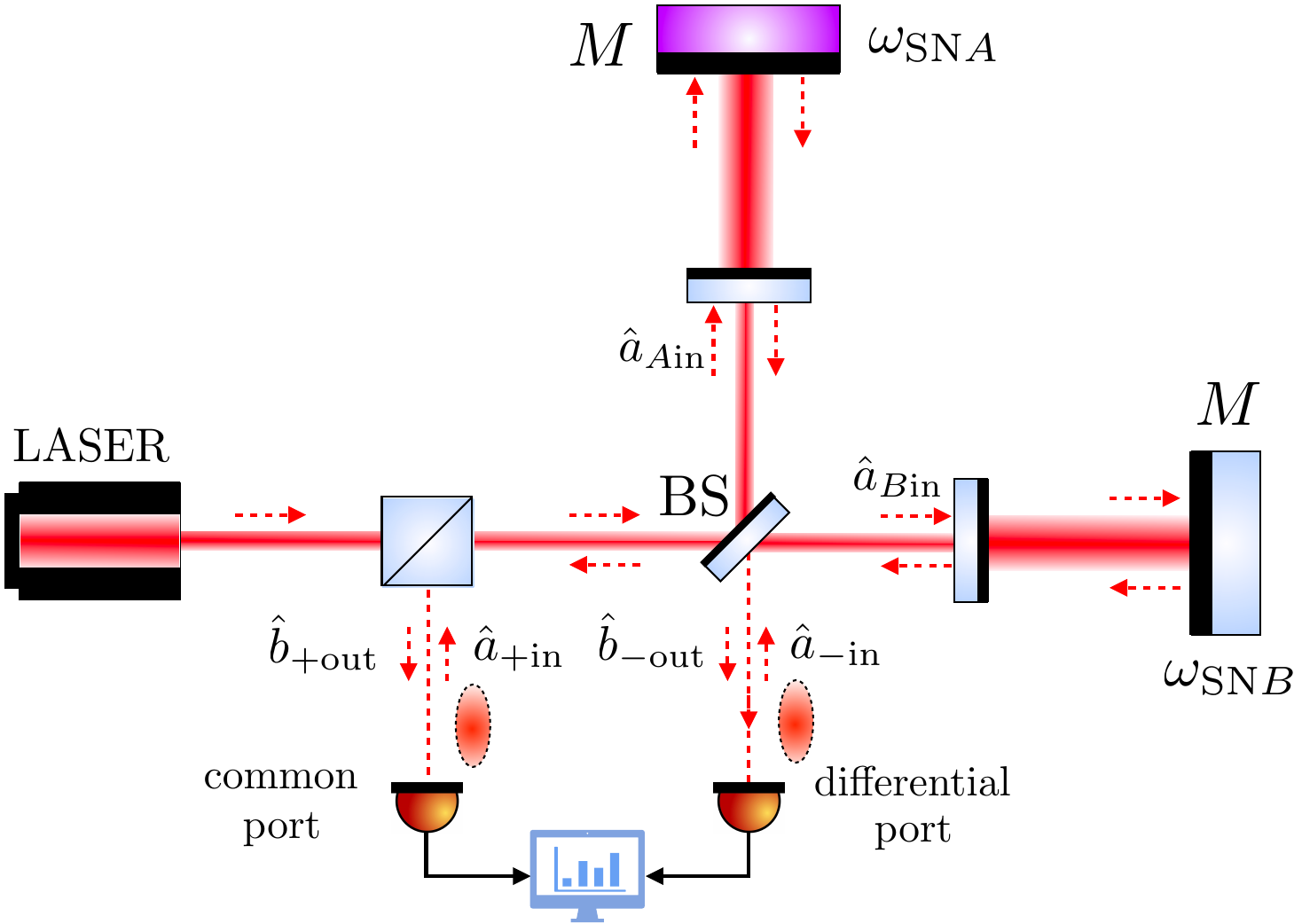}
\caption{Configuration for testing semi-classical gravity. Two end mirrors (equal mass $M$, different Schrödinger-Newton frequencies $\omega_{\rm{SN}A/B}$) and input mirrors form two arm cavities in the bad-cavity limit. Identical squeezed vacuums are injected into common/differential ports, with the SN signature extracted via output quadrature cross-correlation.
}
\label{fig:new_configuration}
\end{figure}

In this paper, we introduce a Michelson interferometer whose two test-mass mirrors have the same mass but different materials, with an asymmetry in $\omega_{\rm SN}$. The SN asymmetry, as well as the simultaneous measurements made in the common and differential modes, make non-trivial modifications to the statistics of the out-going fields; we predict steady-state cross correlations between out-going fields at the common and differential ports, which are absent in QG.   Enhanced by the squeezing injection and common-mode suppression of the laser intensity noise, this interferometric protocol exhibits high experimental feasibility, even at 1\,K temperature. 

\emph{Configuration and Hamiltonian ---}
The proposed interferometric protocol (Fig.\,\ref{fig:new_configuration}) utilizes two movable end mirrors $A$ and $B$ with equal masses $M$ but distinct materials, yielding different SN frequencies ($\omega_{\text{SN}A} \neq \omega_{\text{SN}B}$) due to material-dependent $\omega_{\text{SN}}$. The common and differential mechanical modes $\hat{x}_\pm \equiv (\hat{x}_A \pm \hat{x}_B)/\sqrt{2}$ (with conjugate momenta $\hat{p}_\pm \equiv (\hat{p}_A \pm \hat{p}_B)/\sqrt{2}$ satisfying $[\hat{x}_\pm,\hat{p}_\pm] = i\hbar$) are probed through optomechanical interactions with arm-cavity modes  $\hat a_{A,B}$.  The beam splitter relates the optical fields entering the arm cavity to the input fields at the common and differential ports via $\hat a_{A,B{ \rm in}}=(\hat a_{{\rm in}+}\pm\hat a_{{\rm in}-})/\sqrt{2}$,  driven by \emph{phase-squeezed} vacuum inputs with squeezing degree $r$ . While QG preserves symmetry between the arms (resulting in uncorrelated output fields $\hat{b}_+$, $\hat{b}_-$), SN gravity breaks this symmetry via different SN self-gravity potential $\propto M\omega_{\text{SN}I}^2(\hat{x}_{I} - \langle \hat{x}_{I} \rangle)^2$ ($I=A,B$)  of the two mirrors. 

The system dynamics are governed by the Hamiltonian $\hat{H} = \hat{H}_m + \hat{H}_{\text{cav}} + \hat{H}_{\text{om}} + \hat{H}_{\text{ext}}$, where the mechanical term is: 
\be\label{eq:common_differential_mechanical}
\begin{split}
&\hat H_m=\sum_{i=+,-}\frac{\hat p^2_{i}}{2M}+\frac{1}{2}M\omega_m^2\hat x^2_{i}+\frac{1}{2}M\omega^2_{{\rm SN}}(\hat{x}_{i}-\langle\psi|\hat{x}_{i}|\psi\rangle)^2\\
&\qquad\quad+\frac{1}{2}M{\delta \omega}^2_{\rm SN}(\hat{x}_{+}-\langle\psi|\hat{x}_{+}|\psi\rangle)(\hat{x}_{-}-\langle\psi|\hat{x}_{-}|\psi\rangle),
\end{split}
\ee
with $\omega^2_{\text{SN}} \equiv (\omega_{\text{SN}A}^2 + \omega_{\text{SN}B}^2)/2$, $\delta\omega^2_{\text{SN}} \equiv \omega_{\text{SN}A}^2 - \omega_{\text{SN}B}^2$. The optical and optomechanical terms are:
\begin{equation}
\begin{split}
&\hat{H}_{\text{cav}} = \hbar\omega_c (\hat{a}^\dagger_+ \hat{a}_+ + \hat{a}^\dagger_- \hat{a}_-), \quad \hat{H}_{\text{om}} = -\hbar g (\hat{a}_{1+}\hat{x}_+ + \hat{a}_{1-}\hat{x}_-),\\
&\hat{H}_{\text{ext}} = i\hbar\sqrt{2\gamma} (\hat{a}^\dagger_+ \hat{a}_{+\text{in}} + \hat{a}^\dagger_- \hat{a}_{-\text{in}} - \text{h.c.})
\end{split}
\end{equation}
where $\hat{a}_{\pm 1} \equiv (\hat{a}_{\pm} + \hat{a}^\dagger_{\pm})/\sqrt{2}$ and $\hat{a}_{\pm2} \equiv (\hat{a}_{\pm} - \hat{a}^\dagger_{\pm})/\sqrt{2}i$ are the amplitude and phase quadratures of cavity fields. The optomechanical term $\hat H_{\rm om}$ describes the optomechanical interaction, with the strength $g$ related to the intra-cavity power $P_{\rm cav}$ and cavity length $L$ by $g\equiv(\omega_c P_{\rm cav}/\hbar cL)^{1/2}$. In addition, $\hat{H}_{\text{ext}}$ describes the interaction between the cavity fields and the external optical fields, with $\gamma$ denoting the bandwidth of the arm cavities. The $\delta\omega^2_{\text{SN}}$-induced $\hat{x}_+$-$\hat{x}_-$ coupling (Fig.\,\ref{fig:effective_mode}) generates measurable cross-correlations in $\hat{b}_\pm$, providing a distinct signature of SN gravity. 

\begin{figure}[h]
\centering
\includegraphics[width=0.5\textwidth]{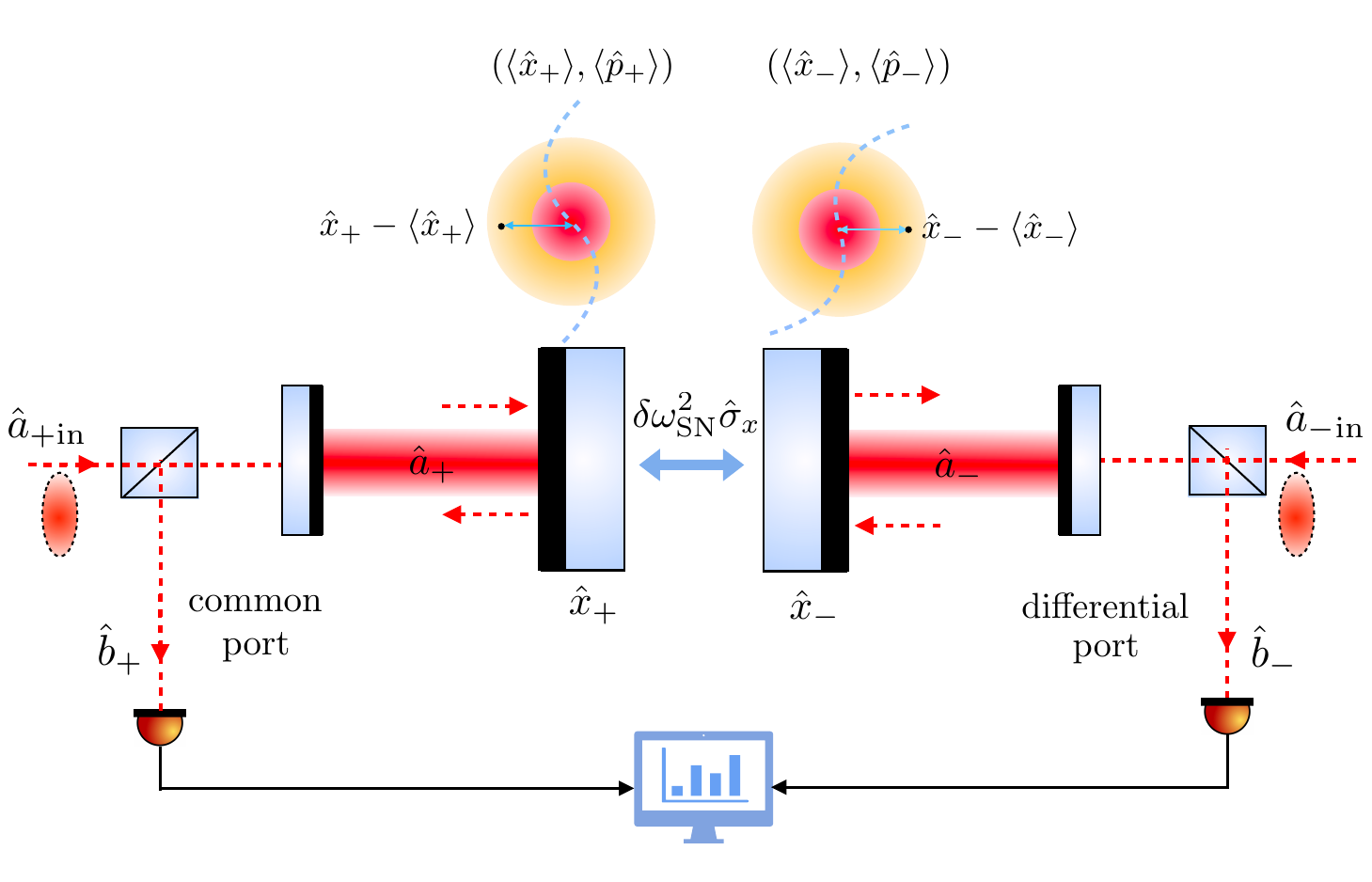}
\caption{Cross-talk between common and differential modes. The SN gravity are sourced by the quantum trajectories of the two mechanical modes under continuous optical measurement, which consequentially affect the test mass mechanical states. Red disks: Wigner function cross-sections with only quantum optical noise. Orange disks: includes environmental bath noise.}
\label{fig:effective_mode}
\end{figure}

\emph{Role of quantum measurement ---}
To predict the statistics of out-going fields, we obtain Heisenberg equations of motion~\cite{Liu2024,Miki2025}.
By assembling $\pm$ variables into vectors and use boldface, i.e., $\mathbf{\hat o}=(\hat o_+, \hat o_-)^T$, we can write
\begin{align}
& D_t\mathbf{\hat x}=-
\underbrace{\left[\omega_{\rm SN}^2+\delta \omega^2_{\rm SN}\hat\sigma_x/2\right]}_{\mathbb{M}_{\rm SN}}(\mathbf{\hat x}-\langle\mathbf{\hat x}\rangle_c)
+\sqrt{{\hbar}{M}}\Lambda\mathbf{\hat a}_{1{\rm in}},\label{eq:eom}\\
& \mathbf{b}_1 =\mathbf{a}_{1\rm in }\,,\quad 
\mathbf{b}_2 =\mathbf{a}_{2\rm in }+\sqrt{ M/\hbar}\Lambda \hat{\mathbf{x}}\,.
\end{align}
Here 
 $D_t=d^2/dt^2+\gamma_md/dt+\omega_m^2$, $\hat \sigma_x$ is the $x$-component of Pauli matrix.  A combination of the out-going field  quadratures $\mathbf{\hat b}_\theta = \mathbf{\hat{b}}_{1}\cos\theta + \mathbf{\hat{b}}_{2}\sin\theta$ is measured via homodyne detection with angle $\theta$.  We have neglected the classical thermal fluctuations, which simply superimpose onto the Heisenberg operators unaffected by SN effect. Cavity modes have also been adiabatically eliminated, with $\Lambda \equiv \sqrt{8\omega_cP_{\rm cav}/(MT_{\rm in}c^2)}$ the optomechanical cooperativity.

In the CCSN formulation $\langle\hat{\mathbf{x}}\rangle_c$ in Eq.~\eqref{eq:eom} is the conditional expectation obtained from measurement results of $\hat{\mathbf{b}}_\theta$.  From Eq.~\eqref{eq:eom}, we can decompose the displacement operator $\hat{\mathbf{x}}=\hat{\mathbf{x}}_q+{\mathbf{x}}_{\rm cl}$ into a quantum piece $\hat{\mathbf{x}}_q= \chi_q \sqrt{M/\hbar}\Lambda\,\hat{\mathbf{a}}_{1\rm in}$ driven by $\hat{\mathbf{a}}_{1\rm in}$ and a classical piece $\mathbf{x}_{\rm cl}=\chi_q \mathbb{M}_{\rm SN}\langle\hat{\mathbf{x}}\rangle_c$ driven by $\langle \mathbf{x} \rangle_c$.  The  outgoing field $\mathbf{\hat{b}}_\theta=\mathbf{\hat{b}}_{q\theta}+\mathbf{b}_{\rm cl \theta}$ inherits this decomposition, with the quantum component given by $
\hat{\mathbf{b}}_{q\theta} = \hat{\mathbf{a}}_{\theta\rm in} + M\Lambda^2\sin\theta\,{\chi}_q \hat{\mathbf{a}}_{1\rm in}$ and the classical component $\mathbf{b}_{\rm cl\theta} =\sqrt{\hbar M}\Lambda \mathbf{x}_{\rm cl}$. 
Here we have introduced the quantum response matrix  $\chi_q = [D_t +\mathbb{M}_{\rm SN}]^{-1}$  which has poles at $\pm\omega_{qI}-i\gamma_m$, with  $\omega_{qI}\equiv\sqrt{\omega_m^2 +\omega_{{\rm SN}I}^2}$, shifted away from $\omega_m$~\footnote{Note that the application of $\chi_q$ involves a time-domain integral.}. By contrast, the response matrix in standard quantum mechanics is given by $\chi_m = D_t^{-1}$ and has poles at $\pm\omega_m -i\gamma_m$; we also have the relation $\mathbb{M}_{\rm SN}=\chi_q^{-1}-\chi_m^{-1}$. 

Measuring the output fields $\hat{\mathbf{b}}_{\theta}$ generates data $\mathbf{y}_{\theta}=\mathbf{y}_{q\theta}+\mathbf{y}_{\rm cl \theta}$, collapses the joint optomechanical states to the conditional Gaussian mechanical states.  The conditional expectation of $\hat{\mathbf{x}}$ is then the sum of the conditional expectation of $\hat{\mathbf{x}}_q$ plus $\mathbf{x}_{\rm cl}$, with the former written as an integral over past values of $\mathbf{y}_{\theta q}$, leading to:
\begin{equation}\label{eq:quantum_trajectory}
\langle \mathbf{x}\rangle_c = \langle\mathbf{\hat{x}}\rangle_c = \langle\mathbf{\hat{x}}_q\rangle_c + \mathbf{x}_{\rm cl}  = \mathbb{K}_{\theta q}\mathbf{y}_{\theta q}+\chi_q\mathbb{M}_{\rm SN}\langle\mathbf{{x}}\rangle_c.
\end{equation}
Here the causal Wiener filter $\mathbb{K}_{q\theta}$ is solved using the spectra of $\hat{\mathbf{x}}_q $ and $\hat{\mathbf{b}}_{\theta q}$ via the Wiener-Hopf method~\cite{Ebhardt2009,Wiener1964Extrapolation}. Taking $\chi_q^{-1}$ on both side of Eq.~\eqref{eq:quantum_trajectory}, we obtain
$\langle \mathbf{x}\rangle _c = \chi_m \chi_q^{-1}\mathbb{K}_{\theta q} \mathbf{y}_{\theta q}$.  The presence of $\chi_m$ and $\chi_q^{-1}$ in   $\langle \mathbf{x}\rangle_c$ replaces the $\omega_{qI}$ poles by $\omega_m$, indicating that the quantum trajectory of the conditional expectations still follow standard quantum mechanics. Substituting Eq.\,\eqref{eq:quantum_trajectory} into  
definitions of $\mathbf{x}_{\rm cl}$ and $\mathbf{b}_{\theta{\rm cl}}$, we obtain:
\be\label{eq:in-out-y}
\begin{split}
\mathbf{y}_{\theta}
=\left[\mathbb{I}+\sqrt{M/\hbar}\Lambda\sin\theta\chi_m\mathbb{M}_{\rm SN}\mathbb{K}_{\theta q}\right]\mathbf{y}_{\theta q},.
\end{split}
\ee
Even though the spectrum of $\mathbf{y}_{\theta q}$ has poles  $\sim\pm\omega_{qI}$, the spectrum of measurement data $\mathbf{y}_\theta$ still peaks at $\omega_m$. This is because $\hat{\mathbf{b}}_{q\theta}$ at these frequencies are dominated by contributions from $\hat{\mathbf{a}}_{1\rm in}$, the Wiener filter $\mathbb{K}_{\theta q} (\pm \omega_{qI}-i\gamma) = [\sqrt{M/\hbar}\Lambda\sin\theta]^{-1}$, leading to $\mathbf{y}_\theta \sim \chi_m \chi_q^{-1} \mathbf{y}_{\theta q}$.  In this way, the injection of classical information into gravity, via the Wiener filter,  eliminates the poles near $\omega_{qI}$ and restores them near $\omega_m$. The physical mechanism is clear:  All SN terms in Eq.\eqref{eq:eom} take the form $\propto\mathbf{\hat{x}} - \langle\mathbf{\hat{x}}\rangle_c=\mathbf{\hat{x}}_q - \langle\mathbf{\hat{x}}_q\rangle_c$, where both terms stem from quantum radiation pressure noise at $\omega_q$: $\mathbf{\hat{x}}_q$ is directly driven by it, while $\langle\mathbf{\hat{x}}_q\rangle_c$ comes from filtering $\mathbf{y}_{\theta q}$ which is dominated by the same noise. The Wiener filter minimizes the error of $\mathbf{\hat{x}}_q - \langle\mathbf{\hat{x}}_q\rangle_c$, suppressing the $\omega_q$-peak in the output.

\emph{Signature ---} 
With the Wiener filter solved in Supplementary Material\,(SM), the covariance matrix elements of $\mathbf{y}_{\theta}$ can be derived as:
\be
\begin{split}
S_{y_{\theta +}y_{\theta \mp}}(\omega)=S_{y_{\theta -}y_{\theta \pm}}(\omega)=\xi|M\chi_{m}(\omega)|^2[\mathcal{F}_A(\omega)\mp\mathcal{F}_B(\omega)],\\
\end{split}
\ee
in which we have $\xi=(\cosh2r+\cos{2\theta}\sinh2r)/2$ and 
\be
\begin{split}
\mathcal{F}_{I}(\omega)=\omega^4_{{\rm SN}I}+2(\omega^2-|\beta_{I}|^2)\omega^2_{{\rm SN}I}+|\omega^2-\beta^2_{I}|^2,\\
\end{split}
\ee
where $\beta_I$ are given by 
\begin{equation}
    \beta_I^2 \approx\omega_{qI}^2 +\Lambda^2/(\cot\theta-ie^{-2r}).
\end{equation}

In QG when  $\omega_{{\rm SN}I}=0$, the symmetric arms leads to $\mathcal{F}_{A}(\omega)=\mathcal{F}_{B}(\omega)$ and a null cross-spectrum $S_{y_{\theta +}y_{\theta \mp}}(\omega)=0$. However, a nonzero $\delta \omega_{\rm SN}^2$ in SN theory breaks the symmetry and induces a nonzero cross-spectrum, providing a distinct SN signature shown in Fig.\,\ref{fig:SN_squeezing_spectrum}.

\begin{figure}
\centering
\includegraphics[width=0.5\textwidth]{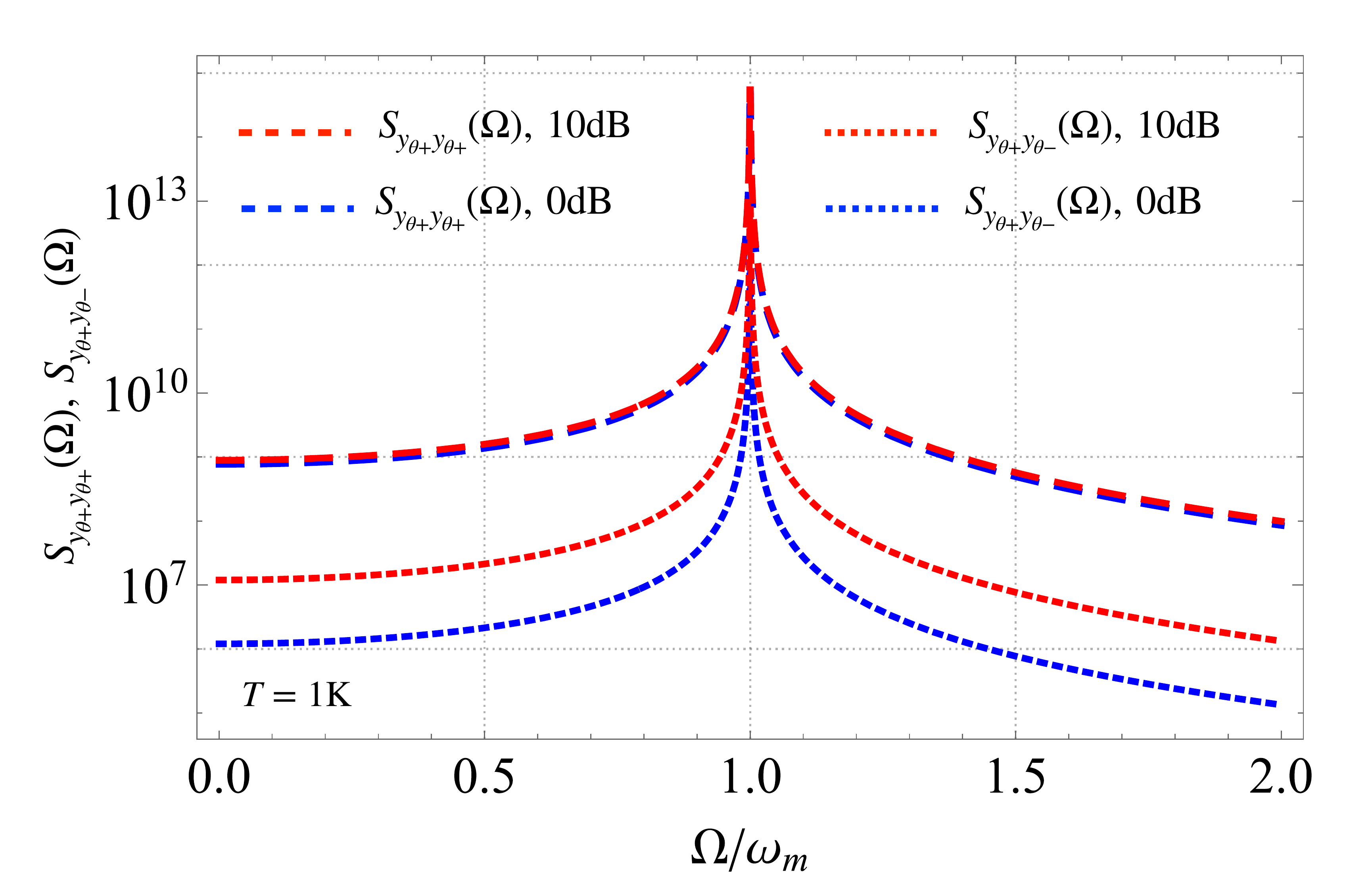}\\
\caption{The diagonal spectrum $S_{y_{\theta \pm}y_{\theta \pm}}(\omega)$\,(dashed) and the cross-spectrum $S_{y_{\theta \pm}y_{\theta \mp}}(\omega)$\,(dotted)  between the common and differential output fields with different input phase-squeezing level, which peak at $\omega_m$. The homodyne measurement angle is $\theta=-0.14\,{\rm rad}$ and $\Lambda/(2\pi)=1\,{\rm Hz}$. }
\label{fig:SN_squeezing_spectrum}
\end{figure}

\begin{table}[h]
    \centering
    \begin{tabular}{|c|c|c|}
    \hline
    \textbf{Parameters} & \textbf{Symbol} & \textbf{Value} \\\hline
    Mirror mass & $M$ & $1$~{\rm kg}\\
    Mirror bare frequency & $\omega_m/2\pi$ & $6$~{\rm mHz}\\
    SN frequency of Mirror A & $\omega_{\rm SNA}/2\pi$ & $7.8$\,mHz\\
    SN frequency of Mirror B & $\omega_{\rm SNB}/2\pi$ & $77$~{\rm mHz}\\
    Mechanical quality factor & $Q_m$ & $10^{7}$\\
    Optical wavelength & $\lambda$ & $1064$~{\rm nm}\\
    Input-mirror power transmissivity & $T_{\rm in}$ & $0.01$\\
    Squeezing level @ mHz &$10\,{\rm Log}[e^{2r}]$& 10\,dB\,\cite{Yaohui2024}\\ 
    Intracavity power & $P_{\rm cav}$ & $4$\,W\\
    Homodyne angle & $\theta$ &  $-0.14$\,rad\\
    Temperature & $T$ & 1\,K\\
    \hline 
    \end{tabular}
  \caption{Sampling parameters for the interferometric protocol targeted at testing SN theory. The material selected for mirrors A/B are silicon/osmium, and the corresponding SN frequencies are computed using $\omega_{\rm SN}= \sqrt{Gm/(6\sqrt{\pi}x_{\rm int}^3})$.}
    \label{tab:parameters}
\end{table}

In practice, the finite measurement time $\mathcal{T}$ introduces the frequency resolution $1/\mathcal{T}$ and spectrum fluctuations among different experimental repetitions. We first convert the phase quadrature $\hat{b}_{\pm\theta}$ to force $\hat{F}_{\pm\theta}(\omega) = \chi^{-1}_m(\omega)\hat{b}_{\pm\theta}(\omega)$ and define the correlation indicator as $\hat C_{\theta}(\omega)=\hat{F}_{+\theta}(\omega)\hat{F}_{-\theta}(\omega)$. The measurement frequency band $[0,\Gamma]$ is chosen to maintain a white spectrum $\hat C_{\theta}(\omega)$, making $\Gamma = \Gamma(\Lambda,\theta,T/Q_m)$ parameter-dependent (see SM). We then define the detection statistic operator $\hat{\chi}_N$ by averaging $\hat{C}_{\theta}(\omega)$ over $N = \Gamma\mathcal{T}$ statistically independent frequency bins within bandwidth $\Gamma$, of which the expectation value and variance are:
\be
\begin{split}
&\langle \hat{\chi}_N \rangle =\frac{1}{N} \sum_{j=1}^N{\mathcal{T}}\int_{\omega_j-\frac{1}{2\mathcal{T}}}^{\omega_j+\frac{1}{2\mathcal{T}}} d\omega|\chi^{-1}_m(\omega)|^{2}S_{y_{\theta +}y_{\theta -}}(\omega),\\
&\text{Var}[\hat{\chi}_N]=\frac{\mathcal{T}^3}{N}\int_{\omega-\frac{1}{2\mathcal{T}}}^{\omega+\frac{1}{2\mathcal{T}}}d\omega|\chi^{-1}_m(\omega)|^{4}S_{y_{\theta +}y_{\theta +}}(\omega)S_{y_{\theta -}y_{\theta -}}(\omega).
\end{split}
\ee
Finally, using this aggregated measurement, the signal-to-noise ratio of SN-induce correlation is defined as  $\text{SNR} = \langle \hat{\chi}_N \rangle/\sqrt{\text{Var}[\hat{\chi}_N]}$. Under the approximation $\Lambda\sin\theta \gg \omega_{qA/B}$, we can simplify the cumbersome relation\,(derived in SM) between the SNR and $\mathcal{T}$ as:
\be\label{eq:time_snr}
\mathcal{T}\approx\frac{(1+\kappa)\sin^2\theta}{{2\Gamma(\Lambda,\theta,T/Q_m)}}\left[\frac{\Lambda^2}{\delta\omega^2_{\rm SN}}+\frac{4n_{\rm th}}{e^{2r}}\frac{\omega^2_m}{\delta\omega^2_{\rm SN}}\right]^2{\rm SNR}^2,
\ee
where $n_{\rm th}=k_BT/(\hbar\omega_mQ_m)$ is the mean thermal occupation.
Eq.~\eqref{eq:time_snr} is derived incorporating both classical thermal noise and classical common-mode noise $S_{f_{p}f_{p}}(\omega)$\,(e.g. radiation pressure noise from laser power fluctuations), characterized by $\kappa \propto S_{f_{p}f_{p}}(\omega)/S_{y_{\theta-}y_{\theta-}}(\omega)$. Using Tab.~\ref{tab:parameters} parameters, Fig.~\ref{fig:time_parameter_dependence} shows $\mathcal{T}$'s parameter dependence (based on exact formulae in SM) at ${\rm SNR}=1$.
\begin{figure}
\centering
\includegraphics[width=0.45\textwidth]{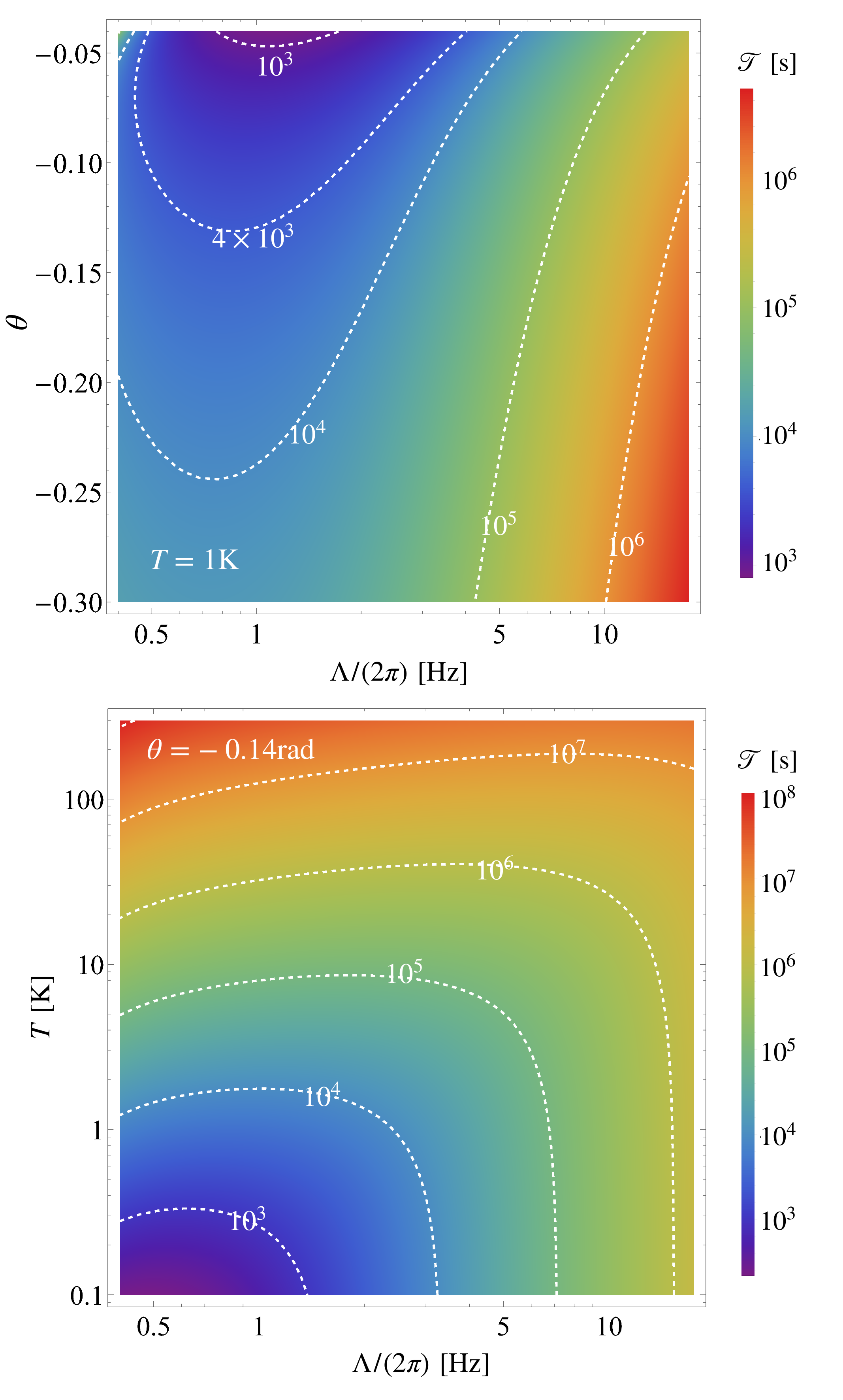}
\caption{Dependence of measurement time $\mathcal{T}$ on the environmental temperature $T$, cooperativity $\Lambda$ and homodyne angle $\theta$, where the signal-to-noise ratio is set to be ${\rm SNR}=1$. The optical input is set to be a 10\,dB phase squeezed vacuum.}
\label{fig:time_parameter_dependence}
\end{figure}

\emph{Features ---}
Methods for enhancing the interferometer’s SN-induced correlation arise from the SN-coupling term in Eq.\,\eqref{eq:eom}: $\propto M\delta\omega_{\rm SN}^2\hat \sigma_x(\mathbf{\hat x}-\langle\mathbf{x}\rangle_c)$. This means selecting appropriate test mass materials and measurement strength\,($\sim\Lambda \sin\theta$) to increase $\delta\omega_{\rm SN}$ and boost the conditional variance  $V^c_{\mathbf{xx}}$, respectively. For instance, we choose silicon and osmium for test masses A and B, resulting in a large $\delta \omega_{\rm SN}/\omega_m\approx 13$. While a higher optical power and a larger homodyne angle (i.e. $\Lambda\sin\theta$) strengthen the measurement, reducing $V^c_{\mathbf{xx}}$ and weakening the SN-coupling. However, the optical power cannot be too weak; otherwise, the displacement signal induced by SN gravity will not imprint on the output optical field, preventing observation of the SN-induced cross-spectrum. Concrete choice of measurement strength depends on the joint optimization considering the parameter-dependent bandwidth and output spectra, shown in Fig.\,\ref{fig:time_parameter_dependence}.  For a fixed measurement strength, phase-squeezed vacuum inputs can amplify quantum radiation pressure noise, thereby increasing $V^c_{\mathbf{xx}}$ and enhancing SN correlations.

Compared to the single-cavity protocol~\cite{Yang2013,Helou2017}, our interferometric approach offers a key advantage by measuring a correlation spectrum rather than an output spectrum. In the single-cavity case, distinguishing between SN gravity and QG requires distinguishing the SN spectrum from a very similar QG spectrum, which requires  highly accurate absolute  calibration\,\cite{Liu2023,Miki2025}.   In contrast, by interferometrically comparing two noise spectra, our protocol provides a clean binary signature: the correlation spectrum exists for classical gravity but vanishes for QG. 

Another feature of the interferometric protocol is the suppression of common-mode noise, which thus does not contaminate the SN-induced correlation signal and the differential-port spectrum $S_{y_{\theta-}y_{\theta-}}(\omega)$. However, it does affect the common-port spectrum as $S_{y_{\theta+}y_{\theta+}}(\omega)\rightarrow (1+\kappa)S_{y_{\theta+}y_{\theta+}}(\omega)$, and will increase the ${\rm Var}[\hat \chi_N]$ by $(1+\kappa)$ (a more complicated configuration for a complete common-mode rejection is presented in SM). While in the single cavity protocol, the ${\rm Var}[\hat \chi_N]$ will be proportional to $(1+\kappa)^2$ since there is only one optical mode. 

\emph{Experimental feasibility ---}
Thermal noise and the interferometer unbalance generated by the instrumental imperfections are key factors that affect the protocol's feasibility. SN theory admits two thermal noise descriptions (see SM) while we adopt the classical prescription in the main text. In this case, the classical thermal force $\mathbf{f}_{\rm th}$ drives both $\mathbf{\hat x}$ and $\langle\mathbf{\hat x}\rangle_c$, canceling in the SN term $\propto \mathbf{\hat x}-\langle\mathbf{\hat x}\rangle_c$. Thus, thermal noise only increases $S_{y_{\theta\pm}y_{\theta\pm}}$ without contributing to the common-differential mode crosstalk, hence degrading the SNR. With the parameters in Tab.\,\ref{tab:parameters}, ${\rm SNR}=1$ requires measurement time $\mathcal{T}\sim 10^4$\,s at temperature $T\approx 1$\,K\,\footnote{For quantum noise prescription, the environmental temperature condition can be further relieved as proved in the Supplementary Material.}, demonstrating a considerably high feasibiltiy. Another challenge in our interferometric protocol arises from arm asymmetries induced by instrumental defects. Given that the SN-induced correlation strength ($\sim 10^{-2}$) is significantly weaker than the background noise (see Fig.~\ref{fig:SN_squeezing_spectrum}), relatively stringent arm-balance tuning is essential to mitigate false positives\,(e.g, for mechanical quality factor $\delta Q/Q<10^{-2}$). In the SM, we provide quantitative estimates of the required parameter fine-tuning precision. Additionally, performing optomechanical experiments at milli-Hertz requires low-frequency noise isolation, while improved configurations (see SM) and ongoing advancements in gravitational wave detection technology make this increasingly attainable\,\cite{Yan2025,Smetana2024,smetana2025,ubhi2022active,prokhorov2023,Mow-Lowry2019,Takano2024,Chua2023}. 

\emph{Conclusion ---}
This work proposes a novel protocol to test the quantumness of gravity by searching for the additional confinement of an object's CoM wavefunction due to classical gravity, an effect described by the Schr\"odinger-Newton equation. In this protocol, SN theory predicts symmetry breaking between common and differential optomechanical modes—a signature absent in quantum gravity. In the quantum gravity case when the A/B arms are completely symmetric, the spectra of common/differential mode output fields $\hat b_\pm=(\hat b_A\pm\hat b_B)/\sqrt{2}$ are in principle identical since the symmetry preserves the independence of $\hat b_A$ and $\hat b_B$. In SN gravity, the symmetry-breaking creates the correlation between the A/B mode, hence manifesting as a non-zero difference between the spectra of the common and differential outputs. Such a spectrum difference serves as a steady binary correlation signature, considerably improving experimental feasibility over prior approaches.  With a boost by 10\,dB input phase squeezed vacuum and experimentally accessible parameters in Tab.\,\ref{tab:parameters}, collecting 3\,hours of aggregated data could achieve sufficient SNR to conclusively test SN gravity, thereby providing strong evidence for the gravity's quantum nature. Moreover, this configuration's similarity to LIGO-type gravitational wave detectors enables leveraging existing interferometer control and noise suppression technologies, facilitating near-future implementation.

\acknowledgements
Y.\,L. and Y.\,M want to thank Dr.\,Haocun Yu for helpful discussion on homodyne measurements. Y.\,M. thanks Shun Wang and Haixing Miao for discussion on the test mass and the laser frequency noise. Y.\,M. and Y.\,L. are supported by the National Key R$\&$D Program of China (2023YFC2205801), National Natural Science Foundation of China under Grant No.12474481, and the start-up funding provided by Huazhong University of Science and Technology. Y. C. is supported by Simons Foundation\,(Award Number 568762).

\bibliographystyle{unsrt}
\bibliography{causal-conditional}

\begin{thebibliography}{10}

\bibitem{Rosenfeld1963}
L.~{Rosenfeld}.
\newblock {On quantization of fields}.
\newblock {\em Nuclear Physics}, 40:353--356, February 1963.

\bibitem{Mueller1962}
{Mueller C}.
\newblock {\em {Les The'ories Relativistes de la Gravitation (Colloques
  Internationaux CNRS), edited by A Lichnerowicz and M-A Tonnelat}}.
\newblock Paris: CNRS, 1962.

\bibitem{Yang2013}
Huan Yang, Haixing Miao, Da-Shin Lee, Bassam Helou, and Yanbei Chen.
\newblock Macroscopic quantum mechanics in a classical spacetime.
\newblock {\em Phys. Rev. Lett.}, 110:170401, Apr 2013.

\bibitem{Bahrami_2014}
Mohammad Bahrami, Andr{\'{e}} Gro{\ss}ardt, Sandro Donadi, and Angelo Bassi.
\newblock The schrödinger{\textendash}newton equation and its foundations.
\newblock {\em New Journal of Physics}, 16(11):115007, nov 2014.

\bibitem{Bose2025}
Sougato Bose, Ivette Fuentes, Andrew~A. Geraci, Saba~Mehsar Khan, Sofia
  Qvarfort, Markus Rademacher, Muddassar Rashid, Marko
  Toro\ifmmode~\check{s}\else \v{s}\fi{}, Hendrik Ulbricht, and Clara~C.
  Wanjura.
\newblock Massive quantum systems as interfaces of quantum mechanics and
  gravity.
\newblock {\em Rev. Mod. Phys.}, 97:015003, Feb 2025.

\bibitem{Chen_2013}
Yanbei Chen.
\newblock Macroscopic quantum mechanics: theory and experimental concepts of
  optomechanics.
\newblock {\em Journal of Physics B: Atomic, Molecular and Optical Physics},
  46(10):104001, may 2013.

\bibitem{Aspelmyer2012}
Markus {Aspelmeyer}.
\newblock {Quantum optomechanics: exploring the interface between quantum
  physics and gravity}.
\newblock In {\em APS Division of Atomic, Molecular and Optical Physics Meeting
  Abstracts}, volume~43 of {\em APS Meeting Abstracts}, page C6.004, June 2012.

\bibitem{Aspelmyer2014}
Markus {Aspelmeyer}, Tobias~J. {Kippenberg}, and Florian {Marquardt}.
\newblock {Cavity optomechanics}.
\newblock {\em Reviews of Modern Physics}, 86(4):1391--1452, October 2014.

\bibitem{Schnabel2015}
Roman Schnabel.
\newblock Einstein-podolsky-rosen--entangled motion of two massive objects.
\newblock {\em Phys. Rev. A}, 92:012126, Jul 2015.

\bibitem{Matsumoto2020}
Seth~B. Cata\~no Lopez, Jordy~G. Santiago-Condori, Keiichi Edamatsu, and
  Nobuyuki Matsumoto.
\newblock High-$q$ milligram-scale monolithic pendulum for quantum-limited
  gravity measurements.
\newblock {\em Phys. Rev. Lett.}, 124:221102, Jun 2020.

\bibitem{Yu2020}
Haocun Yu, L.~McCuller, M.~Tse, N.~Kijbunchoo, L.~Barsotti, N.~Mavalvala, and
  members of~the LIGO Scientific~Collaboration.
\newblock Quantum correlations between light and the kilogram-mass mirrors of
  ligo.
\newblock {\em Nature}, 583(7814):43--47, 2020.

\bibitem{Hoang2016}
Thai~M. Hoang, Yue Ma, Jonghoon Ahn, Jaehoon Bang, F.~Robicheaux, Zhang-Qi Yin,
  and Tongcang Li.
\newblock Torsional optomechanics of a levitated nonspherical nanoparticle.
\newblock {\em Phys. Rev. Lett.}, 117:123604, Sep 2016.

\bibitem{Ando2010}
Masaki Ando, Koji Ishidoshiro, Kazuhiro Yamamoto, Kent Yagi, Wataru Kokuyama,
  Kimio Tsubono, and Akiteru Takamori.
\newblock Torsion-bar antenna for low-frequency gravitational-wave
  observations.
\newblock {\em Phys. Rev. Lett.}, 105:161101, Oct 2010.

\bibitem{ubhi2022active}
Amit~Singh Ubhi, Leonid Prokhorov, Sam Cooper, Chiara~Di Fronzo, John Bryant,
  David Hoyland, Alexandra Mitchell, Jesse Van~Dongen, Conor Mow-Lowry, Alan
  Cumming, et~al.
\newblock Active platform stabilization with a 6d seismometer.
\newblock {\em Applied Physics Letters}, 121(17), 2022.

\bibitem{Mason2019}
David Mason, Junxin Chen, Massimiliano Rossi, Yeghishe Tsaturyan, and Albert
  Schliesser.
\newblock Continuous force and displacement measurement below the standard
  quantum limit.
\newblock {\em Nature Physics}, 15(8):745--749, 2019.

\bibitem{Rossi2018}
Massimiliano Rossi, David Mason, Junxin Chen, Yeghishe Tsaturyan, and Albert
  Schliesser.
\newblock Measurement-based quantum control of mechanical motion.
\newblock {\em Nature}, 563(7729):53--58, 2018.

\bibitem{Aggarwal2020}
Nancy Aggarwal, Torrey~J. Cullen, Jonathan Cripe, Garrett~D. Cole, Robert
  Lanza, Adam Libson, David Follman, Paula Heu, Thomas Corbitt, and Nergis
  Mavalvala.
\newblock Room-temperature optomechanical squeezing.
\newblock {\em Nature Physics}, 16(7):784--788, 2020.

\bibitem{Bose2017}
Sougato Bose, Anupam Mazumdar, Gavin~W. Morley, Hendrik Ulbricht, Marko
  Toro\ifmmode~\check{s}\else \v{s}\fi{}, Mauro Paternostro, Andrew~A. Geraci,
  Peter~F. Barker, M.~S. Kim, and Gerard Milburn.
\newblock Spin entanglement witness for quantum gravity.
\newblock {\em Phys. Rev. Lett.}, 119:240401, Dec 2017.

\bibitem{Marletto2017Gravitationally}
C.~Marletto and V.~Vedral.
\newblock Gravitationally induced entanglement between two massive particles is
  sufficient evidence of quantum effects in gravity.
\newblock {\em Phys. Rev. Lett.}, 119:240402, Dec 2017.

\bibitem{Miao2020}
Haixing Miao, Denis Martynov, Huan Yang, and Animesh Datta.
\newblock Quantum correlations of light mediated by gravity.
\newblock {\em Phys. Rev. A}, 101:063804, Jun 2020.

\bibitem{Christodoulou2023}
Marios Christodoulou, Andrea Di~Biagio, Markus Aspelmeyer, \ifmmode
  \check{C}\else~\v{C}\fi{}aslav Brukner, Carlo Rovelli, and Richard Howl.
\newblock Locally mediated entanglement in linearized quantum gravity.
\newblock {\em Phys. Rev. Lett.}, 130:100202, Mar 2023.

\bibitem{krisnanda2020}
Tanjung Krisnanda, Guo~Yao Tham, Mauro Paternostro, and Tomasz Paterek.
\newblock Observable quantum entanglement due to gravity.
\newblock {\em npj Quantum Information}, 6(1):12, 2020.

\bibitem{Carney2021Using}
Daniel Carney, Holger M\"uller, and Jacob~M. Taylor.
\newblock Using an atom interferometer to infer gravitational entanglement
  generation.
\newblock {\em PRX Quantum}, 2:030330, Aug 2021.

\bibitem{Fujita_2023}
Tomohiro {Fujita}, Youka {Kaku}, Akira {Matumura}, and Yuta {Michimura}.
\newblock {Inverted Oscillators for Testing Gravity-induced Quantum
  Entanglement}.
\newblock {\em arXiv e-prints}, page arXiv:2308.14552, August 2023.

\bibitem{Helou2017}
Bassam Helou, Jun Luo, Hsien-Chi Yeh, Cheng-gang Shao, B.~J.~J. Slagmolen,
  David~E. McClelland, and Yanbei Chen.
\newblock Measurable signatures of quantum mechanics in a classical spacetime.
\newblock {\em Phys. Rev. D}, 96:044008, Aug 2017.

\bibitem{Gan2016Optomechanical}
C.~C. Gan, C.~M. Savage, and S.~Z. Scully.
\newblock Optomechanical tests of a schr\"odinger-newton equation for
  gravitational quantum mechanics.
\newblock {\em Phys. Rev. D}, 93:124049, Jun 2016.

\bibitem{Grossardt2016}
Andr\'e Gro\ss{}ardt, James Bateman, Hendrik Ulbricht, and Angelo Bassi.
\newblock Optomechanical test of the schr\"odinger-newton equation.
\newblock {\em Phys. Rev. D}, 93:096003, May 2016.

\bibitem{Datta_2021}
Animesh Datta and Haixing Miao.
\newblock Signatures of the quantum nature of gravity in the differential
  motion of two masses.
\newblock {\em Quantum Science and Technology}, 6(4):045014, aug 2021.

\bibitem{giulini2011}
Domenico Giulini and Andr{\'e} Gro{\ss}ardt.
\newblock Gravitationally induced inhibitions of dispersion according to the
  schr{\"o}dinger--newton equation.
\newblock {\em Classical and Quantum Gravity}, 28(19):195026, 2011.

\bibitem{giulini2014}
Domenico Giulini and Andr{\'e} Gro{\ss}ardt.
\newblock Centre-of-mass motion in multi-particle schr{\"o}dinger--newton
  dynamics.
\newblock {\em New Journal of Physics}, 16(7):075005, 2014.

\bibitem{scully2022}
Sabina Scully.
\newblock {\em Semiclassical Pictures of Gravity: Investigating and Testing
  Non-Linear Theories of Quantum Gravity}.
\newblock PhD thesis, The Australian National University (Australia), 2022.

\bibitem{Nimmrichter2015}
Stefan Nimmrichter and Klaus Hornberger.
\newblock Stochastic extensions of the regularized schr\"odinger-newton
  equation.
\newblock {\em Phys. Rev. D}, 91:024016, Jan 2015.

\bibitem{Kafri_2014}
D~Kafri, J~M Taylor, and G~J Milburn.
\newblock A classical channel model for gravitational decoherence.
\newblock {\em New Journal of Physics}, 16(6):065020, jun 2014.

\bibitem{Diosi1989}
L.~Di\'osi.
\newblock Models for universal reduction of macroscopic quantum fluctuations.
\newblock {\em Phys. Rev. A}, 40:1165--1174, Aug 1989.

\bibitem{Diosi1998}
Lajos Di\'osi and Jonathan~J. Halliwell.
\newblock Coupling classical and quantum variables using continuous quantum
  measurement theory.
\newblock {\em Phys. Rev. Lett.}, 81:2846--2849, Oct 1998.

\bibitem{kryhin2025}
Serhii Kryhin and Vivishek Sudhir.
\newblock Distinguishable consequence of classical gravity on quantum matter.
\newblock {\em Physical Review Letters}, 134(6):061501, 2025.

\bibitem{oppenheim2023}
Jonathan Oppenheim, Carlo Sparaciari, Barbara {\v{S}}oda, and Zachary
  Weller-Davies.
\newblock Gravitationally induced decoherence vs space-time diffusion: testing
  the quantum nature of gravity.
\newblock {\em Nature Communications}, 14(1):7910, 2023.

\bibitem{Oppenheim2023A}
Jonathan Oppenheim.
\newblock A postquantum theory of classical gravity?
\newblock {\em Phys. Rev. X}, 13:041040, Dec 2023.

\bibitem{Tilloy2016}
Antoine Tilloy and Lajos Di\'osi.
\newblock Sourcing semiclassical gravity from spontaneously localized quantum
  matter.
\newblock {\em Phys. Rev. D}, 93:024026, Jan 2016.

\bibitem{Miki2025}
Daisuke Miki, Youka Kaku, Yubao Liu, Yiqiu Ma, and Yanbei Chen.
\newblock The role of quantum measurements when testing the quantum nature of
  gravity.
\newblock {\em arXiv: 2503.11882}, 2025.

\bibitem{Tilloy2024}
Antoine Tilloy.
\newblock {General quantum-classical dynamics as measurement based feedback}.
\newblock {\em SciPost Phys.}, 17:083, 2024.

\bibitem{Doherty1999}
A.~C. Doherty and K.~Jacobs.
\newblock Feedback control of quantum systems using continuous state
  estimation.
\newblock {\em Phys. Rev. A}, 60:2700--2711, Oct 1999.

\bibitem{Liu2023}
Yubao Liu, Haixing Miao, Yanbei Chen, and Yiqiu Ma.
\newblock Semiclassical gravity phenomenology under the causal-conditional
  quantum measurement prescription.
\newblock {\em Phys. Rev. D}, 107:024004, Jan 2023.

\bibitem{Liu2024}
Yubao Liu, Wenjie Zhong, Yanbei Chen, and Yiqiu Ma.
\newblock Semiclassical gravity phenomenology under the causal-conditional
  quantum measurement prescription. ii. heisenberg picture and apparent optical
  entanglement.
\newblock {\em Phys. Rev. D}, 111:062004, Mar 2025.

\bibitem{Diosi2025}
Lajos Diósi.
\newblock Causality violation of schr\"{o}dinger-newton equation: direct test
  on the horizon?, 2025.

\bibitem{Note1}
Note that the application of $\chi _q$ involves a time-domain integral.

\bibitem{Ebhardt2009}
Helge M\"uller-Ebhardt, Henning Rehbein, Chao Li, Yasushi Mino, Kentaro Somiya,
  Roman Schnabel, Karsten Danzmann, and Yanbei Chen.
\newblock Quantum-state preparation and macroscopic entanglement in
  gravitational-wave detectors.
\newblock {\em Phys. Rev. A}, 80:043802, Oct 2009.

\bibitem{Wiener1964Extrapolation}
Norbert Wiener.
\newblock {\em Extrapolation, Interpolation, and Smoothing of Stationary Time
  Series}.
\newblock The MIT Press, 1964.

\bibitem{Yaohui2024}
Li~Gao, Li-ang Zheng, Bo~Lu, Shaoping Shi, Long Tian, and Yaohui Zheng.
\newblock Generation of squeezed vacuum state in the millihertz frequency band.
\newblock {\em Light: Science \& Applications}, 13(1):294, 2024.

\bibitem{Note2}
For quantum noise prescription, the environmental temperature condition can be
  further relieved as proved in the Supplementary Material.

\bibitem{Yan2025}
Tianliang Yan, Leonid Prokhorov, Jiri Smetana, Vincent Boyer, Denis Martynov,
  Yubao Liu, Yiqiu Ma, and Haixing Miao.
\newblock First result for testing semiclassical gravity effect with a torsion
  balance.
\newblock {\em Phys. Rev. D}, 111:082007, Apr 2025.

\bibitem{Smetana2024}
Jiri Smetana, Tianliang Yan, Vincent Boyer, and Denis Martynov.
\newblock A high-finesse suspended interferometric sensor for macroscopic
  quantum mechanics with femtometre sensitivity.
\newblock {\em Sensors}, 24(7), 2024.

\bibitem{smetana2025}
Jiri Smetana, Amit~Singh Ubhi, Emilia Chick, Leonid Prokhorov, John Bryant,
  Artemiy Dmitriev, Alex Gill, Lari Koponen, Haixing Miao, Alan~V. Cumming,
  Giles Hammond, Valery Frolov, Richard Mittleman, Peter Fritchel, and Denis
  Martynov.
\newblock Sensitivity and control of a 6-axis fused-silica seismometer, 2025.

\bibitem{prokhorov2023}
Leonid Prokhorov, Sam Cooper, Amit~Singh Ubhi, Conor Mow-Lowry, John Bryant,
  Artemiy Dmitriev, Chiara~Di Fronzo, Christopher~J. Collins, Alex Gill,
  Alexandra Mitchell, Joscha Heinze, Jiri Smetana, Tianliang Yan, Alan~V.
  Cumming, Giles Hammond, and Denis Martynov.
\newblock Design and sensitivity of a 6-axis seismometer for gravitational wave
  observatories, 2023.

\bibitem{Mow-Lowry2019}
C~M Mow-Lowry and D~Martynov.
\newblock A 6d interferometric inertial isolation system.
\newblock {\em Classical and Quantum Gravity}, 36(24):245006, nov 2019.

\bibitem{Takano2024}
Satoru {Takano}, Tomofumi {Shimoda}, Yuka {Oshima}, Ching~Pin {Ooi}, Perry
  William~Fox {Forsyth}, Mengdi {Cao}, Kentaro {Komori}, Yuta {Michimura},
  Ryosuke {Sugimoto}, Nobuki {Kame}, Shingo {Watada}, Takaaki {Yokozawa},
  Shinji {Miyoki}, Tatsuki {Washimi}, and Masaki {Ando}.
\newblock {TOrsion-Bar Antenna: A Ground-Based Detector for Low-Frequency
  Gravity Gradient Measurement}.
\newblock {\em Galaxies}, 12(6):78, November 2024.

\bibitem{Chua2023}
S.~S.~Y. Chua, N.~A. Holland, P.~W.~F. Forsyth, A.~Kulur~Ramamohan, Y.~Zhang,
  J.~Wright, D.~A. Shaddock, D.~E. McClelland, and B.~J.~J. Slagmolen.
\newblock {The torsion pendulum dual oscillator for low-frequency Newtonian
  noise detection}.
\newblock {\em Appl. Phys. Lett.}, 122(20):201102, 2023.

\bibitem{Carney2022}
Daniel Carney.
\newblock Newton, entanglement, and the graviton.
\newblock {\em Phys. Rev. D}, 105:024029, Jan 2022.

\bibitem{Danilishin2013}
Stefan~L. Danilishin, Haixing Miao, Mueller~Joerg Helge, and Yanbei Chen.
\newblock Optomechanical entanglement: How to prepare, verify and "steer" a
  macroscopic mechanical quantum state?
\newblock 2013.

\end{thebibliography}

\clearpage

\suppmaterial

\begin{suppabstract}
In this Supplementary Material, we will present a detailed analysis of interferometric configuration for probing the quantum nature of gravity via testing phenomenology predicted by semi-classical gravity theory.  We will first compare our interferometric protocol with the protocol that is used to probe gravity-induced entanglement, which share some appearing similarities but are fundamentally different. Then we present the formalism of our analysis, which is called the causal conditional formalism, and the multichannel Wiener filter to solve the causal conditional dynamics in the Schr{\"o}dinger-Newton theory. We also discuss in detail the aggregated measurement scheme for extracting the Schr{\"o}dinger-Newton signature. The thermal noise effect in our protocol is also presented with two different prescriptions\,(classical vs quantum)  in the SN theory. Finally, we also present the analysis of the influence of the interferometer unbalance due to the instrumental imperfections.
\end{suppabstract}

\section{Comparison of interferometric configuration with the protocols for probing gravity-induced entanglement}
The system consists of two mechanical degrees of freedom\,(d.o.f)  optomechanically interact with two optical cavities, the Hamiltonian can be written as:
\be
\begin{split}
&\hat H=\hat H_m+\hat H_{\rm cav}+\hat H_{\rm om}+\hat H_{\rm ext},\\
&\hat H_m=\sum_{i=A,B}\frac{\hat p^2_{i}}{2M}+\frac{1}{2}M\omega_m^2\hat x^2_{i}+\frac{1}{2}M\omega^2_{{\rm SN}i}(\hat{x}_{i}-\langle\psi|\hat{x}_{i}|\psi\rangle)^2,\\
&\hat H_{\rm cav}=\hbar \omega_c(\hat a^\dag_{A}\hat a_{A}+\hat a^\dag_{B}\hat a_{B}),\,\hat H_{\rm om}=-\hbar g(\hat a_{1A}\hat x_A+\hat a_{1B}\hat x_B),\\
&\hat H_{\rm ext}=i\hbar\sqrt{2\gamma}(\hat a^\dag_A\hat a_{A{\rm in}}+\hat a^\dag_B\hat a_{B{\rm in}}-h.c),
\end{split}
\ee
where $\hat a_{A/B}, \hat a_{A/B\rm in}$ are the annihilation operators of two arm cavity fields and two input fields after the beam splitter\,(BS). The $\gamma$ denotes the bandwidth of the arm cavity, and $g$ is the optomechanical interaction strength in the arm A/B. The $\hat a_{A/B 1}=(\hat a_{A/B}+\hat a^\dag_{A/B})/\sqrt{2},\hat a_{A/B 2}=(\hat a_{A/B}-\hat a^\dag_{A/B})/\sqrt{2}i$ are the amplitude and phase quadrature operators of the cavity fields. The BS relates the $\hat a_{A/B \rm in},\hat a_{A/B \rm out}$  with optical fields at the common and differential ports as $\hat a_{A/B{ \rm in}}=(\hat a_{{\rm in}+}\pm\hat a_{{\rm in}-})/\sqrt{2}$.

The system dynamical d.o.f.s can be transformed into the common and differential modes, and the mechanical Hamiltonian can be rewritten as:
\be\label{eq:Hamiltonian_CD}
\hat H_m=\sum_{i=+,-}\left[\frac{\hat p^2_{i}}{2M}+\frac{1}{2}M\omega_m^2\hat x^2_{i}+\frac{1}{2}M\omega^2_{{\rm SN}}(\hat{x}_{i}-\langle\psi|\hat{x}_{i}|\psi\rangle)^2\right]+\frac{1}{2}M{\delta \omega}^2_{\rm SN}(\hat{x}_{+}-\langle\psi|\hat{x}_{+}|\psi\rangle)(\hat{x}_{-}-\langle\psi|\hat{x}_{-}|\psi\rangle),
\ee
where $\hat x_\pm=(\hat x_A\pm \hat x_B)/\sqrt{2}$ and $\hat p_\pm=(\hat p_A\pm \hat p_B)/\sqrt{2}$ are the displacement and momentum operators of the common/differential mechanical modes, preserving the $[\hat x_{\pm},\hat p_{\pm}]=i\hbar$. For the SN term, we define $\omega_{\rm SN}^2\equiv(\omega_{{\rm SN} A}^2+\omega_{{\rm SN} B}^2)/2$ and $\delta\omega_{\rm SN}^2\equiv\omega_{{\rm SN} A}^2-\omega_{{\rm SN} B}^2$. 
The optical and optomechanical Hamiltonian can be similarly revised as:
\be
\begin{split}
&\hat H_{\rm cav}=\hbar\omega_c(\hat a^\dag_+\hat a_++\hat a^\dag_-\hat a_-),\, \hat H_{\rm om}
=-\hbar g(\hat a_{1+}\hat x_++\hat a_{1-}\hat x_-),\\
&\hat H_{\rm ext}=i\hbar\sqrt{2\gamma}(\hat a^\dag_+\hat a_{+{\rm in}}+\hat a^\dag_-\hat a_{-{\rm in}}-h.c).
\end{split}
\ee

\begin{figure}[h]
\centering
\includegraphics[width=0.6\textwidth]{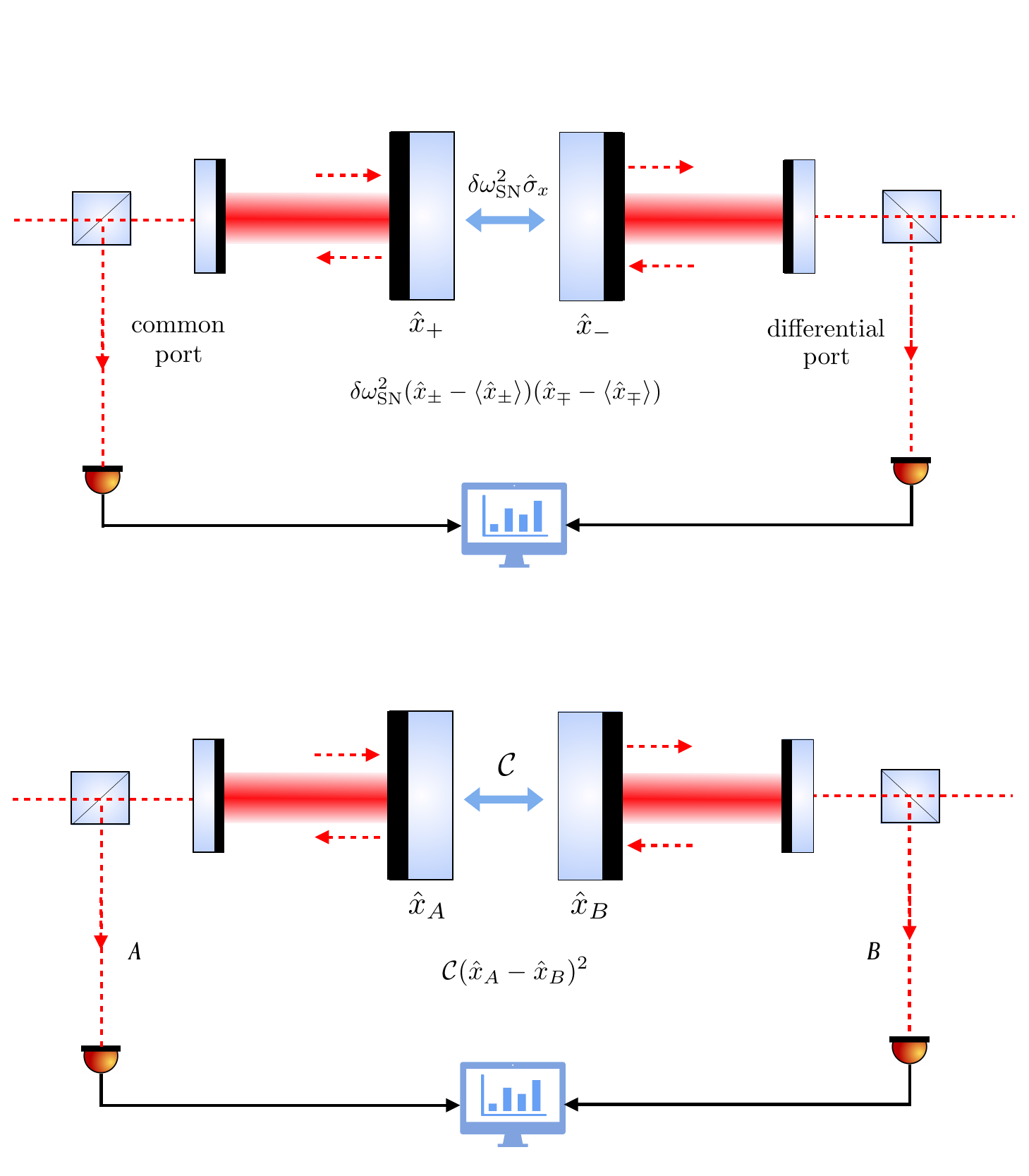}
\caption{Similarity and difference between the optomechanical protocols for testing semi-classical gravity and probing the gravity-induced entanglement.  }
\label{fig:effective_mode_SM}
\end{figure}

This Hamiltonian reminds of the Hamiltonian that describes the protocol for probing the gravity-induced entanglement discussed in\,\cite{Miao2020,Carney2022,Liu2023}:
\be
\hat H_{m,\rm QG}=\sum_{A/B}\left[\frac{\hat p^2_{A/B}}{2M}+\frac{M\omega_m^2}{2}\hat x^2_{A/B}-\mathcal{C}(\hat x_{A/B}-\hat x_{B/A})^2\right],
\ee
or its Sch{\"o}dinger-Newton version:
\be
\hat H_{m,\rm SN}=\sum_{A/B}\left[\frac{\hat p^2_{A/B}}{2M}+\frac{M\omega_m^2}{2}\hat x^2_{A/B}-\mathcal{C}(\hat x_{A/B}-\langle\hat x_{B/A}\rangle)^2\right],
\ee

Note that although these two protocols share some similarities, particularly in Fig.\,\ref{fig:effective_mode_SM}, the interactions between the mechanical degrees of freedom in these two protocols are fundamentally different.
In our interferometric protocol, the test mass interactions are contributed by the asymmetry of the SN terms, which are essentially the Newtonian \emph{self-gravity} of the two test masses. In contrast, the Newtonian \emph{mutual-gravity} contributes to the test mass interactions between these two masses in the protocol for testing gravity-induced entanglement. Therefore these two interactions are physically distinct from each other.

The differences also manifest in the mathematical form of the interactions. Firstly, the ``interaction strength" In our interferometric protocol is the SN frequency difference $\delta \omega^2_{\rm SN}$, while the interaction strength in the GIE protocol depends on the mutual Newtonian gravity. Secondly, a particular feature in our interferometric protocol is the $\delta \omega^2_{\rm SN}$ couples the two modes' deviations from the conditional expectation values of themselves, respectively. However,  only the differences between the displacement of the two modes are relevant in the interaction process of the GIE protocol.

These differences create sharp differences between the GIE protocol and our interferometric protocol, which is essentially summarised as follows. 

1) Our interferometric protocol is designed to test the SN theory. The predictions of SN theory are the SN terms of the test mass dynamics and the test mass mode coupling induced by the asymmetric SN terms. 
The detection or nondetection of this coupling proves or rules out the semi-classical gravity theory, which is somewhat of a proof-by-contradiction method to probe the quantum nature of gravity.  In distinct, the GIE protocol uses gravity-induced entanglement to directly probe the quantum nature of gravity, although the experimental condition is much more stringent and demanding. 

2) Concretely, semi-classical gravity or quantum gravity decides the \emph{existence or non-existence} of the coupling term in our interferometric protocol. The coupling term in the GIE protocol always exists no matter whether the gravity is a quantum entity or not.

3) Unmeasured environmental effects, such as thermal noise from the quantum origin\,(for detail, see Section III in this supplementary material), that can increase the uncertainty of $\hat x-\langle \hat x\rangle$ will contaminate the signals via quantum decoherence for probing the gravity-induced entanglement. However, these noises would enhance the signals for testing SN theory in our interferometric protocol since the magnitude of the interaction term will increase with the larger uncertainty of $\hat x-\langle \hat x\rangle$.

An illustrative example for demonstrating the correlation between the common and differential modes induced by SN gravity is an extremely asymmetric case, in which we assume that $\omega_m=0$ and $\omega_{\rm SN A}\gg\omega_{\rm SN B}$, we therefore have: $\omega_{\rm SN}=\delta\omega_{\rm SN}$.
Then the equations of motion have the following form:
\be
\begin{split}
&\dot{\hat p}_+=-M\omega^2_{\rm SN}(\hat x_B-\langle \hat x_B\rangle)+\hbar g\hat a_{1+},\\
&\dot{\hat p}_-=+M\omega^2_{\rm SN}(\hat x_B-\langle \hat x_B\rangle)+\hbar g\hat a_{1-},
\end{split}
\ee 
which means that both the common and differential mechanical motion are governed by the same $(\hat x_B-\langle \hat x_B\rangle)$-term. Consequently, there will be strong correlations between these mechanical modes. Hence there will also be strong correlation between the outgoing fields from common and differential ports.

\section{Causal-conditional dynamics of SN theory}
Solving the causal-conditional dynamics of SN theory leads to the dynamics of the displacement operators:
\be
\begin{split}
\hat x_+(\omega)=&\hat x_{q+}(\omega)+\frac{M\omega^2_{\rm SN}\chi^{-1}_q(\omega)-M^2\delta\omega^4_{\rm SN}/4}{\chi^{-1}_{A}(\omega)\chi^{-1}_{B}(\omega)}\langle \hat x_+(\omega)\rangle+\frac{\chi^{-1}_m(\omega)M\delta\omega^2_{\rm SN}/2}{\chi^{-1}_{A}(\omega)\chi^{-1}_{B}(\omega)}\langle \hat x_-(\omega)\rangle,\\
&\hat x_{q+}(\omega)=\frac{\chi^{-1}_q(\omega)}{\chi^{-1}_{A}(\omega)\chi^{-1}_{B}(\omega)}\hbar g \hat a_{1+}(\omega)-\frac{M\delta\omega^2_{\rm SN}/2}{\chi^{-1}_{A}(\omega)\chi^{-1}_{B}(\omega)}\hbar g\hat a_{1-}(\omega),\\
\hat x_-(\omega)=&\hat x_{q-}(\omega)+\frac{M\omega^2_{\rm SN}\chi^{-1}_q(\omega)-M^2\delta\omega^4_{\rm SN}/4}{\chi^{-1}_{A}(\omega)\chi^{-1}_{B}(\omega)}\langle \hat x_-(\omega)\rangle+\frac{\chi^{-1}_m(\omega)M\delta\omega^2_{\rm SN}/2}{\chi^{-1}_{A}(\omega)\chi^{-1}_{B}(\omega)}\langle \hat x_+(\omega)\rangle,\\
&\hat x_{q-}(\omega)=\frac{\chi^{-1}_q(\omega)}{\chi^{-1}_{A}(\omega)\chi^{-1}_{B}(\omega)}\hbar g \hat a_{1-}(\omega)-\frac{M\delta\omega^2_{\rm SN}/2}{\chi^{-1}_{A}(\omega)\chi^{-1}_{B}(\omega)}\hbar g\hat a_{1+}(\omega),
\end{split}
\ee
where the response functions are defined as:
\be
\begin{split}
&\chi_{A}(\omega)=\frac{1}{-M(\omega^2-\omega_m^2-\omega^2_{\rm SNA}+i\omega\gamma_m)},\\
&\chi_{B}(\omega)=\frac{1}{-M(\omega^2-\omega_m^2-\omega^2_{\rm SNB}+i\omega\gamma_m)},\\
&\chi_{q}(\omega)=\frac{1}{-M(\omega^2-\omega_m^2-\omega^2_{\rm SN}+i\omega\gamma_m)}.
\end{split}
\ee
Taking the conditional mean of both sides, we have:
\be\label{eq:conditional_mean}
\begin{split}
\langle\hat x_+(\omega)\rangle=&\langle\hat x_{q+}(\omega)\rangle+\frac{M\omega^2_{\rm SN}\chi^{-1}_q(\omega)-M^2\delta\omega^4_{\rm SN}/4}{\chi^{-1}_{A}(\omega)\chi^{-1}_{B}(\omega)}\langle \hat x_+(\omega)\rangle+\frac{\chi^{-1}_m(\omega)M\delta\omega^2_{\rm SN}/2}{\chi^{-1}_{A}(\omega)\chi^{-1}_{B}(\omega)}\langle \hat x_-(\omega)\rangle,\\
\langle\hat x_-(\omega)\rangle=&\langle\hat x_{q-}(\omega)\rangle+\frac{M\omega^2_{\rm SN}\chi^{-1}_q(\omega)-M^2\delta\omega^4_{\rm SN}/4}{\chi^{-1}_{A}(\omega)\chi^{-1}_{B}(\omega)}\langle \hat x_-(\omega)\rangle+\frac{\chi^{-1}_m(\omega)M\delta\omega^2_{\rm SN}/2}{\chi^{-1}_{A}(\omega)\chi^{-1}_{B}(\omega)}\langle \hat x_+(\omega)\rangle,\\
\end{split}
\ee
from which we can solve the $\langle\hat x_+(\omega)\rangle$ from $\langle\hat x_{q+}(\omega)\rangle$.
The input-output relations for the common and differential detection channels can also be solved, where the quantum part is:
\be
\begin{split}
\hat b_{\theta_+ q}(\omega)=\sin\theta_+\hat a_{+2}(\omega)+\cos\theta_+ \hat a_{+1}(\omega)+\sqrt{M/\hbar}\Lambda\sin\theta_+\hat x_{q+}(\omega) ,\\
\hat b_{\theta_- q}(\omega)= \sin\theta_-\hat a_{-2}(\omega)+\cos\theta_- \hat a_{-1}(\omega)+\sqrt{M/\hbar}\Lambda\sin\theta_-\hat x_{q-}(\omega).
\end{split}
\ee
where $\Lambda=\sqrt{2\hbar g^2/(M\gamma)}=\sqrt{8\omega_cP_{\rm cav}/(MT_{\rm in}c^2)}$ here is defined as optomechanical cooperativity, where $T_{\rm in}$ is the input-mirror power transmissivity of cavity. The classical part is:
\be
\begin{split}
&\hat b_{+\theta c}(\omega)=\sqrt{\frac{M^3}{\hbar}}\Lambda\sin\theta_+\left[\frac{\omega^2_{\rm SN}\chi^{-1}_q(\omega)-M\delta\omega^4_{\rm SN}/4}{\chi^{-1}_{A}(\omega)\chi^{-1}_{B}(\omega)}\langle \hat x_+(\omega)\rangle+\frac{\chi^{-1}_m(\omega)\delta\omega^2_{\rm SN}/2}{\chi^{-1}_{A}(\omega)\chi^{-1}_{B}(\omega)}\langle \hat x_-(\omega)\rangle\right],\\
&\hat b_{-\theta c}(\omega)=\sqrt{\frac{M^3}{\hbar}}\Lambda\sin\theta_-\left[\frac{\omega^2_{\rm SN}\chi^{-1}_q(\omega)-M\delta\omega^4_{\rm SN}/4}{\chi^{-1}_{A}(\omega)\chi^{-1}_{B}(\omega)}\langle \hat x_-(\omega)\rangle+\frac{\chi^{-1}_m(\omega)\delta\omega^2_{\rm SN}/2}{\chi^{-1}_{A}(\omega)\chi^{-1}_{B}(\omega)}\langle \hat x_+(\omega)\rangle\right].
\end{split}
\ee
Using the solutions of Eq.\,\eqref{eq:conditional_mean}, we can represent the $\hat b_{\pm\theta c}(\omega)$ by the $\langle\hat x_{q+}(\omega)\rangle$. Then the key now is to obtain  $\langle\hat x_{q+}(\omega)\rangle$, which can be solved using the Wiener-filter approach.

\section{Wiener filter}\label{sec:Wiener filter}
The causal conditional expectation value of displacement operators in SN theory represents the theory's non-linear feature. To tackle this nonlinearity in the Heisenberg picture, we developed a Wiener filter method which is extensively discussed in previous work, in which we refer the reader to the reference\,\cite{Liu2024} for details. Applying the method to our configuration here needs multi-channel Wiener filtering since we have two independent measurement channels for the common and differential modes while determining the causal conditional dynamics of each mechanical mode needs information from both measurement channels. In\,\cite{Liu2024}, the single-channel Wiener filter is sufficient since the quantum dynamics of the two optomechanical systems coupled via Newtonian mutual gravity are independent. A similar treatment of multi-channel Wiener filtering can also be found in\,\cite{Danilishin2013}. 

The basic idea of this Wiener filtering method is to solve the following problem:
\be
\left[
\begin{array}{c}
\hat x_{q+}(t)\\
\hat x_{q-}(t)\\
\end{array}
\right]=
\int^t_{-\infty}dt'
\left[
\begin{array}{cc}
K_{q\theta_+\theta_+}(t-t')&K_{q\theta_+\theta_-}(t-t')\\
K_{q\theta_-\theta_+}(t-t')&K_{q\theta_-\theta_-}(t-t')\\
\end{array}
\right]
\left[
\begin{array}{c}
 y_{\theta_+ q}(t')\\
 y_{\theta_- q}(t')\\
\end{array}
\right]+
\left[
\begin{array}{c}
\hat R_{\theta_+q}(t)\\
\hat R_{\theta_-q}(t)\\
\end{array}
\right],
\ee
and
\be
\langle y_{\theta_+\pm q}(t')\hat R_{\theta_+q}(t)\rangle=0,\quad \langle y_{\theta_\pm q}(t')\hat R_{\theta_-q}(t)\rangle=0,
\ee
with $\theta_\pm$ are the homodyne angle for the common/differential mode. The second equation means that the filter provides the best estimation for the conditional mean of the test mass displacement.

Combining the above two equations leads to the Wiener-Hopf equation:
\be
\left[
\begin{array}{cc}
C_{y_{\theta_+ q} x_{q+}}(t)&C_{y_{\theta_- q} x_{q+}}(t)\\
C_{y_{\theta_+ q} x_{q-}}(t)&C_{y_{\theta_- q} x_{q-}}(t)\\
\end{array}
\right]=\int^t_{-\infty}dt'
\left[
\begin{array}{cc}
K_{q\theta_+\theta_+}(t-t')&K_{q\theta_+\theta_-}(t-t')\\
K_{q\theta_-\theta_+}(t-t')&K_{q\theta_-\theta_-}(t-t')\\
\end{array}
\right]
\left[
\begin{array}{cc}
C_{y_{\theta_+ q} y_{\theta_+ q}}(t')&C_{y_{\theta_+ q} y_{\theta_- q}}(t')\\
C_{y_{\theta_- q}y_{\theta_+ q}}(t')&C_{y_{\theta_- q} y_{\theta_- q}}(t')\\
\end{array}
\right],
\ee
and our target is to solve this Wiener-Hopf equation to obtain the optimal filter matrix.
The method to solve this Wiener-Hopf equation is the so-called Wiener-Hopf method in the frequency domain, which is given by:
\be
\left[\left[
\begin{array}{cc}
S_{y_{\theta_+ q} x_{q+}}(\omega)&S_{y_{\theta_- q} x_{q+}}(\omega)\\
S_{y_{\theta_+ q} x_{q-}}(\omega)&S_{y_{\theta_- q} x_{q-}}(\omega)\\
\end{array}
\right]-
\left[
\begin{array}{cc}
K_{q\theta_+\theta_+}(\omega)&K_{q\theta_+\theta_-}(\omega)\\
K_{q\theta_-\theta_+}(\omega)&K_{q\theta_-\theta_-}(\omega)\\
\end{array}
\right]
\left[
\begin{array}{cc}
S_{y_{\theta_+ q} y_{\theta_+ q}}(\omega)&S_{y_{\theta_+ q} y_{\theta_- q}}(\omega)\\
S_{y_{\theta_- q}y_{\theta_+ q}}(\omega)&S_{y_{\theta_- q} y_{\theta_- q}}(\omega)\\
\end{array}
\right]\right]_-=0,
\ee
where $[...]_-$ means that the function inside the curved bracket has no poles on the upper-half complex plane of $\omega$.  
In the component form:
\be\label{eq:Wiener-Hopf_component}
\begin{split}
\left\{
\begin{split}
[S_{y_{\theta_+ q} x_{q+}}(\omega)-K_{q\theta_+\theta_+}(\omega)S_{y_{\theta_+ q} y_{\theta_+ q}}(\omega)-K_{q\theta_+\theta_-}(\omega)S_{y_{\theta_- q}y_{\theta_+ q}}(\omega)]_-=0,\\
[S_{y_{\theta_- q} x_{q+}}(\omega)-K_{q\theta_+\theta_+}(\omega)S_{y_{\theta_+ q} y_{\theta_- q}}(\omega)-K_{q\theta_+\theta_-}(\omega)S_{y_{\theta_- q}y_{\theta_- q}}(\omega)]_-=0;\\
\end{split}\right.\\
\left\{
\begin{split}
[S_{y_{\theta_+ q} x_{q-}}(\omega)-K_{q\theta_-\theta_+}(\omega)S_{y_{\theta_+ q} y_{\theta_+ q}}(\omega)-K_{q\theta_-\theta_-}(\omega)S_{y_{\theta_- q}y_{\theta_- q}}(\omega)]_-=0,\\
[S_{y_{\theta_- q} x_{q-}}(\omega)-K_{q\theta_-\theta_+}(\omega)S_{y_{\theta_+ q} y_{\theta_- q}}(\omega)-K_{q\theta_-\theta_-}(\omega)S_{y_{\theta_- q}y_{\theta_- q}}(\omega)]_-=0;\\
\end{split}\right.
\end{split}
\ee
The equations for $(K_{q\theta_+\theta_+}(\omega),K_{q\theta_+\theta_-}(\omega))$ and that for $(K_{q\theta_-\theta_+}(\omega),K_{q\theta_-\theta_-}(\omega))$ are decoupled and therefore these two sets of filter function can be solved separately. The symmetry of our system indicates that $S_{y_{\theta_+ q} y_{\theta_+ q}}(\omega)=S_{y_{\theta_- q} y_{\theta_- q}}(\omega), S_{y_{\theta_- q} y_{\theta_- q}}(\omega)=S_{y_{\theta_- q}y_{\theta_+ q}}(\omega)$, with $K_{q\theta_+\theta_-}(\omega)=K_{q\theta_-\theta_+}(\omega), K_{q\theta_+\theta_+}(\omega)=K_{q\theta_-\theta_-}(\omega)$. Therefore,  we only need to solve one set of Eq.\,\eqref{eq:Wiener-Hopf_component} to obtain the full solution of the Wiener filter matrix. In the following subsection, we will discuss the results for two different cases: optical quadrature measurement at equal and unequal homodyne angles.

\subsection{Wiener filter with equal measurement angles}\label{sec:analytical_method}
In the case of the equal homodyne angle $\theta_+=\theta_-=\theta$, and making use of the symmetry, we have:
\be\label{eq:Wiener-Hopf_equal}
\left\{
\begin{split}
[S_{y_{\theta q+} x_{q+}}(\omega)-K_{q\theta++}(\omega)S_{y_{\theta q+} y_{\theta q+}}(\omega)-K_{q\theta+-}(\omega)S_{y_{\theta q-}y_{\theta q+}}(\omega)]_-=0,\\
[S_{y_{\theta q-} x_{q+}}(\omega)-K_{q\theta++}(\omega)S_{y_{\theta q-} y_{\theta q+}}(\omega)-K_{q\theta+-}(\omega)S_{y_{\theta q+}y_{\theta q+}}(\omega)]_-=0;\\
\end{split}\right.
\ee
and we then add and subtract the above two equations:
\be
\left\{
\begin{split}
[[S_{y_{\theta q+} x_{q+}}(\omega)+S_{y_{\theta q-} x_{q+}}(\omega)]-[K_{q\theta++}(\omega)+K_{q\theta+-}(\omega)][S_{y_{\theta q+} y_{\theta q+}}(\omega)+S_{y_{\theta q-} y_{\theta q+}}(\omega)]]_-=0,\\
[[S_{y_{\theta q+} x_{q+}}(\omega)-S_{y_{\theta q-} x_{q+}}(\omega)]-[K_{q\theta++}(\omega)-K_{q\theta+-}(\omega)][S_{y_{\theta q+} y_{\theta q+}}(\omega)-S_{y_{\theta q-} y_{\theta q+}}(\omega)]]_-=0.
\end{split}\right.
\ee
These two equations are the Wiener-Hopf equation for a single measurement channel discussed in\,\cite{Liu2024}. Finally, the Wiener filter can be written as:
\be\label{eq:wiener_filter_quantum}
\begin{split}
&K_{q\theta++}(\omega)+K_{q\theta+-}(\omega)=\sqrt{\frac{\hbar}{M}}\frac{(\omega-\beta_A)(\omega+\beta_A^*)+(\omega^2_{qA}-\omega^2-i\gamma_m\omega)}{\Lambda\sin\theta(\omega-\beta_A)(\omega+\beta_A^*)}=K_{q\theta A}(\omega),\\
&K_{q\theta++}(\omega)-K_{q\theta+-}(\omega)=\sqrt{\frac{\hbar}{M}}\frac{(\omega-\beta_B)(\omega+\beta_B^*)+(\omega^2_{qB}-\omega^2-i\gamma_m\omega)}{\Lambda\sin\theta(\omega-\beta_B)(\omega+\beta_B^*)}=K_{q\theta B}(\omega),
\end{split}
\ee
where,$\omega_{qA/B}=\sqrt{\omega_m^2+\omega^2_{\rm SNA/B}}$ and the $\beta_{A/B}$,  including the influence of squeezed light with squeezing degree $r$ and squeezing angle $\phi$, are:
\begin{equation}\label{eq:squeezing_beta}
\begin{split}
&\beta_{A/B}=\sqrt{\frac{c_1+\sqrt{c_2}}{4\sqrt{\xi}}}+i\sqrt{\frac{c_1+\sqrt{c_2}}{4\sqrt{\xi}}}\times\\
&\frac{\cosh2r[(\gamma_m^2-2\omega^2_{qA/B})\sqrt{\xi}-\Lambda^2\sin2\theta\sqrt{\xi}+\sqrt{c_2}]+\sinh2r[\cos2(\theta-\phi)[(\gamma_m^2-\omega^2_{qA/B})\sqrt{\xi}+\sqrt{c_2}]-2\Lambda^2\cos(\theta-2\phi)\sin\theta\sqrt{\xi}]}{\sqrt{\xi c_3}},
\end{split}
\end{equation}
where
\begin{equation}\label{eq:squeezing_beta1}
\begin{split}
\xi&=\cosh2r+\cos2(\theta-\phi)\sinh2r,\quad c_1=-\gamma_m^2+2\omega_{qA}^2+\Lambda^2\sin2\theta+\frac{2}{\xi}\Lambda^2\sin^2\theta\sin2(\theta-\phi)\sinh2r,\\
c_2&=2\cosh2r(\Lambda^4+2\omega^4_{qA/B}-\Lambda^4\cos2\theta+2\Lambda^2\omega^2_{qA/B}\sin2\theta)+2\left(\cos2\phi(\Lambda^4-(\Lambda^4-2\omega^4_{qA/B})\cos2\theta)+2\Lambda^2\omega^2_{qA/B}\sin2\theta\right.\\&\left.+4\omega^2_{qA/B}\sin\theta(\omega^2_{qA/B}\cos\theta+\Lambda^2\sin\theta)\sin2\phi\right)\sinh2r,\\
c_3&=-\gamma_m^4+6\Lambda^4+4\gamma^2_m\omega^2_{qA/B}+2\Lambda^4(\cos4\theta-4\cos2\theta)+\gamma_m^2\left[2\Lambda^2\sin2\theta-3\cosh4r(\gamma_m^2-4\omega^2_{qA/B}-2\Lambda^2\sin2\theta)\right.\\&\left.+(2(4\omega^2_{qA/B}-\gamma_m^2)\cos4(\theta-\phi)+8\Lambda^2\cos(3\theta-4\phi)\sin\theta)\sinh^22r+4((4\omega^2_{qA/B}-\gamma_m^2)\cos2(\theta-\phi)\right.\\&\left.+\Lambda^2(3\cos(\theta-2\phi)+\cos(3\theta-2\phi))\sin\theta)\sinh4r\right].
\end{split}
\end{equation}
For a detailed derivation of the Wiener filter function, we refer the reader to the reference\,\cite{Liu2024}.

\subsection{Wiener filter with unequal measurement angles}\label{sec:numerical_method}
There are no analytical expressions for the Wiener filter with unequal measurement angles, our calculation follows the method presented in\,\cite{Danilishin2013}, and sketched as follows. 
We will start from Eq.\,\eqref{eq:Wiener-Hopf_component}. These two equations are decoupled from each other, therefore, we only focus on solving the first pair of equations. 
\begin{align}
[S_{y_{\theta_+ q} x_{q+}}(\omega)-K_{q\theta_+\theta_+}(\omega)S_{y_{\theta_+ q} y_{\theta_+ q}}(\omega)-K_{q\theta_+\theta_-}(\omega)S_{y_{\theta_- q}y_{\theta_+ q}}(\omega)]_-=0,\,\label{eq:Syplusx}\\
[S_{y_{\theta_- q} x_{q+}}(\omega)-K_{q\theta_+\theta_+}(\omega)S_{y_{\theta_+ q} y_{\theta_- q}}(\omega)-K_{q\theta_+\theta_-}(\omega)S_{y_{\theta_- q}y_{\theta_- q}}(\omega)]_-=0;\label{eq:Syminusx}
\end{align}
Because $C_{y_{\theta_+ q} y_{\theta_+ q}}(\tau)=C_{y_{\theta_+ q} y_{\theta_+ q}}(-\tau)$, its power spectrum can be factorized as:
\be
S_{y_{\theta_+ q} y_{\theta_+ q}}(\omega)=\phi_+(\omega)\phi_-(\omega),
\ee
where the $\phi_\pm(\omega)$ is analytical in the lower or upper half-plane. 

Multiplying Eq.\,\eqref{eq:Syplusx} by a function analytic on the lower-half plane will not change the equality, then we can multiply it by $\phi^{-1}_+(\omega)$ and obtain:
\be
K_{q\theta_+\theta_+}(\omega)=\frac{1}{\phi_-(\omega)}\left[\frac{S_{y_{\theta_+ q} x_{q+}}(\omega)}{\phi_+(\omega)}-\frac{K_{q\theta_+\theta_-}(\omega)S_{y_{\theta_- q}y_{\theta_+ q}}(\omega)}{\phi_+(\omega)}\right]_-,
\ee
where we have made use of the fact that $K_{q\theta_+\theta_+}(\omega)(\tau)\propto\Theta(\tau)$ in the Wiener-Hopf method.
Substituting this $K_{q\theta_+\theta_+}(\omega)$ into the Eq.\,\eqref{eq:Syminusx}:
\be\label{eq:Kplusplus_substitute}
\left[-S_{y_{\theta_- q} x_{q+}}(\omega)+\frac{S_{y_{\theta_+ q} y_{\theta_- q}}(\omega)}{\phi_-(\omega)}\left[\frac{S_{y_{\theta_+ q} x_{q+}}(\omega)}{\phi_+(\omega)}\right]_-
+\frac{S_{y_{\theta_+ q} y_{\theta_- q}}(\omega)}{\phi_-(\omega)}\left[\frac{K_{q\theta_+\theta_-}(\omega)S_{y_{\theta_- q}y_{\theta_+ q}}(\omega)}{\phi_+(\omega)}\right]_+
-K_{q\theta_+\theta_-}(\omega)S^{\rm cond}_{y_{\theta_- q}y_{\theta_- q}}(\omega)\right]_-,
\ee
where
\be
S^{\rm cond}_{y_{\theta_- q}y_{\theta_- q}}(\omega)=S_{y_{\theta_- q}y_{\theta_- q}}(\omega)-
\frac{S_{y_{\theta_+ q} y_{\theta_- q}}(\omega)S_{y_{\theta_- q} y_{\theta_+ q}}(\omega)}{S_{y_{\theta_+ q}y_{\theta_+ q}}(\omega)}=S_{y_{\theta_- q}y_{\theta_- q}}(\omega)-
\frac{S_{y_{\theta_+ q} y_{\theta_- q}}(\omega)S_{y_{\theta_- q} y_{\theta_+ q}}(\omega)}{\phi_+(\omega)\phi_-(\omega)}.
\ee
Since the Fourier transformation of $S^{\rm cond}_{y_{\theta_- q}y_{\theta_- q}}(\omega)$ is also time-symmetric, then we can also decompose it as 
\be
S^{\rm cond}_{y_{\theta_- q}y_{\theta_- q}}(\omega)=\psi_+(\omega)\psi_-(\omega).
\ee
Then if we multiply the Eq.\,\eqref{eq:Kplusplus_substitute} by $\psi^{-1}_+(\omega)$, we obtain:
\be
\begin{split}
&K_{q\theta_+\theta_-}(\omega)=\\
&\frac{1}{\psi_-(\omega)}\left[\frac{1}{\psi_+(\omega)}\left(S_{y_{\theta_- q} x_{q+}}(\omega)-\frac{S_{y_{\theta_+ q} y_{\theta_- q}}(\omega)}{\phi_-(\omega)}\left[\frac{S_{y_{\theta_+ q} x_{q+}}(\omega)}{\phi_+(\omega)}\right]_-
-\frac{S_{y_{\theta_+ q} y_{\theta_- q}}(\omega)}{\phi_-(\omega)}\left[\frac{K_{q\theta_+\theta_-}(\omega)S_{y_{\theta_- q}y_{\theta_+ q}}(\omega)}{\phi_+(\omega)}\right]_+\right)\right]_-,
\end{split}
\ee
which is an equation for $K_{q\theta_+\theta_-}(\omega)$.
\begin{figure*}
\centering
\includegraphics[width=1.0\textwidth]{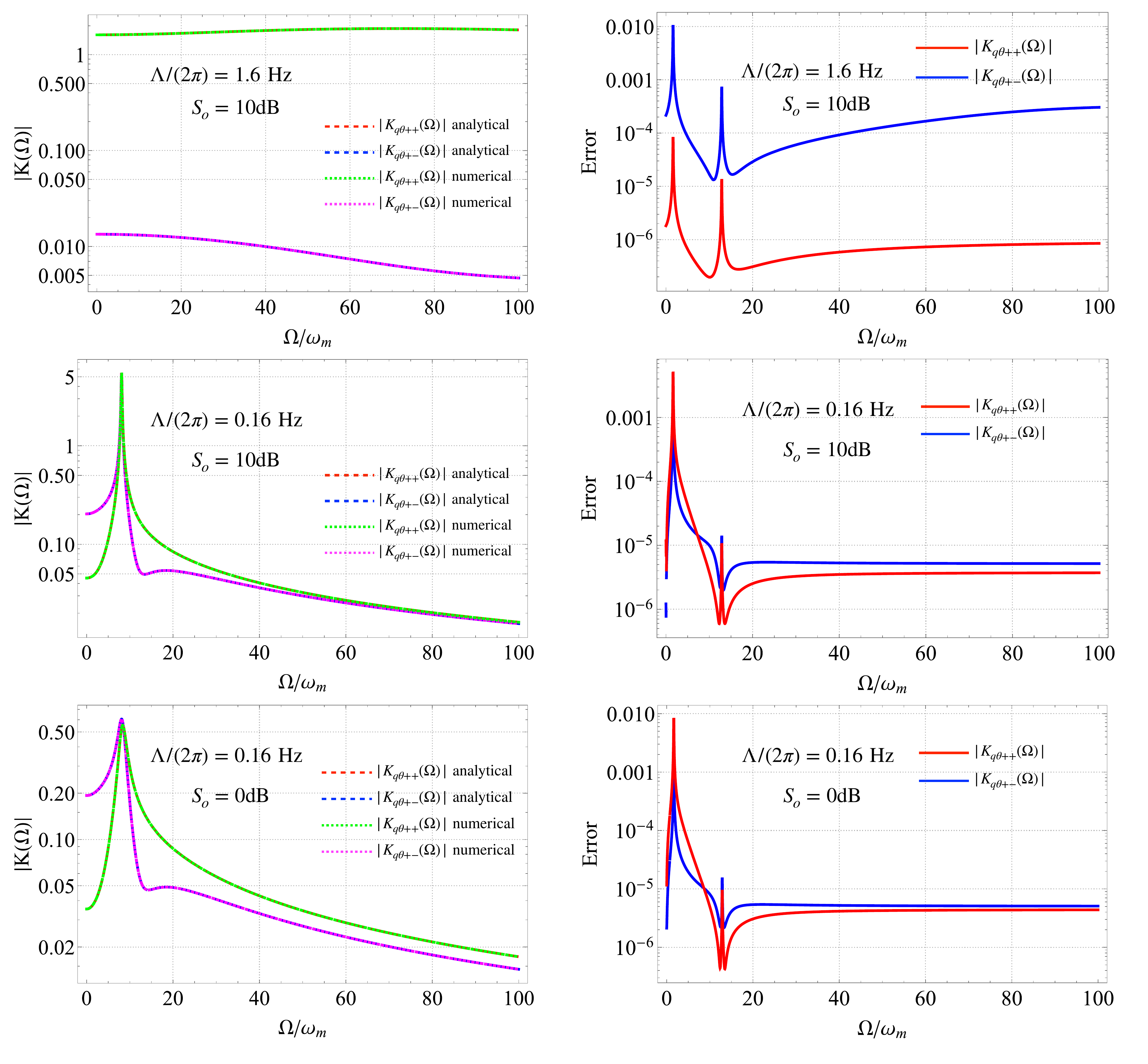}
\caption{Left half-plane: Displays results for several sets of parameters of the filter function obtained using both the analytical method with $\theta_{+}=\theta_{-}$ and a general numerical simulation method. The consistency between the numerical method and the analytical method is evident. Right half-plane: Illustrates the relative error between the two methods, which remains below $10^{-2}$, demonstrating the effectiveness and consistency of both approaches.}
\label{fig:SN_Wiener_filter}
\end{figure*}

The filter functions in the case of equal homodyne angles are illustrated in Fig.\,\ref{fig:SN_Wiener_filter} using the above two methods. The left panel presents the filter function for several sets of system parameters obtained using the analytical method in the Sec.\,\ref{sec:analytical_method} and a general numerical simulation method in the Sec.\,\ref{sec:numerical_method}. This shows that the results of the numerical and analytical methods are consistent. This consistency is manifested by the small relative error between the two methods:
\begin{equation}
{\rm Error}(\Omega)=\left|\frac{K_{\rm ana}(\Omega)-K_{\rm num}(\Omega)}{K_{\rm ana}(\Omega)}\right|<10^{-2},
\end{equation}
where $K_{\rm ana}(\Omega)$ is the filter function solved by the analytical method, and $K_{\rm num}(\Omega)$ is the filter function obtained by the numerical method.


\section{The noise spectrum in the interferometric confuguration}\label{sec:the_spectrum_IV}
The key to calculating the noise spectrum is to connect the expectation value $\langle\hat{x}_{\pm}\rangle$ to the output fields. The Eq.\,\ref{eq:conditional_mean} can rewritten as:

\begin{equation}
\begin{split}
\langle\hat{x}_{+}(\omega)\rangle&=\frac{1}{2}\left[\frac{\chi_m(\omega)}{\chi_{A}(\omega)}+\frac{\chi_m(\omega)}{\chi_{B}(\omega)}\right]\langle\hat{x}_{q+}(\omega)\rangle+\frac{1}{2}\left[\frac{\chi_m(\omega)}{\chi_{A}(\omega)}-\frac{\chi_m(\omega)}{\chi_{B}(\omega)}\right]\langle\hat{x}_{q-}(\omega)\rangle,\\
\langle\hat{x}_{-}(\omega)\rangle&=\frac{1}{2}\left[\frac{\chi_m(\omega)}{\chi_{A}(\omega)}-\frac{\chi_m(\omega)}{\chi_{B}(\omega)}\right]\langle\hat{x}_{q+}(\omega)\rangle+\frac{1}{2}\left[\frac{\chi_m(\omega)}{\chi_{A}(\omega)}+\frac{\chi_m(\omega)}{\chi_{B}(\omega)}\right]\langle\hat{x}_{q-}(\omega)\rangle.
\end{split}
\end{equation}
Using the Weiner filter function, the expectation value can be represented by the quantum part of outgoing fields:
\begin{equation} 
\begin{split}
\langle\hat{x}_{+}(\omega)\rangle&=\left[\frac{\chi_m(\omega)K_{q\theta A}(\omega)}{2\chi_A(\omega)}+\frac{\chi_m(\omega)K_{q\theta B}(\omega)}{2\chi_B(\omega)}\right]y_{\theta +q}+\left[\frac{\chi_m(\omega)K_{q\theta A}(\omega)}{2\chi_A(\omega)}-\frac{\chi_m(\omega)K_{q\theta B}(\omega)}{2\chi_B(\omega)}\right]y_{\theta -q},\\
\langle\hat{x}_{-}(\omega)\rangle&=\left[\frac{\chi_m(\omega)K_{q\theta A}(\omega)}{2\chi_A(\omega)}-\frac{\chi_m(\omega)K_{q\theta B}(\omega)}{2\chi_B(\omega)}\right]y_{\theta +q}+\left[\frac{\chi_m(\omega)K_{q\theta A}(\omega)}{2\chi_A(\omega)}+\frac{\chi_m(\omega)K_{q\theta B}(\omega)}{2\chi_B(\omega)}\right]y_{\theta -q}.
\end{split}
\end{equation} 
The outgoing field can be represented by:
\be
\begin{split}
&\hat b_{+\theta}(\omega)=\hat b_{+\theta q}(\omega)+\sqrt{\frac{M^3}{\hbar}}\Lambda\sin\theta\left[\frac{\omega^2_{\rm SN}\chi^{-1}_q(\omega)-M\delta\omega^4_{\rm SN}/4}{\chi^{-1}_{A}(\omega)\chi^{-1}_{B}(\omega)}\langle \hat x_+(\omega)\rangle+\frac{\chi^{-1}_m(\omega)\delta\omega^2_{\rm SN}/2}{\chi^{-1}_{A}(\omega)\chi^{-1}_{B}(\omega)}\langle \hat x_-(\omega)\rangle\right],\\
&\hat b_{-\theta}(\omega)=\hat b_{-\theta q}(\omega)+\sqrt{\frac{M^3}{\hbar}}\Lambda\sin\theta\left[\frac{\omega^2_{\rm SN}\chi^{-1}_q(\omega)-M\delta\omega^4_{\rm SN}/4}{\chi^{-1}_{A}(\omega)\chi^{-1}_{B}(\omega)}\langle \hat x_-(\omega)\rangle+\frac{\chi^{-1}_m(\omega)\delta\omega^2_{\rm SN}/2}{\chi^{-1}_{A}(\omega)\chi^{-1}_{B}(\omega)}\langle \hat x_+(\omega)\rangle\right].
\end{split}
\ee
The recording data can be calculated as:
\begin{equation}
\begin{split}
y_{\theta +}&=\left[1+\frac{\Lambda\sin\theta}{\chi^{-1}_m(\omega)}\sqrt{\frac{M^3}{4\hbar}}\left(\omega^2_{\rm SNA}K_{q\theta A}(\omega)+\omega^2_{\rm SNB}K_{q\theta B}(\omega)\right)\right]y_{\theta +q}+\left[\frac{\Lambda\sin\theta}{\chi^{-1}_m(\omega)}\sqrt{\frac{M^3}{4\hbar}}\left(\omega^2_{\rm SNA}K_{q\theta A}(\omega)-\omega^2_{\rm SNB}K_{q\theta B}(\omega)\right)\right]y_{\theta -q},\\
y_{\theta -}&=\left[\frac{\Lambda\sin\theta}{\chi^{-1}_m(\omega)}\sqrt{\frac{M^3}{4\hbar}}\left(\omega^2_{\rm SNA}K_{q\theta A}(\omega)-\omega^2_{\rm SNB}K_{q\theta B}(\omega)\right)\right]y_{\theta +q}+\left[1+\frac{\Lambda\sin\theta}{\chi^{-1}_m(\omega)}\sqrt{\frac{M^3}{4\hbar}}\left(\omega^2_{\rm SNA}K_{q\theta A}(\omega)+\omega^2_{\rm SNB}K_{q\theta B}(\omega)\right)\right]y_{\theta -q}.
\end{split}
\end{equation}

Here, we define,
\begin{equation}
C_A(\omega)=\sqrt{\frac{M}{\hbar}}\frac{\Lambda\sin\theta\omega^2_{\rm SNA}K_{q\theta A}(\omega)}{2\chi^{-1}_m(\omega)},\quad C_B(\omega)=\sqrt{\frac{M}{\hbar}}\frac{\Lambda\sin\theta\omega^2_{\rm SNB}K_{q\theta B}(\omega)}{2\chi^{-1}_m(\omega)},
\end{equation}
The diagonal noise spectrum and the cross-spectrum can be obtained as
\begin{equation}\label{eq:spectrum_quantum_spectrum}
\begin{split}
S_{y_{\theta \pm}y_{\theta \pm}}(\omega)=[|1+C_A(\omega)+C_B(\omega)|^2+&|C_A(\omega)-C_B(\omega)|^2]S_{y_{\theta +q}y_{\theta +q}}(\omega)\\&+2{\rm Re}[(1+C_A(\omega)+C_B(\omega))(C^{*}_A(\omega)-C^{*}_B(\omega))]S_{y_{\theta +q}y_{\theta -q}}(\omega),\\
S_{y_{\theta \pm}y_{\theta \mp}}(\omega)=[|1+C_A(\omega)+C_B(\omega)|^2+&|C_A(\omega)-C_B(\omega)|^2]S_{y_{\theta +q}y_{\theta -q}}(\omega)\\
&+2{\rm Re}[(1+C_A(\omega)+C_B(\omega))(C^{*}_A(\omega)-C^{*}_B(\omega))]S_{y_{\theta +q}y_{\theta +q}}(\omega).
\end{split}
\end{equation}
Using the properties
\be
\begin{split}
&S_{y_{\theta q+} y_{\theta q+}}(\omega)+S_{y_{\theta q-} y_{\theta q+}}(\omega)=S_{y_{\theta qA} y_{\theta qA}}(\omega),\\
&S_{y_{\theta q+} y_{\theta q+}}(\omega)-S_{y_{\theta q-} y_{\theta q+}}(\omega)=S_{y_{\theta qB} y_{\theta qB}}(\omega),
\end{split} 
\ee
we can obtain,
\begin{equation}\label{eq:total_spectrum}
\begin{split}
S_{y_{\theta \pm}y_{\theta \pm}}(\omega)&=\left(\frac{1}{2}+2{\rm Re}[C_A(\omega)]+2|C_A(\omega)|^2\right)S_{y_{\theta qA}y_{\theta qA}}(\omega)+\left(\frac{1}{2}+2{\rm Re}[C_B(\omega)]+2|C_B(\omega)|^2\right)S_{y_{\theta qB}y_{\theta qB}}(\omega),\\
S_{y_{\theta \pm}y_{\theta \mp}}(\omega)&=\left(\frac{1}{2}+2{\rm Re}[C_A(\omega)]+2|C_A(\omega)|^2\right)S_{y_{\theta qA}y_{\theta qA}}(\omega)-\left(\frac{1}{2}+2{\rm Re}[C_B(\omega)]+2|C_B(\omega)|^2\right)S_{y_{\theta qB}y_{\theta qB}}(\omega).
\end{split}
\end{equation}

Substituting the Wiener filter derived in the previous section, we can calculate the noise spectrum and cross spectrum of the common and differential modes of the outgoing field:
\begin{equation}
\begin{split}\label{eq:output_spectrum_diag_nondiag}
S_{y_{\theta \pm}y_{\theta \mp}}(\omega)&=M^2\xi|\chi_m(\omega)|^2\frac{\mathcal{F}_A(\omega)-\mathcal{F}_B(\omega)}{2},\\
S_{y_{\theta \pm}y_{\theta \pm}}(\omega)&=M^2\xi|\chi_m(\omega)|^2\frac{\mathcal{F}_A(\omega)+\mathcal{F}_B(\omega)}{2},
\end{split}
\end{equation}
where 
\begin{equation}\label{eq:definition_F}
F_{A/B}(\omega)=(\omega-\beta_{A/B})(\omega+\beta^{*}_{A/B})(\omega+\beta_{A/B})(\omega-\beta^{*}_{A/B})+2(\omega^2-|\beta_{A/B}|^2)\omega^2_{\rm SNA/B}+\omega^4_{\rm SNA/B},
\end{equation}

Using Eq.\,\ref{eq:squeezing_beta} and  Eq.\,\ref{eq:squeezing_beta1}, $|\beta_{A/B}|^2$ can be simplified as:
\begin{equation}
|\beta_{A/B}|^2=\sqrt{\frac{c_2}{4\xi}}=\sqrt{\frac{\sin^2\theta\zeta}{\xi}\Lambda^4+\frac{\tilde{\zeta}}{\xi}\Lambda^2\omega^2_{qA/B}+\omega^4_{qA/B}}.
\end{equation}
where $\zeta=\cosh 2r+\cos 2\phi\sinh 2r$ and $\tilde{\zeta}=\cosh2r\sin2\theta+2\cos(\theta-2\phi)\sin\theta\sinh2r$.
When $\sin\theta\Lambda^2\gg(\sqrt{1+\cos^2\theta}+\cos^2\theta)\omega^2_{qA/B}$, we have the approximated formula:
\begin{equation}
|\beta_{A/B}|^2\approx|\sin\theta|\Lambda^2\sqrt{\frac{\zeta}{\xi}}\left(1+\frac{\tilde{\zeta}}{2\sin^2\theta\zeta}\frac{\omega^2_{qA/B}}{\Lambda^2}+\frac{\xi}{2\sin^2\theta\zeta}\frac{\omega^4_{qA/B}}{\Lambda^4}\right).
\end{equation}

In this case, the cross-spectrum can be written as: 
\begin{equation}\label{eq:cross_spectrum_optical}
\begin{split}
S_{y_{\theta \pm}y_{\theta \mp}}(\omega)&=M^2\xi\frac{\omega^2_{\rm SNA}(\omega^2_{qA}-|\beta_{A}|^2)-\omega^2_{\rm SNB}(\omega^2_{qB}-|\beta_{B}|^2))}{|\chi^{-1}_m(\omega)|^2}+\frac{M^2\delta\omega^2_{\rm SN}\Lambda^2\tilde{\zeta}}{2|\chi^{-1}_m(\omega)|^2}\\&=\frac{M^2\delta\omega^2_{\rm SN}\Lambda^2(\tilde{\zeta}-2|\sin\theta|\sqrt{\zeta\xi})}{2|\chi^{-1}_m(\omega)|^2}+\frac{\left(1-\sqrt{\tilde{\zeta}^2/(\sin^2\theta\zeta\xi)}\right)(\omega^2_{\rm SNA}\omega^2_{qA}-\omega^2_{\rm SNB}\omega^2_{qB})}{2|\chi^{-1}_m(\omega)|^2}\\&-\frac{\sqrt{\xi/(\sin^2\theta\zeta)}}{2|\chi^{-1}_m(\omega)|^2}(\omega^2_{\rm SNA}\frac{\omega^4_{qA}}{\Lambda^2}-\omega^2_{\rm SNB}\frac{\omega^4_{qB}}{\Lambda^2}).
\end{split}
\end{equation}

As discussed in the main text, we should choose $|\sin\theta|\ll1$ to decrease the component of test mass in the outgoing field, which leads to larger conditional variance $V^c_{xx}$ and enhances the SN gravity effect. In this way, we expand $\xi$ and $\tilde{\zeta}$ as:
\begin{equation}
\begin{split}
\xi&=\cosh2r+\cos2(\theta-\phi)\sinh2r\approx(\cosh2r+\cos2\phi\sinh2r)+(2\sin\phi\sinh2r)\sin\theta-(2\cos\phi\sinh2r)\sin^2\theta+{\rm o}(\sin^2\theta)\\
\tilde{\zeta}&=\cosh2r\sin2\theta+2\cos(\theta-2\phi)\sin\theta\sinh2r\approx2(\cosh2r+\cos2\phi\sinh2r)\sin\theta+(2\sin\phi\sinh2r)\sin^2\theta+{\rm o}(\sin^3\theta)
\end{split}
\end{equation}

Substituting the above $\xi$ and $\tilde{\zeta}$ into Eq.\,\ref{eq:cross_spectrum_optical} and keep upto the $\sin^2\theta$ order, we have:

\begin{equation}\label{eq:cross_spectrum_optical}
\begin{split}
S_{y_{\theta \pm}y_{\theta \mp}}(\omega)&=\frac{2M^2\delta\omega^2_{\rm SN}\Lambda^2}{2|\chi^{-1}_m(\omega)|^2}(\sin\theta-|\sin\theta|)(\zeta-\sin\phi\sinh2r\sin\theta)+\frac{\left(\sin^2\theta/\zeta^2-1\right)}{2|\chi^{-1}_m(\omega)|^2}(\omega^2_{SNA}\omega^2_{qA}-\omega^2_{SNB}\omega^2_{qB})\\&-\frac{1}{2|\sin\theta||\chi^{-1}_m(\omega)|^2}\left(\omega^2_{SNA}\frac{\omega^4_{qA}}{\Lambda^2}-\omega^2_{SNB}\frac{\omega^4_{qB}}{\Lambda^2}\right).
\end{split}
\end{equation}
Note that $\sin\theta-|\sin\theta|=0$ for $\theta\in[0,\pi/2]$, hence the $\Lambda^2$-term diminishes when $\theta>0$. It is important to bear in mind that the diminishing of the $\Lambda^2$-term in this case is simply a result of the small-$\theta$ approximation. If we use the exact formula Eq.\,\eqref{eq:output_spectrum_diag_nondiag}, the $\Lambda^2$-term  only has a negligibly small but non-zero contribution. To keep the $\sim\Lambda^2$ term dominated, we choose the homodyne angle $\theta\in[-\pi/2,0]$ (for instance, we choose $\theta$ is around $-0.14\,{\rm rad}$ to keep $|\sin\theta|\ll1$). In addition, when  $|\sin\theta\Lambda^2|\gg\omega^2_{qA/B}$, we can simplify Eq.\,\ref{eq:cross_spectrum_optical} as:

\begin{equation}
S_{y_{\theta \pm}y_{\theta \mp}}(\omega)=\frac{2M^2\delta\omega^2_{\rm SN}\Lambda^2\zeta\sin\theta}{|\chi^{-1}_m(\omega)|^2}.
\end{equation}
 
\begin{figure}
\centering
\includegraphics[width=0.49\textwidth]{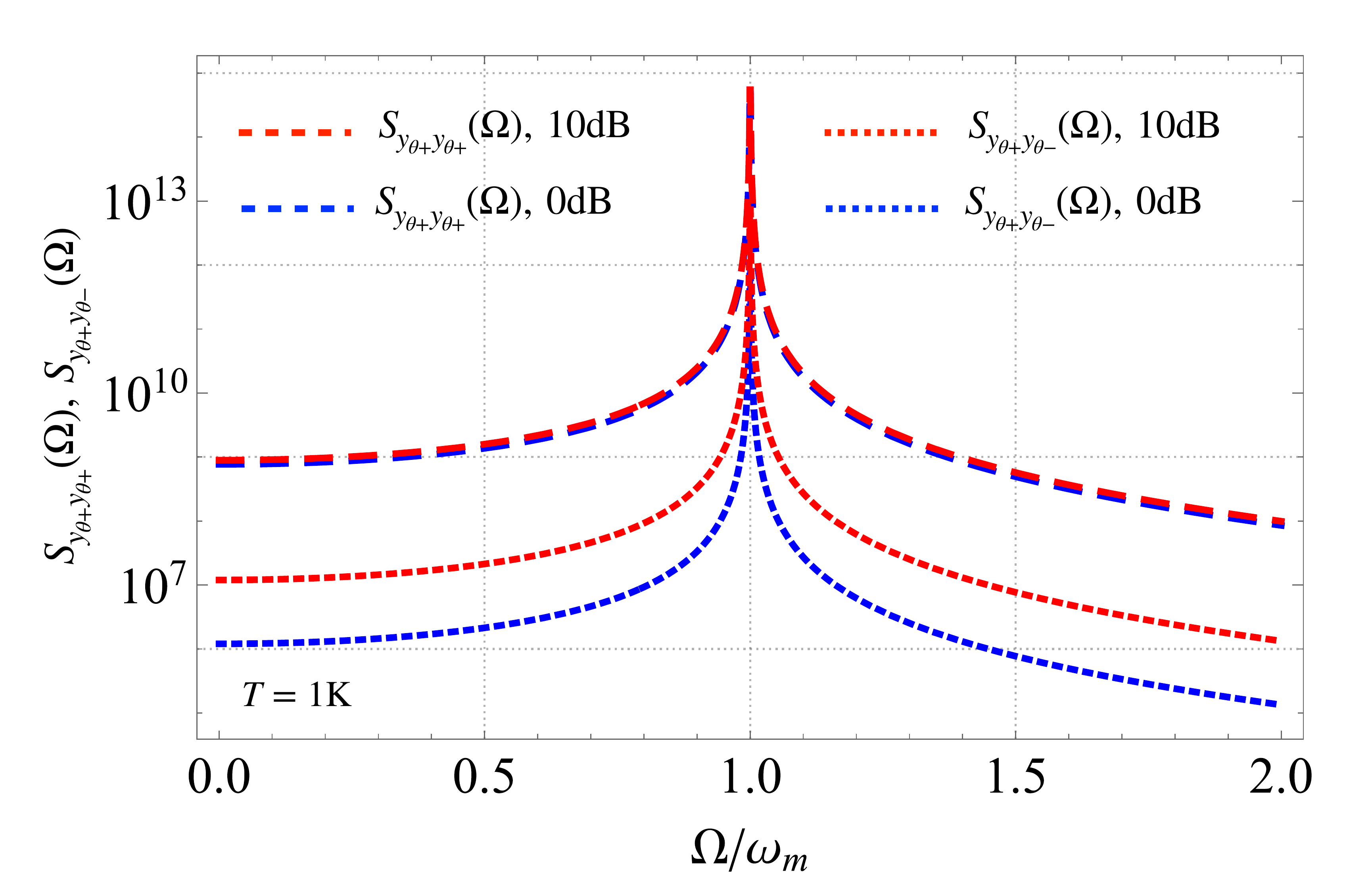}
\includegraphics[width=0.49\textwidth]{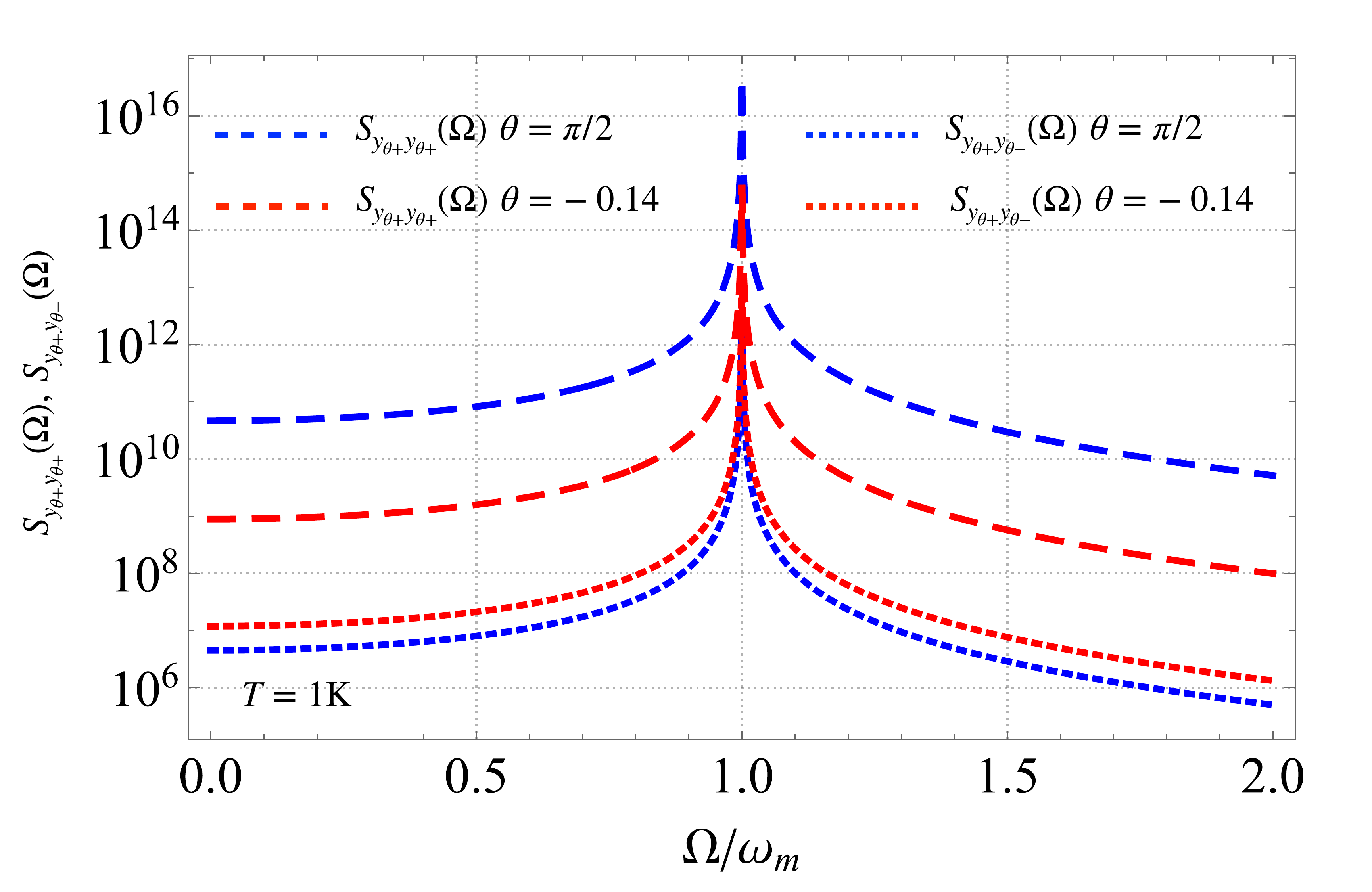}\\
\caption{Noise spectrum of the outgoing field in the interferometric protocol for the SN gravity case. The left panel displays the diagonal spectrum and the cross-spectrum of the light between the common mode and the differential mode with different squeezing of input optical field, each exhibiting a peak located at $\omega_m$. The right panel shows the diagonal spectrum and the cross-spectrum with different homodyne measurement angle $\theta$. }
\label{fig:spectrum_squeese_SM}
\end{figure}

\subsection*{The unequal measurement angles case}
While we chose the equal measurement angle $\theta_{+}=\theta_{-}$ in the above analysis with an analytically solvable Wiener filter function and the noise spectrum, this subsection is devoted to the unequal measurement angles $\theta_{+}\neq\theta_{-}$ case, which is not analytically solvable and can only be numerically analyzed. Since we are interested in the SN gravity-induced correlation between the two output fields in this interferometric protocol, we define the following normalized correlation $\epsilon(\omega)$ as: 
\begin{equation}
\epsilon(\omega)=\sqrt{\frac{S_{y_{\theta +}y_{\theta -}}(\omega)S_{y_{\theta -}y_{\theta +}}(\omega)}{S_{y_{\theta +}y_{\theta +}}(\omega)S_{y_{\theta -}y_{\theta -}}(\omega)}}.
\end{equation}

\begin{figure}
\centering
\includegraphics[width=0.48\textwidth]{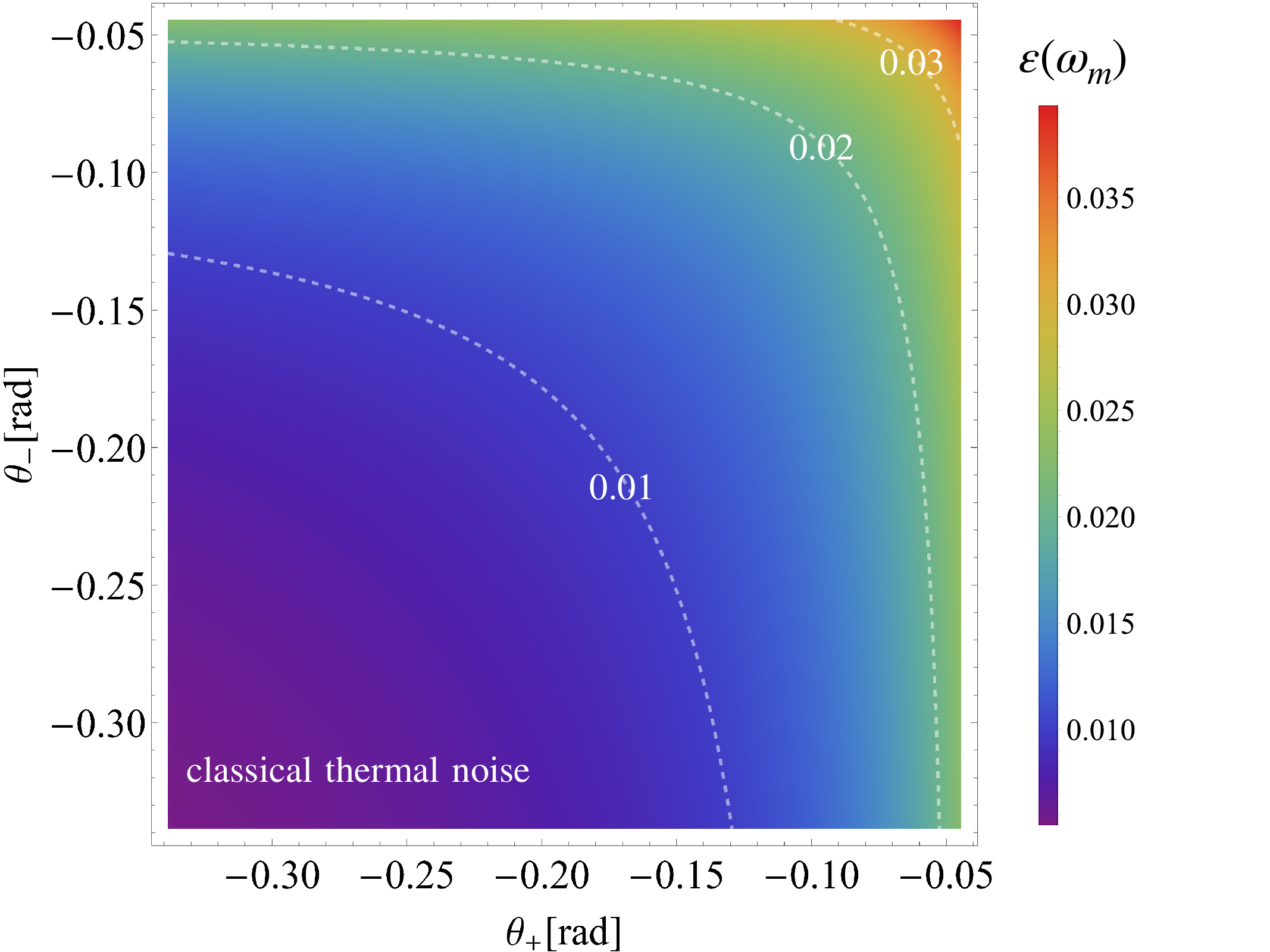}
\includegraphics[width=0.48 \textwidth]{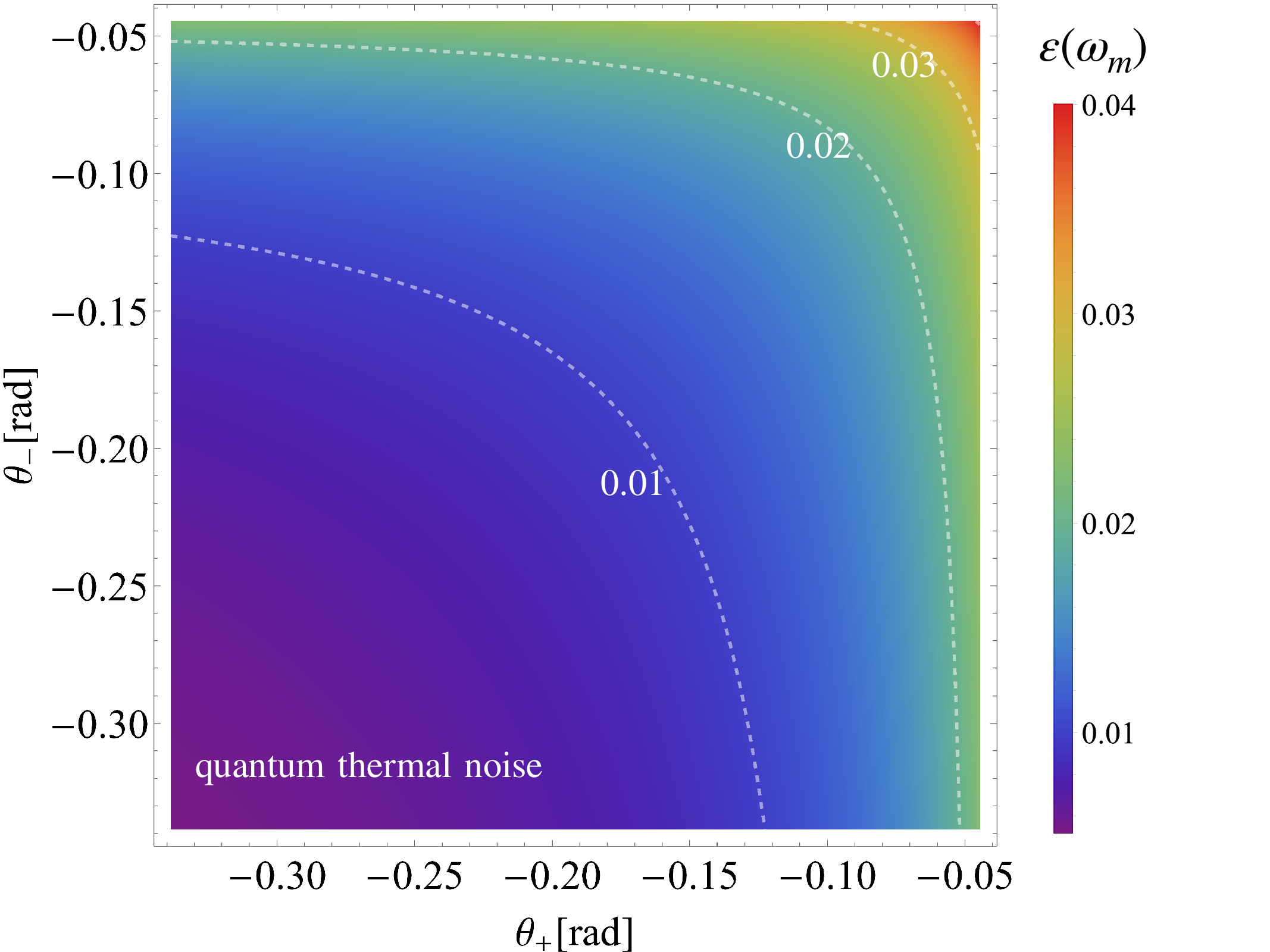}\\
\caption{The normalized correlation level $\epsilon(\omega_m)$ as a function of the homodyne measurement angles $\theta_+,\theta_-$. The contour lines of the $\epsilon(\omega_m)$ intersect along  $\theta_{+}=\theta_{-}$, indicating that the optimal $\epsilon(\omega=\omega_m)$ can be achieved using identical homodyne measurement angles. As an example, we choose $\Lambda/(2\pi)=1\,{\rm Hz}$ and $T=1\,{\rm K}$, for the other $\Lambda$ and $T$, and the normalized correlation level has the same feature.}
\label{fig:theta_theta_squeezing}
\end{figure}

Using the numerical result of the Wiener filter in case $\theta_+\neq\theta_-$, we generate the phase diagram presented in Fig.\,\ref{fig:theta_theta_squeezing}, which illustrates the dependence of the normalized correlation level $\epsilon(\omega_m)$ on the two homodyne angles $\theta_{+}$ and $\theta_{-}$. It shows that all contours of the normalized correlation level intersect the diagonal line $\theta_{+}=\theta_{-}$, indicating that optimal performance of the interferometric protocol can be achieved when $\theta_{+}=\theta_{-}=\theta$. In the subsequent section, we focus on this equal homodyne angle strategy, as it is analytically tractable.



\section{The conditional state of test masses}
The quadratic SN potential's validity limits the test masses' conditional displacement, which is determined by the interaction between the optical fields and the test masses. This limitation will further constrain the parameter space of our proposed interferometric protocol. In this section, the conditional states of the test masses are calculated, which will be used to calculate the boundary of the parameter phase diagram.

\subsection{The conditional covariance of the test masses}
Linear optomechanical interactions and the linear optical quadrature measurement preserve the Gaussianity of the initially prepared quantum Gaussian state. This section discusses the conditional covariance of the test mass displacements.


For example, the conditional variance of the displacement is:
\begin{equation}
V^c_{xx}=\langle(\hat{x}-\langle\hat{x}\rangle_c)^2\rangle,
\end{equation}
which can be rewritten, using the relation $\hat{x}=\hat{x}_q+x_{\rm cl}$ and $\langle\hat{x}\rangle_c=\langle\hat{x}_q\rangle_c+x_{\rm cl}$, as:
\begin{equation}
V^c_{xx}=\langle(\hat{x}_q-\langle\hat{x}_q\rangle_c)^2\rangle.
\end{equation}
For mirror A, we have:
\begin{equation}\label{eq:xAxA}
\begin{split}
V^c_{x_Ax_A}=\langle(\hat{x}_{qA}-\langle\hat{x}_{qA}\rangle_c)^2\rangle&=\frac{1}{2}\left(\langle(\hat{x}_{q+}-\langle\hat{x}_{q+}\rangle_c)^2\rangle+\langle(\hat{x}_{q+}-\langle\hat{x}_{q-}\rangle_c)^2\rangle+\langle(\hat{x}_{q-}-\langle\hat{x}_{q+}\rangle_c)^2\rangle+\langle(\hat{x}_{q-}-\langle\hat{x}_{q-}\rangle_c)^2\rangle\right)\\
&=\frac{1}{2}\left(V^c_{x_{+}x_{+}}+V^c_{x_{+}x_{-}}+V^c_{x_{-}x_{+}}+V^c_{x_{-}x_{-}}\right),
\end{split}
\end{equation}
where 
\begin{equation}
V^c_{x_{+}x_{+}}=\left\langle\left(\hat{x}_{+}(t)-\int_{-\infty}^tK_{q\theta{++}}(t-t')\hat{b}_{q\theta{+}}(t')dt'-\int_{-\infty}^tK_{q\theta{+-}}(t-t')\hat{b}_{q\theta{-}}(t')dt'\right)^2\right\rangle,
\end{equation}
Changing the variable $t'-t=\tau$, the $V_{x_{+}x_{+}}$ can be written as,
\begin{equation}\label{eq:V_x+x+}
V^c_{x_{+}x_{+}}=\left\langle\left(\hat{x}_{+}(t)-\int_{-\infty}^0K_{q\theta{++}}(-\tau)\hat{b}_{q\theta{+}}(t+\tau)d\tau-\int_{-\infty}^0K_{q\theta{+-}}(-\tau)\hat{b}_{q\theta{-}}(t+\tau)d\tau\right)^2\right\rangle.
\end{equation}
Since the Wiener filter $K_{q\theta}(\Omega)$ is analytic in the upper-half complex plane, we can extend the upper limit of the integral from $0$ to $\infty$.  Then in the Fourier domain, we have: 
\begin{equation}\label{eq:Vxpxp}
\begin{split}
V^c_{x_{+}x_{+}}=\int_{-\infty}^{\infty}d\Omega &\left(S_{x_{+}x_{+}}(\Omega)\mathbb{I}
-2[
\begin{array}{cc}
K_{q\theta{++}}(\Omega)&K_{q\theta{+-}}(\Omega)
\end{array}
]
\left[
\begin{array}{c}
S_{x_{+}b^q_{\theta{+}}}(\Omega)\\
S_{x_{+}b^q_{\theta{-}}}(\Omega)
\end{array}
\right]+\right.\\
&\left.\qquad\qquad\qquad\left[
\begin{array}{cc}
K_{q\theta{++}}(\Omega)&K_{q\theta{+-}}(\Omega)
\end{array}
\right]
\left[
\begin{array}{cc}
S_{b^q_{\theta{+}}b^q_{\theta{+}}}(\Omega)&S_{b^q_{\theta{+}}b^q_{\theta{-}}}(\Omega)\\
S_{b^q_{\theta{-}}b^q_{\theta{+}}}(\Omega)&S_{b^q_{\theta{-}}b^q_{\theta{-}}}(\Omega)
\end{array}
\right]
\left[
\begin{array}{c}
K^*_{q\theta{++}}(\Omega)\\
K^*_{q\theta{+-}}(\Omega)
\end{array}
\right]\right),
\end{split}
\end{equation}
and similarly for $V^c_{x_{-}x_{-}},V^c_{x_{+}x_{-}},V^c_{x_{-}x_{+}}$:
\begin{equation}\label{eq:Vxmxm}
\begin{split}
V^c_{x_{-}x_{-}}=\int_{-\infty}^{\infty}d\Omega&\left( S_{x_{-}x_{-}}(\Omega)\mathbb{I}
-2[
\begin{array}{cc}
K_{q\theta{--}}(\Omega)&K_{q\theta{-+}}(\Omega)
\end{array}
]
\left[
\begin{array}{c}
S_{x_{-}b^q_{\theta{-}}}(\Omega)\\
S_{x_{-}b^q_{\theta{+}}}(\Omega)
\end{array}
\right]+\right.\\
&\left.\qquad\qquad\qquad\left[
\begin{array}{cc}
K_{q\theta{--}}(\Omega)&K_{q\theta{-+}}(\Omega)
\end{array}
\right]
\left[
\begin{array}{cc}
S_{b^q_{\theta{-}}b^q_{\theta{-}}}(\Omega)&S_{b^q_{\theta{-}}b^q_{\theta{+}}}(\Omega)\\
S_{b^q_{\theta{+}}b^q_{\theta{-}}}(\Omega)&S_{b^q_{\theta{+}}b^q_{\theta{+}}}(\Omega)
\end{array}
\right]
\left[
\begin{array}{c}
K^*_{q\theta{--}}(\Omega)\\
K^*_{q\theta{-+}}(\Omega)
\end{array}
\right]\right).
\end{split}
\end{equation}

\begin{equation}\label{eq:Vxpxm}
\begin{split}
V^c_{x_{+}x_{-}}=\int_{-\infty}^{\infty}d\Omega &\left( S_{x_{+}x_{-}}(\Omega)\mathbb{I}
-[
\begin{array}{cc}
K_{q\theta{++}}(\Omega)&K_{q\theta{+-}}(\Omega)
\end{array}
]
\left[
\begin{array}{c}
S_{x_{-}b^q_{\theta{+}}}(\Omega)\\
S_{x_{-}b^q_{\theta{-}}}(\Omega)
\end{array}
\right]-[
\begin{array}{cc}
K_{q\theta{-+}}(\Omega)&K_{q\theta{--}}(\Omega)
\end{array}
]
\left[
\begin{array}{c}
S_{x_{+}b^q_{\theta{+}}}(\Omega)\\
S_{x_{+}b^q_{\theta{-}}}(\Omega)
\end{array}
\right]\right.\\
&\left.\qquad\qquad\qquad\left[
\begin{array}{cc}
K_{q\theta{--}}(\Omega)&K_{q\theta{-+}}(\Omega)
\end{array}
\right]
\left[
\begin{array}{cc}
S_{b^q_{\theta{-}}b^q_{\theta{-}}}(\Omega)&S_{b^q_{\theta{-}}b^q_{\theta{+}}}(\Omega)\\
S_{b^q_{\theta{+}}b^q_{\theta{-}}}(\Omega)&S_{b^q_{\theta{+}}b^q_{\theta{+}}}(\Omega)
\end{array}
\right]
\left[
\begin{array}{c}
K^*_{q\theta{+-}}(\Omega)\\
K^*_{q\theta{++}}(\Omega)
\end{array}
\right]\right),
\end{split}
\end{equation}

\begin{equation}\label{eq:Vxmxp}
\begin{split}
V^c_{x_{-}x_{+}}=\int_{-\infty}^{\infty}d\Omega& \left(S_{x_{-}x_{+}}(\Omega)\mathbb{I}
-[
\begin{array}{cc}
K_{q\theta{--}}(\Omega)&K_{q\theta{-+}}(\Omega)
\end{array}
]
\left[
\begin{array}{c}
S_{x_{+}b^q_{\theta{-}}}(\Omega)\\
S_{x_{+}b^q_{\theta{+}}}(\Omega)
\end{array}
\right]-[
\begin{array}{cc}
K_{q\theta{+-}}(\Omega)&K_{q\theta{++}}(\Omega)
\end{array}
]
\left[
\begin{array}{c}
S_{x_{-}b^q_{\theta{-}}}(\Omega)\\
S_{x_{-}b^q_{\theta{+}}}(\Omega)
\end{array}
\right]\right.\\
&\left.\qquad\qquad\qquad+\left[
\begin{array}{cc}
K_{q\theta{++}}(\Omega)&K_{q\theta{+-}}(\Omega)
\end{array}
\right]
\left[
\begin{array}{cc}
S_{b^q_{\theta{+}}b^q_{\theta{+}}}(\Omega)&S_{b^q_{\theta{+}}b^q_{\theta{-}}}(\Omega)\\
S_{b^q_{\theta{-}}b^q_{\theta{+}}}(\Omega)&S_{b^q_{\theta{-}}b^q_{\theta{-}}}(\Omega)
\end{array}
\right]
\left[
\begin{array}{c}
K^*_{q\theta{-+}}(\Omega)\\
K^*_{q\theta{--}}(\Omega)
\end{array}
\right]\right),
\end{split}
\end{equation}

Using previous calculations as shown in Eq.\,\eqref{eq:spectrum_quantum_spectrum}-\,\eqref{eq:output_spectrum_diag_nondiag}, we can find the following relation,
\begin{equation}\label{eq:symetry}
\begin{split}
&S_{x_{+}x_{+}}(\Omega)+S_{x_{+}x_{-}}(\Omega)=S_{x_{-}x_{-}}(\Omega)+S_{x_{-}x_{+}}(\Omega)=S_{x_Ax_A}(\Omega),\\
&S_{x_{+}b^q_{\theta{+}}}(\Omega)+S_{x_{-}b^q_{\theta{+}}}(\Omega)=S_{x_{-}b^q_{\theta{-}}}(\Omega)+S_{x_{+}b^q_{\theta{-}}}(\Omega)=S_{x_Ab^q_{\theta A}}(\Omega),\\
&S_{b^q_{\theta{+}}b^q_{\theta{+}}}(\Omega)+S_{b^q_{\theta{-}}b^q_{\theta{+}}}(\Omega)=S_{b^q_{\theta{-}}b^q_{\theta{-}}}(\Omega)+S_{b^q_{\theta{+}}b^q_{\theta{-}}}(\Omega)=S_{b^q_{\theta A}b^q_{\theta A}}(\Omega),
\end{split}
\end{equation}
and 
\begin{equation}
\begin{split}
K_{q\theta{++}}(\Omega)=K_{q\theta{--}}(\Omega),\quad K_{q\theta{+-}}(\Omega)=K_{q\theta{-+}}(\Omega),\quad K_{q\theta{++}}(\Omega)+K_{q\theta{+-}}(\Omega)=K_{q\theta A}(\Omega).
\end{split}
\end{equation}

Substituting the Eqs.\,\ref{eq:Vxpxp}-\ref{eq:Vxmxm} into the Eq.\,\ref{eq:xAxA},

\begin{equation}\label{eq:VxAxA_2}
V^c_{x_Ax_A}=\int_{-\infty}^{\infty}d\Omega[S_{x_Ax_A}(\Omega)-2K_{q\theta A}(\Omega)S_{x_Ab^q_{\theta A}}(\Omega)+K_{q\theta A}(\Omega)K^{*}_{q\theta A}(\Omega)S_{b^q_{\theta A}b^q_{\theta A}}(\Omega)]
\end{equation}
where
\begin{equation}
K_{q\theta A}(\Omega)=\frac{1}{S^{-}_{b^q_{\theta A}b^q_{\theta A}}(\Omega)}\left[\frac{S_{x_Ab^q_{\theta A}}(\Omega)}{S^{+}_{b^q_{\theta A}b^q_{\theta A}}(\Omega)}\right]_{-}.
\end{equation}
The second term of the integrand in Eq.\,\ref{eq:VxAxA_2} can be written as,
\begin{equation}
\begin{split}
K_{q\theta A}(\Omega)S_{x_Ab^q_{\theta A}}(\Omega)=\frac{S_{x_Ab^q_{\theta A}}(\Omega)}{S^{-}_{b^q_{\theta A}b^q_{\theta A}}(\Omega)}\left[\frac{S_{x_Ab^q_{\theta A}}(\Omega)}{S^{+}_{b^q_{\theta A}b^q_{\theta A}}(\Omega)}\right]_{-}=\left[\frac{S_{x_Ab^q_{\theta A}}(\Omega)}{S^{-}_{b^q_{\theta A}b^q_{\theta A}}(\Omega)}\right]_{+}\left[\frac{S_{x_Ab^q_{\theta A}}(\Omega)}{S^{+}_{b^q_{\theta A}b^q_{\theta A}}(\Omega)}\right]_{-}+\left[\frac{S_{x_Ab^q_{\theta A}}(\Omega)}{S^{-}_{b^q_{\theta A}b^q_{\theta A}}(\Omega)}\right]_{-}\left[\frac{S_{x_Ab^q_{\theta A}}(\Omega)}{S^{+}_{b^q_{\theta A}b^q_{\theta A}}(\Omega)}\right]_{-}.
\end{split}
\end{equation}
The last term of the above equation is analytic in the upper-half complex plane, and the power index of $\Omega$ in the denominator exceeds that in the numerator by two. Therefore, 
\begin{equation}
\int_{-\infty}^{\infty}\left[\frac{S_{x_Ab^q_{\theta A}}(\Omega)}{S^{-}_{b^q_{\theta A}b^q_{\theta A}}(\Omega)}\right]_{-}\left[\frac{S_{x_Ab^q_{\theta A}}(\Omega)}{S^{+}_{b^q_{\theta A}b^q_{\theta A}}(\Omega)}\right]_{-}d\Omega=0.
\end{equation}
The Eq.\,\ref{eq:VxAxA_2} can be simplified as,
\begin{equation}\label{eq:V_xAxA_filter}
V^c_{x_Ax_A}=\int_{-\infty}^{\infty}d\Omega\left(S_{x_Ax_A}(\Omega)-\left|\left[\frac{S_{x_Ab^q_{\theta A}}(\Omega)}{S^{+}_{b^q_{\theta A}b^q_{\theta A}}(\Omega)}\right]_{-}\right|^2\right),
\end{equation}
which has the same form as if the mirrors A and B are measured independently.

Now, the quadratic SN potential condition can be represented by:
\begin{equation}
\Delta x_{A/B}=\sqrt{V^c_{x_{A/B}x_{A/B}}}\ll\Delta x_{A/B\rm int},
\end{equation}
where $\Delta x_{A\rm int}=4.92\times10^{-12}{\rm m}$ for Silicon and $\Delta x_{B\rm int}=2.023\times10^{-12}{\rm m}$ for Osmium. The results of $\Delta x_{A/B}$ are shown in Fig.\,\ref{fig:SM_conditional}, demonstrating that the quadratic SN potential approximation is valid within our parameter space.

\begin{figure*}
\centering
\includegraphics[width=0.49\textwidth]{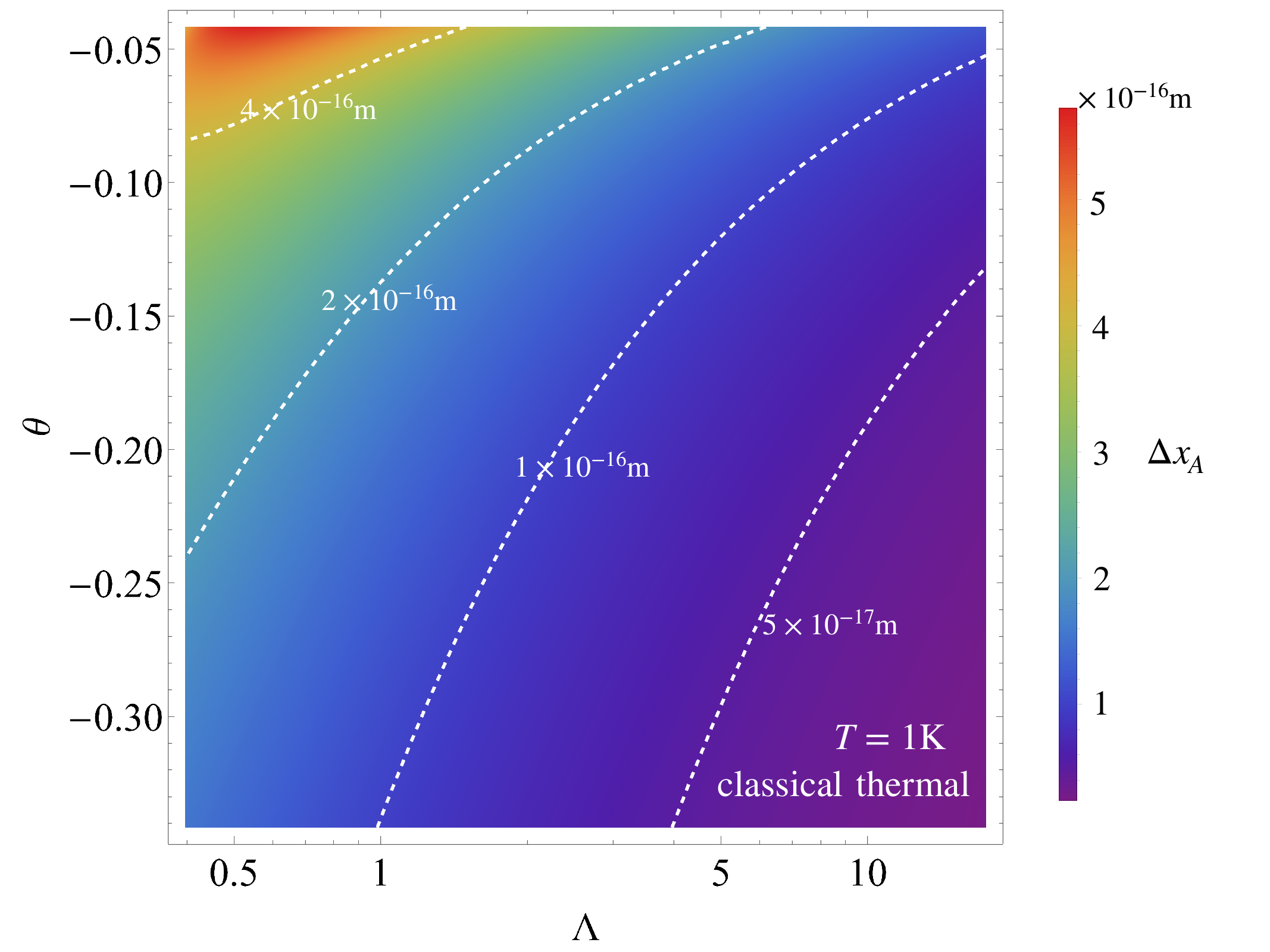}
\includegraphics[width=0.49\textwidth]{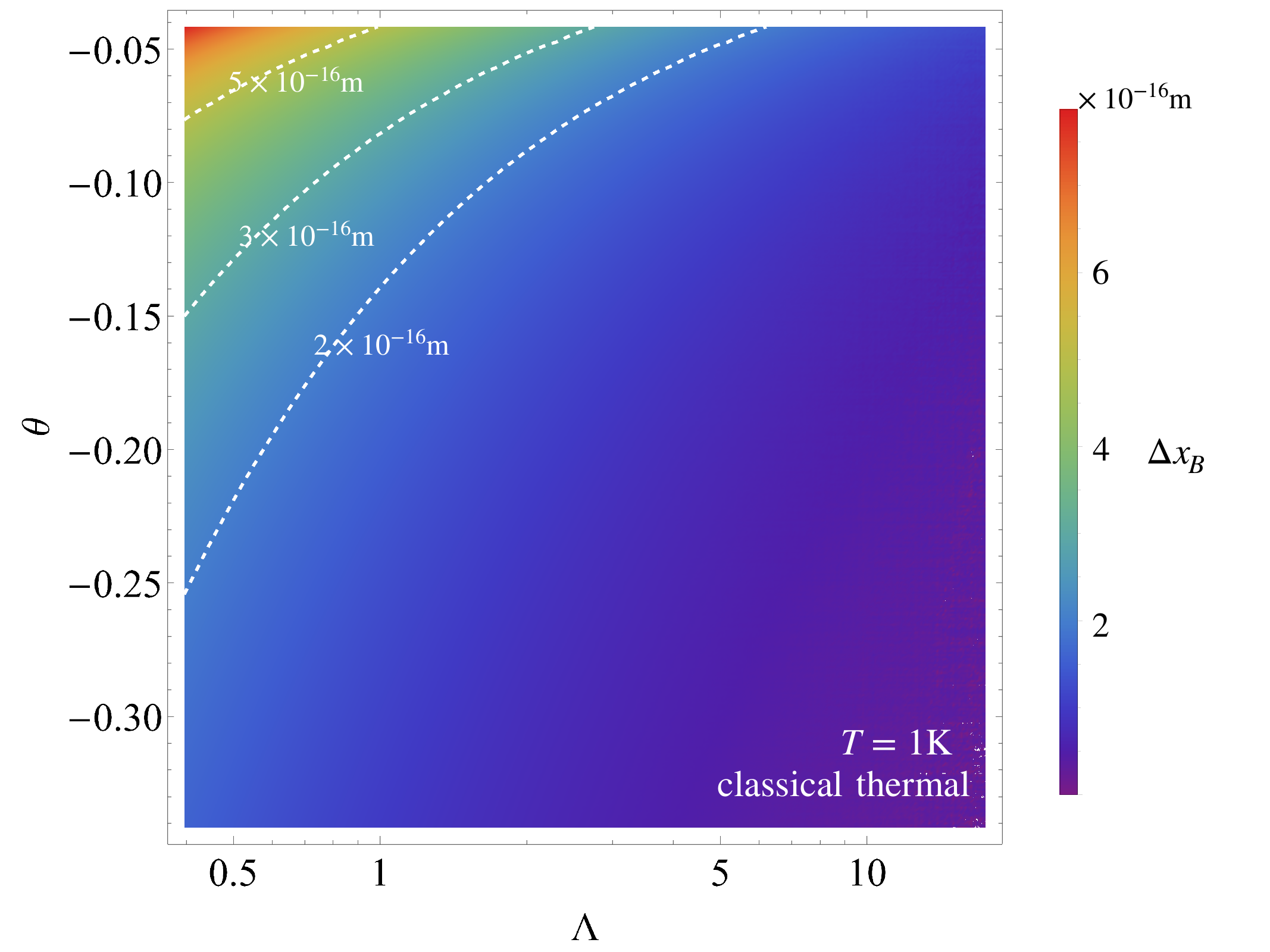}\\
\includegraphics[width=0.49\textwidth]{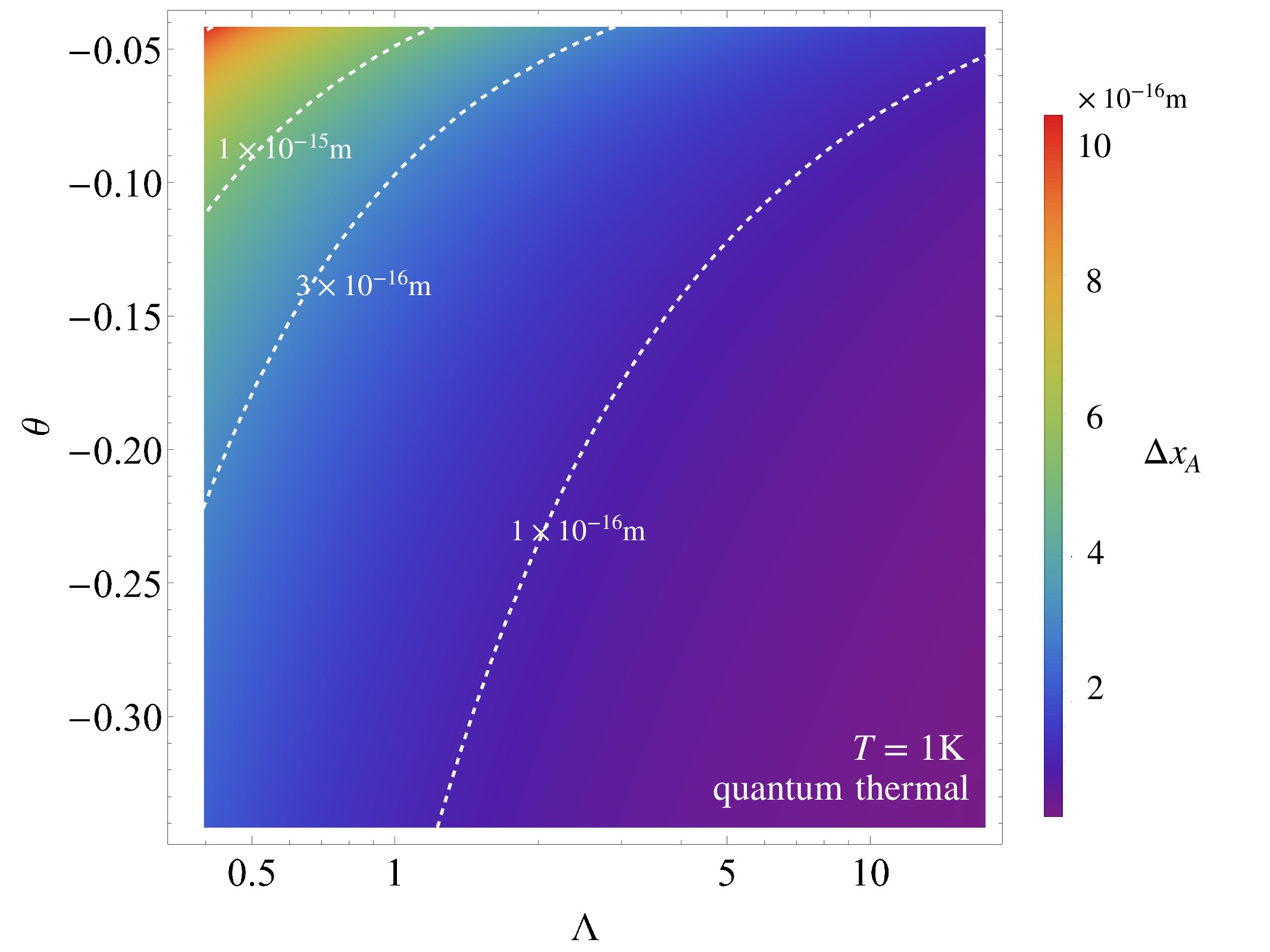}
\includegraphics[width=0.49\textwidth]{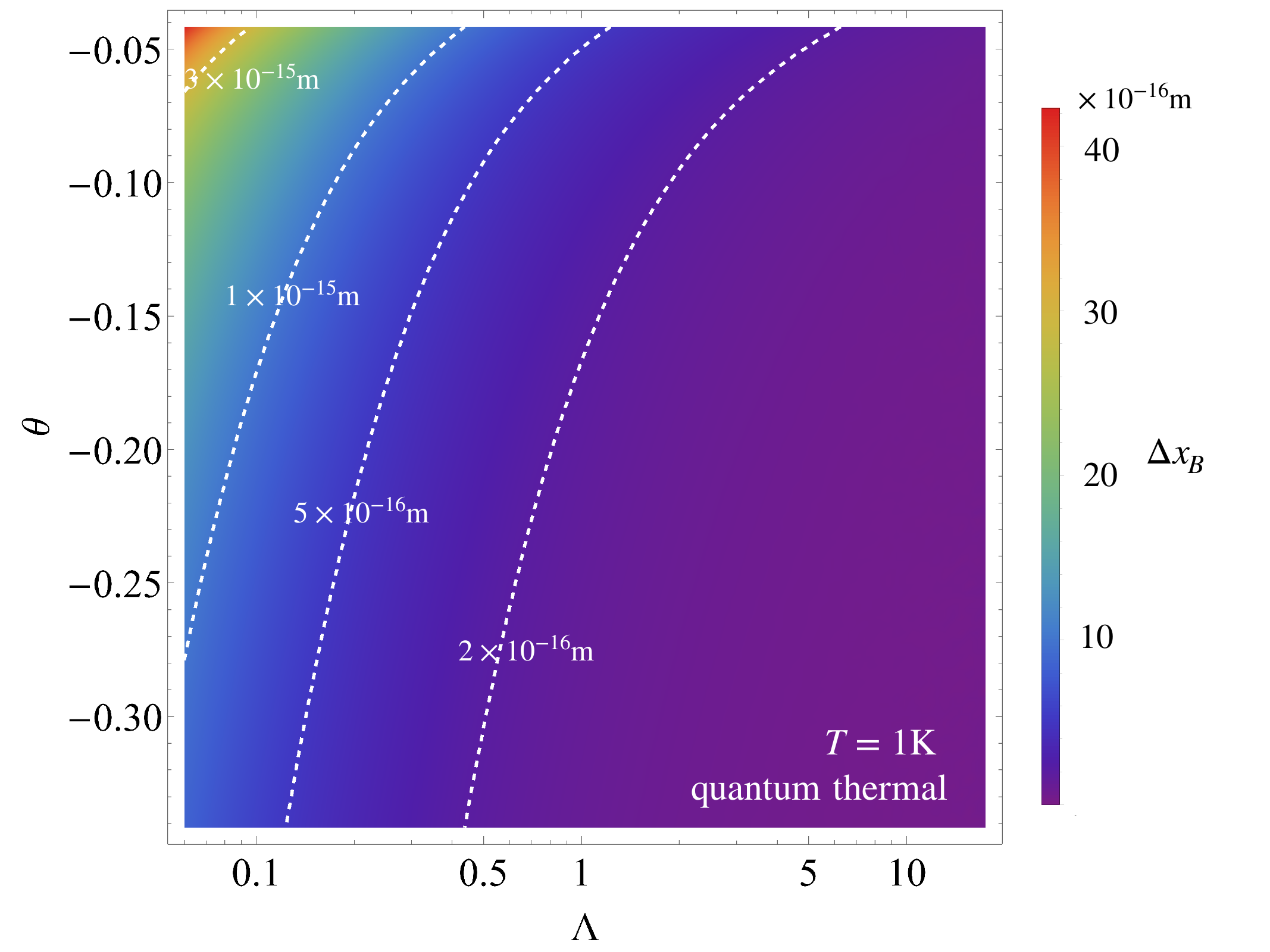}
\caption{Quantum uncertainty in mirror displacement with different thermal noise prescription. The left column represents the displacement uncertainty of mirror A, and the right column represents the displacement uncertainty of mirror B. Both satisfy $\Delta x_{A/B}\ll\Delta x_{A/B\rm int}$ within the parameter regime studied in this work. Therefore, the quadratic potential of SN gravity holds. }
\label{fig:SM_conditional}
\end{figure*}

\subsection{The conditional cross correlation between the displacements of the two mirrors}

Now, we discuss the cross-correlation between mirror A and mirror B $V^c_{x_Ax_B}$:
\begin{equation}
V^c_{x_Ax_B}=V^c_{x_{+}x_{+}}+V^c_{x_{-}x_{+}}-V^c_{x_{+}x_{-}}-V^c_{x_{-}x_{-}}.
\end{equation}

When we choose the same homodyne angle $\theta_{+}=\theta_{-}$, due to the symmetry of the equations of motion of the common and differential mode, the cross-correlation between $x_{+}$ and $x_{-}$ satisfy,
\begin{equation}
V^c_{x_{+}x_{+}}=V^c_{x_{-}x_{-}},\quad V^c_{x_{+}x_{-}}=V^c_{x_{-}x_{+}},
\end{equation}
thereby $V^c_{x_Ax_B}=0$. 

Therefore, although the asymmetry in SN self-gravity of mirrors A and B leads to the coupling between common and differential motions,  this asymmetry can not create a correlation between the conditional state of mirror A and mirror B. Differently, in the configuration targeted at probing gravity-induced entanglement\,\cite{Miao2020} and the configuration target at preparing the entanglement state of macroscopical mirrors\,\cite{Ebhardt2009},  the dynamical equation of the common and differential motion is decoupled during the whole process. 

\section{Aggregated Measurements}
Now we develop methods to extract the observational signature that can distinguish between classical gravity and quantum gravity, using the correlation spectrum derived in the above sections. In principle, the SN theory based on classical gravity predicts a non-zero correlation spectrum of the common and differential output fields, and the measurement of this correlation spectrum can verify or rule out the classical gravity theory. However, any measurement process has a limited measurement time $\mathcal{T}$ and hence a non-zero frequency resolution $1/\mathcal{T}$. Therefore,  there exist fluctuations of the measured spectrum among many experimental repetitions.  

\begin{figure}
\centering
\includegraphics[width=0.7\textwidth]{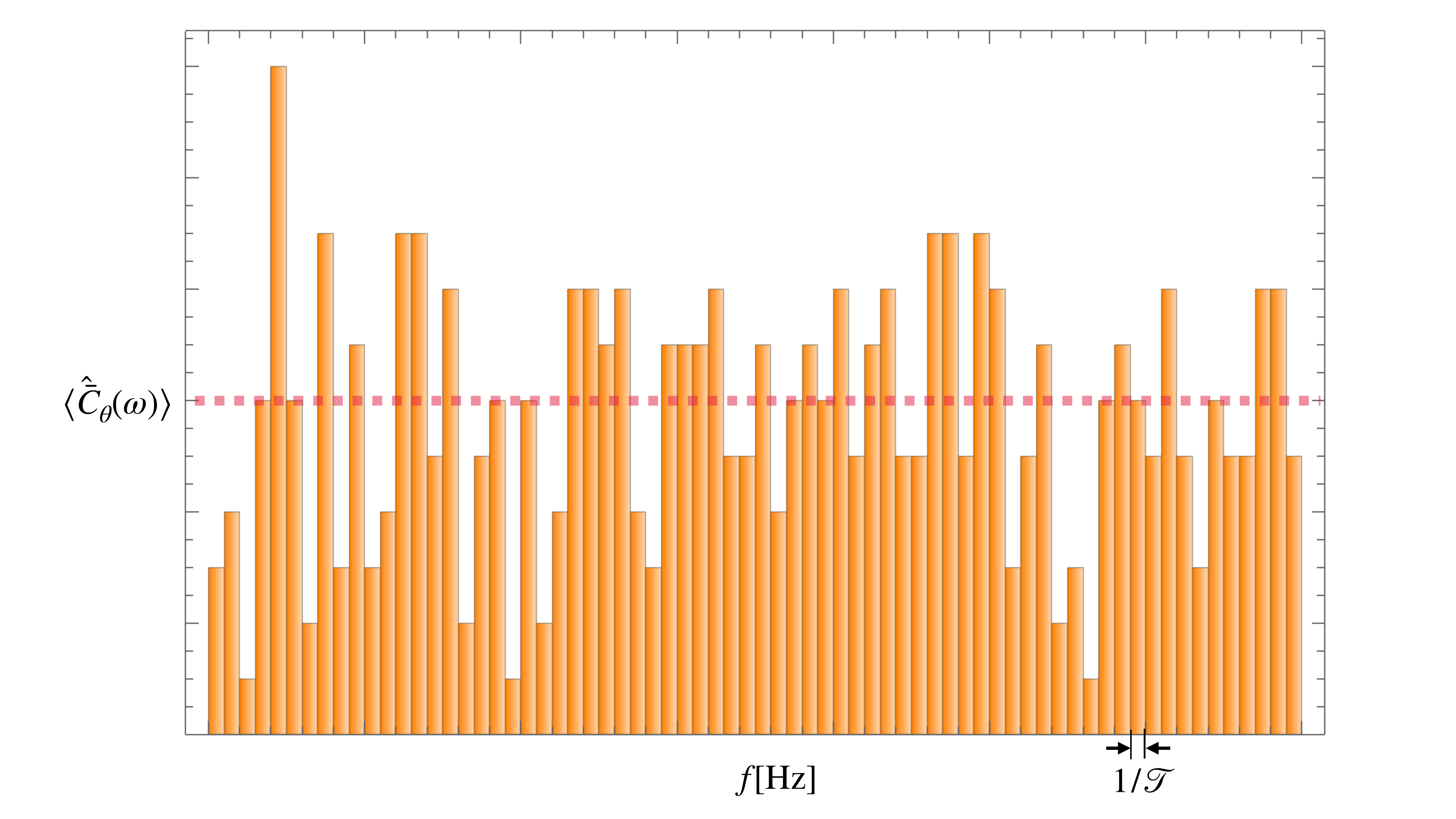}
\caption{The schematic diagram of the aggregated measurement method.}
\label{fig:histogram}
\end{figure}

To be more concrete, we transfer the ``phase quadrature data" into ``force data", using the mechanical response function, defined as: $\hat{F}_{\pm\theta}(\omega)=\chi^{-1}_m(\omega)\hat{b}_{\pm\theta}(\omega)$. In the languge of this transfered data,  we define the correlation indicator as: $\hat{C}_{\theta}(\omega)=\hat{F}_{+\theta}(\omega)\hat{F}^{\dagger}_{-\theta}(\omega)$, since the signature of this interferometric configuration is the correlation between the common and differential modes of the outgoing fields. We then define the average correlation indicator at $\omega$ within the frequency bin $1/\mathcal{T}$ as:
\begin{equation}
\hat{\bar{C}}_\theta(\omega)={\mathcal{T}}\int_{\omega-\frac{1}{2\mathcal{T}}}^{\omega+\frac{1}{2\mathcal{T}}}\hat{C}_\theta(\omega')d\omega'.
\end{equation}
The expectation value of the correlation indicator $\hat{\bar{C}}_\theta(\omega)$ can calculated as:
\begin{equation}\label{expectation_correlation_1}
\langle\hat{\bar{C}}_\theta(\omega)\rangle={\mathcal{T}}\int_{\omega-\frac{1}{2\mathcal{T}}}^{\omega+\frac{1}{2\mathcal{T}}}\langle\hat{C}_\theta(\omega')\rangle d\omega'.
\end{equation}
The variance of $\hat{\bar{C}}_\theta(\omega)$ is given by:
\begin{equation}\label{eq:variance_indicator}
\begin{split}
{\rm Var}[\hat{\bar{C}}_\theta(\omega)]&=\langle\hat{\bar{C}}^2_\theta(\omega)\rangle-\langle\hat{\bar{C}}_\theta(\omega)\rangle^2\\&=\mathcal{T}^2\int^{\omega+\frac{1}{2\mathcal{T}}}_{\omega-\frac{1}{2\mathcal{T}}}\int^{\omega+\frac{1}{2\mathcal{T}}}_{\omega-\frac{1}{2\mathcal{T}}}\left\langle\hat{F}_{+\theta}(\tilde{\omega})\hat{F}^{\dagger}_{-\theta}(\tilde{\omega})\hat{F}_{-\theta}(\tilde{\omega}')\hat{F}^{\dagger}_{+\theta}(\tilde{\omega}')\right\rangle-\left\langle\hat{F}_{+\theta}(\tilde{\omega})\hat{F}^{\dagger}_{-\theta}(\tilde{\omega})\right\rangle\left\langle\hat{F}_{+\theta}(\tilde{\omega}')\hat{F}^{\dagger}_{-\theta}(\tilde{\omega}')\right\rangle d\tilde{\omega}d\tilde{\omega}'
\end{split}
\end{equation}
To calculate the first kernel function $\left\langle\hat{F}_{+\theta}(\tilde{\omega}_1)\hat{F}^{\dagger}_{-\theta}(\tilde{\omega}_2)\hat{F}_{-\theta}(\tilde{\omega}_3)\hat{F}^{\dagger}_{+\theta}(\tilde{\omega}_4)\right\rangle$, we make use of the fact that the outgoing fields meets the commutation relation $\left[\hat{b}_{\pm}(t),\hat{b}_{\pm}(t')\right]=0$ and $\left[\hat{b}_{\pm}(t),\hat{b}_{\mp}(t')\right]=0$ and hence the above force operators are also mutually commute. Furthermore, considering the Gaussianity of our system, the above four-point correlation function can be expanded as the multiplication of the two-point correlation function using the Wick theorem:
\begin{equation}
\begin{split}
\left\langle\hat{F}_{+\theta}(\tilde{\omega}_1)\hat{F}^{\dagger}_{-\theta}(\tilde{\omega}_2)\hat{F}_{-\theta}(\tilde{\omega}_3)\hat{F}^{\dagger}_{+\theta}(\tilde{\omega}_4)\right\rangle&=\left\langle\hat{F}_{+\theta}(\tilde{\omega}_1)\hat{F}^{\dagger}_{-\theta}(\tilde{\omega}_2)\right\rangle\left\langle\hat{F}_{-\theta}(\tilde{\omega}_3)\hat{F}^{\dagger}_{+\theta}(\tilde{\omega}_4)\right\rangle\\&+\left\langle\hat{F}_{+\theta}(\tilde{\omega}_1)\hat{F}^{\dagger}_{+\theta}(\tilde{\omega}_4)\right\rangle\left\langle\hat{F}_{-\theta}(\tilde{\omega}_3)\hat{F}^{\dagger}_{-\theta}(\tilde{\omega}_2)\right\rangle\\&+\left\langle\hat{F}_{+\theta}(\tilde{\omega}_1)\hat{F}_{-\theta}(\tilde{\omega}_3)\right\rangle\left\langle\hat{F}^{\dagger}_{-\theta}(\tilde{\omega}_2)\hat{F}^{\dagger}_{+\theta}(\tilde{\omega}_4)\right\rangle,
\end{split}
\end{equation}
where the first term will be canceled with the second term in Eq.\,\eqref{eq:variance_indicator}.
Hence we have:
\begin{equation}
\begin{split}
\left\langle\hat{F}_{+\theta}(\tilde{\omega}_1)\hat{F}^{\dagger}_{-\theta}(\tilde{\omega}_2)\hat{F}_{-\theta}(\tilde{\omega}_3)\hat{F}^{\dagger}_{+\theta}(\tilde{\omega}_4)\right\rangle&=|\chi^{-1}_m(\omega')|^2|\chi^{-1}_m(\tilde{\omega}_3)|^2S_{y_{+\theta}y_{-\theta}}(\tilde{\omega}_1)S_{y_{-\theta}y_{-\theta}}(\tilde{\omega}_3)\delta(\tilde{\omega}_1-\tilde{\omega}_2)\delta(\tilde{\omega}_3-\tilde{\omega}_4)\\&+\mathcal{G}_1(\tilde{\omega}_1)\mathcal{G}_2(\tilde{\omega}_2)\delta(\tilde{\omega}_1+\tilde{\omega}_3)\delta(\tilde{\omega}_2+\tilde{\omega}_4),
\end{split}
\end{equation}
where $\mathcal{G}_1(\tilde{\omega}_1)$ and $\mathcal{G}_2(\tilde{\omega}_2)$ meet: $\left\langle\hat{F}_{+\theta}(\tilde{\omega}_1)\hat{F}_{-\theta}(\tilde{\omega}_3)\right\rangle=\mathcal{G}_1(\tilde{\omega}_1)\delta(\tilde{\omega}_1+\tilde{\omega}_3)$ and $\left\langle\hat{F}^{\dagger}_{-\theta}(\tilde{\omega}_2)\hat{F}^{\dagger}_{+\theta}(\tilde{\omega}_4)\right\rangle=\mathcal{G}_2(\tilde{\omega}_2)\delta(\tilde{\omega}_2+\tilde{\omega}_4)$.

When the total measurement time is $\mathcal{T}$ and the resolution of frequency is $\delta f=1/\mathcal{T}$, in this case, $\delta(\tilde{\omega}_1-\tilde{\omega}_2)|_{\tilde{\omega}_1=\tilde{\omega}_2=\tilde{\omega}}=1/{\delta f}=\mathcal{T}$.  Moreover, the delta function in the $\mathcal{G}$- terms are all zero since  $\tilde \omega_{3/4}$ are around $\tilde{\omega}_1=\tilde{\omega}_2=\tilde{\omega}$. Therefore, the variance of $\hat{\bar{C}}_\theta(\omega)$ can simplified as:
\begin{equation}\label{eq:variance}
\begin{split}
{\rm Var}[\hat{\bar{C}}_\theta(\omega)]&=\mathcal{T}^2\int^{\omega+\frac{1}{2\mathcal{T}}}_{\omega-\frac{1}{2\mathcal{T}}}\int^{\omega+\frac{1}{2\mathcal{T}}}_{\omega-\frac{1}{2\mathcal{T}}}|\chi^{-1}_m(\tilde{\omega})|^2|\chi^{-1}_m(\tilde{\omega}')|^2S_{y_{+\theta}y_{+\theta}}(\tilde{\omega})S_{y_{-\theta}y_{-\theta}}(\tilde{\omega}')\delta(\tilde{\omega}-\tilde{\omega}')^2d\tilde{\omega} d\tilde{\omega}'\\
&=\frac{\mathcal{T}^3}{4}\int^{\omega+\frac{1}{2\mathcal{T}}}_{\omega-\frac{1}{2\mathcal{T}}}\xi^2(\mathcal{F}_A(\tilde{\omega})+\mathcal{F}_B(\tilde{\omega}))^2 d\tilde{\omega}.
\end{split}
\end{equation}

Suppose the spectrum has a bandwidth $\Gamma$, there will be in total $N=\Gamma\mathcal{T}$ different frequency bins. To cover the bandwidth, we define the SN detection statistics  as the sum of the correlation indicators of the different independent frequencies:
\begin{equation}
\hat{\chi}_N=\frac{1}{N}\sum_{j=1}^N\hat{\bar{C}}_\theta(\omega_j),
\end{equation}
where $\omega_j=j/\Gamma \mathcal{T}$. As we shall see later, the expectation $\langle \hat{\bar{C}}_\theta(\omega_i)\rangle$ is actually frequency-independent. Then the expectation value of the $\hat{\chi}_N$ satisfies $\langle\hat{\chi}_N\rangle=\langle\hat{\bar{C}}_\theta(\omega_i)\rangle$, which is defined as the signal of the SN-induced correlation. The signal-to-noise ratio\,(SNR) of the SN induced correlation can be defined as:
\begin{equation}
{\rm SNR}=\frac{\langle\hat{\chi}_N\rangle}{\sqrt{{\rm Var}[\hat{\chi}_N]}}.
\end{equation}
In the following, we will establish the relationship between the measurement time $\mathcal{T}$ and the SNR, considering the influence of different noise sources.

\subsection{The statistics of the correlation indicator: the observation bandwidth}
Let us discuss the method to determine the observation bandwidth $\Gamma$ in the aggregated measurements method. From Eq.\,\ref{eq:definition_F}, we can find that $\langle\hat{\bar{C}}_\theta\rangle$ is frequency-independent, while ${\rm Var} [\hat{\bar{C}}]$ is frequency-dependent and the dependence of ${\rm Var} [\hat{\bar{C}}]$ on frequency is schematically shown in the upper panel of Fig.\,\ref{fig:bandwidth}. When $\omega<\omega_{\rm th}$, ${\rm Var} [\hat{\bar{C}}]$ is frequency-independent which is dominated by white force noise\,(e.g. thermal force and quantum radiation pressure force), while in the shot noise dominated region\,($\omega>\omega_{\rm th}$), ${\rm Var} [\hat{\bar{C}}]$ increase with frequency.  The extraction of the SN gravity signal becomes inefficient when ${\rm Var} [\hat{\bar{C}}]$ is large. Therefore, in this work, we  only use the $\omega<\omega_{\rm th}$ data extract the signature of SN gravity. Operationally, we can mathmatically separate the ${\rm Var} [\hat{\bar{C}}]$ into two terms, one is frequency-independent term ${\rm Var} [\hat{\bar{C}}]_1$ and the other one is frequency-dependent term ${\rm Var} [\hat{\bar{C}}]_2(\omega)$. Because  ${\rm Var} [\hat{\bar{C}}]_2(\omega)$ increases monotonically as the frequency increases, we can determine the $\omega_{\rm th}$ by 
\begin{equation}
\frac{{\rm Var} [\hat{\bar{C}}]_1}{{\rm Var} [\hat{\bar{C}}]_2(\omega_{\rm th})}=10,
\end{equation}
In this case, the observation frequency regime is $\omega\in[0,\omega_{\rm th}]$ and $\Gamma=\omega_{\rm th}$.  In the lower panel of Fig.\,\ref{fig:bandwidth}, we show the dependence of bandwidth $\Gamma$ on the $\Lambda$ and temperature $T$ with different thermal noise prescriptions.

\begin{figure}
\centering
\includegraphics[width=0.92\textwidth]{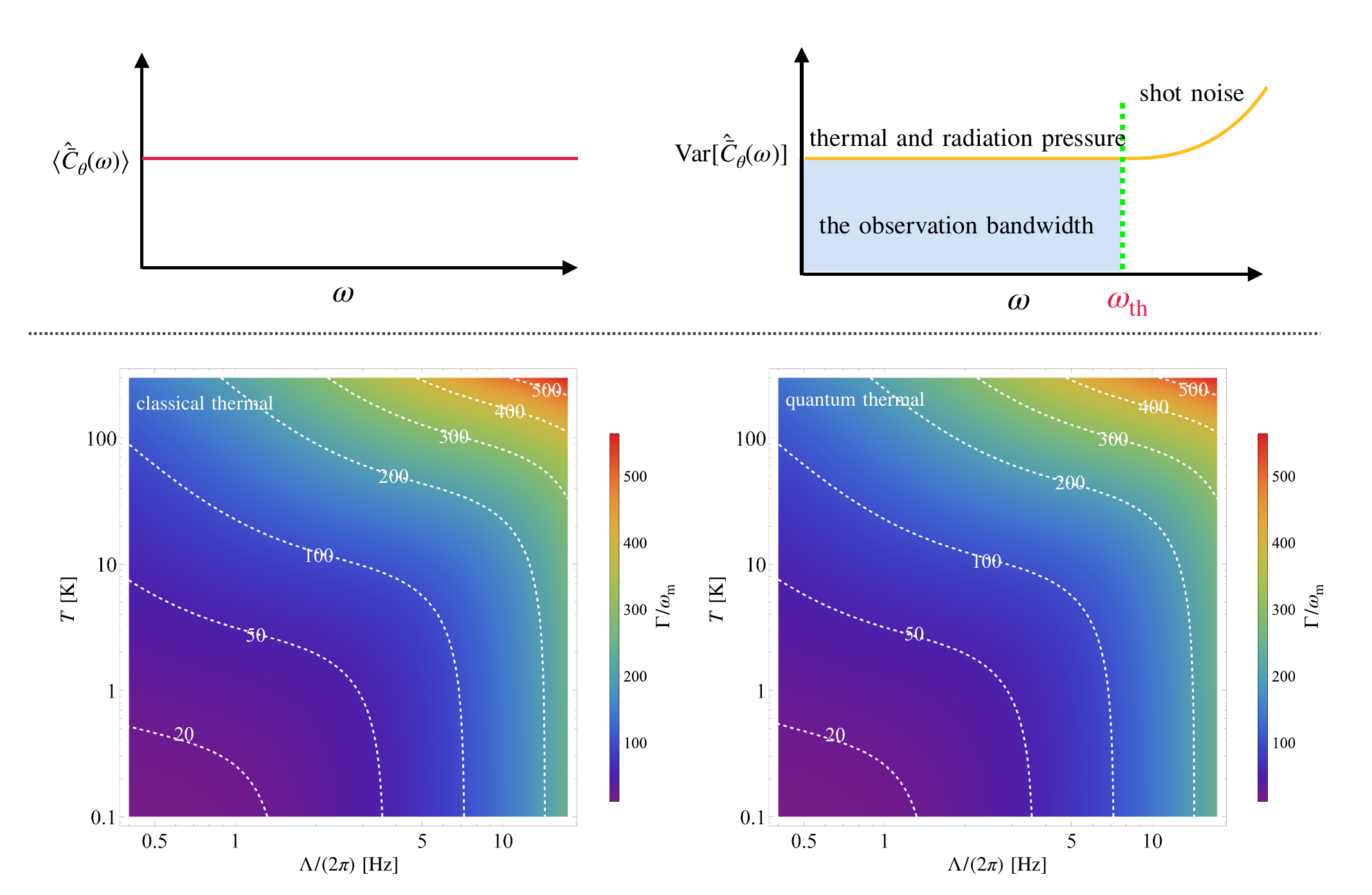}
\caption{The upper panel schematically shows the dependence of $\langle\hat{\bar{C}}_\theta\rangle$ and ${\rm Var}[\hat{\bar{C}}_\theta(\omega)]$ on the  frequency. The lower panel describes the observation bandwidth $\Gamma$ as a function of the cooperativity $\Lambda$ and temperature $T$, where we choose the squeezing of input optical field is 10{\rm dB} and the homodyne measurement angle $\theta=-0.14\,{\rm rad}$.}
\label{fig:bandwidth}
\end{figure}

\subsection{The statistics of the correlation indicator: quantum radiation pressure noise}
Let us first investigate the case when the force noise in the system is dominated by quantum radiation pressure noise. In this case, the expectation value of the correlation indicator $\hat{\bar{C}}_\theta(\omega)$ can be calculated as:
\begin{equation}\label{expectation_correlation_1}
\langle\hat{\bar{C}}_\theta(\omega)\rangle={\mathcal{T}}\int_{\omega-\frac{1}{2\mathcal{T}}}^{\omega+\frac{1}{2\mathcal{T}}}\langle\hat{C}_\theta(\omega')\rangle d\omega'
={\mathcal{T}}\delta \omega^2_{\rm SN}\Lambda^2\zeta\left(\sin\theta-|\sin\theta|\right).
\end{equation}
For calculating the variance, we can use the Eq.\,\ref{eq:squeezing_beta} and Eq.\,\ref{eq:definition_F}, where the $\mathcal{F}_{A/B}$ can be represented as:
\begin{equation}
\begin{split}
&\mathcal{F}_{A/B}=\\
&\xi[(\omega^2-\omega^2_{qA/B})^2+\gamma_m^2\omega^2]+\Lambda^2(\omega^2_{qA}-\omega^2)(\cosh2r\sin2\theta+2\cos(\theta-2\phi)\sin\theta\sinh2r)+\sin^2\theta\Lambda^4\zeta,
\end{split}
\end{equation}
If we choose the bandwidth $\Gamma$ where the frequency-independent component of $\mathcal{F}_{A/B}$ dominates and $\sin\theta\Lambda^2\gg(\sqrt{1+\cos^2\theta}+\cos^2\theta)\omega^2_{qA/B}$, $\mathcal{F}_{A/B}$ can be simplified as:
\begin{equation}
\mathcal{F}_{A/B}=\sin^2\theta\Lambda^4\zeta.
\end{equation}
Substituting the above equation into Eq.\,\ref{eq:variance} leads to the variance of the correlation indicator:
\begin{equation}
{\rm Var}[\hat{\bar{C}}_\theta(\omega)]=\mathcal{T}^2\sin^4\theta\Lambda^8\zeta^2,
\end{equation}
hence the variance of the detection statistics $\hat \chi_N$:
\be
{\rm Var}[\hat \chi_N]=\frac{1}{N}{\rm Var}[\hat{\bar{C}}_\theta(\omega)]=\frac{1}{\Gamma\mathcal{T}}{\rm Var}[\hat{\bar{C}}_\theta(\omega)]
\ee
Finally, we have the relationship between measurement time and the SNR as:
\be
\mathcal{T}=\frac{\sin^2\theta\Lambda^4}{4\Gamma \delta\omega^4_{\rm SN}}{\rm SNR}^2.
\ee

\subsection{The statistics of the correlation indicator: the influence of classical laser power noise}
In the optomechanical experiment, the classical laser power noise is an important and unavoidable noise raised from the fluctuating pumping light. In our interferometric configuration, the laser power noise generates a classical radiation pressure force $f_{p}$ that only drives the common motion of the test masses and does not affect the differential motion. Therefore, the laser power noise modifies the input-output relation of the common mode with an additional term:
\begin{equation}
b_{+p}=\sqrt{\frac{M}{\hbar}}\frac{\Lambda\sin\theta_+}{\chi_m(\omega)}f_{p}(\omega).
\end{equation}
 The above term only appears in the common mode of the output field, contributing to the diagonal spectrum of the common output field $S_{y_{\theta +}y_{\theta +}}$ with an additional term:
 \begin{equation}
S^{p}_{y_{\theta +}y_{\theta +}}(\omega)=\frac{M}{\hbar}\frac{\Lambda^2\sin\theta^2_{+}}{|\chi_m(\omega)|^2}S_{f_{p}f_{p}}(\omega),
\end{equation}
where $S_{f_{p}f_{p}}(\omega)$ is the power density spectrum of classical radiation pressure force. 
The laser power noise does not increase the correlation spectrum between common and differential output fields $S_{y_{\theta\pm}y_{\theta\mp}}$ and also does not affect the diagonal spectrum of the differential output field $S_{y_{\theta -}y_{\theta -}}$. The effect of the classical laser power noise on the final detection statistics will be discussed together with the thermal noise.

\subsection{The statistics of the correlation indicator: the influence of classical thermal noise}
Now we discuss the influence of classical thermal noise on the $\langle\hat{\bar{C}}_\theta(\omega)\rangle$ and ${\rm Var}[\hat{\bar{C}}_\theta(\omega)]$. Firstly, we introduced the classical thermal noise prescription brifely.  A quantum state affected by the thermal environment is a mixed state that can be expressed as a density matrix:
\begin{equation}\label{eq:density_matrix}
\rho=\sum_iP_i|\psi_i\rangle\langle\psi_i|.
\end{equation}
In standard quantum mechanics, this mixed-state density matrix can be obtained by tracing out the environmental degrees of freedom of a system-environment pure joint state.  When applying to the SN theory, a question naturally arises: should the expectation in the Hamiltonian of SN theory be averaged over each pure system ``branch state"  $|\psi_i\rangle$, or the whole system-environment state?   

One viewpoint is based directly on the form of the density matrix Eq.\,\eqref{eq:density_matrix}, that is, the mixed state is interpreted as the ensemble of pure branch states $|\psi_i\rangle$ with classical probabilities $P_i$. In this case, for one experiment realization, only one branch quantum state $|\psi_i\rangle$ will be selected with probability $P_i$, and different experiment realizations will independently select different branch quantum states. Therefore, the Hamiltonian should be averaged over the selected branch states $|\psi_i\rangle$. This prescription of thermal noise has a subtlety that the nonlinear Hamiltonian will depend on the different decomposition basis $|\psi\rangle$, which is elaborated in \,\cite{Helou2017}.


 For example,  the quantum state of the thermal bath can be represented in the \emph{coherent state basis} by the P-distribution:
\begin{equation}
\rho_{\rm th}=\int\frac{1}{\pi\langle n\rangle}e^{-|\alpha|^2/\langle n\rangle}|\alpha\rangle\langle\alpha|d^2\alpha,
\end{equation}
where $\langle n\rangle$ is the mean occupation number. As the quantum fluctuations of different coherent states all match that of the vacuum state, we assume a zero-point fluctuation follows the quantum probability of the pure state $|\alpha\rangle$. Essentially, it means that all branch states $|\alpha\rangle$ share the same quantum fluctuations, while the selection of the branch state follows the classical probability density $e^{-|\alpha|^2/\langle n\rangle}/(\pi\langle n\rangle)$. Finally, following\,\cite{Helou2017}, we can decompose thermal noise into quantum and classical parts:
\begin{equation}
\hat{F}_{\rm th}(t)=f_{\rm cl}(t)+\hat{f}_{\rm zp}(t),
\end{equation}
where
\begin{equation}
f_{\rm cl}(t)=\langle\hat{F}_{\rm th}(t)\rangle,\quad \hat{f}_{\rm zp}(t)=\hat{F}_{\rm th}(t)-\langle\hat{F}_{\rm th}(t)\rangle.
\end{equation}
The total thermal noise spectrum is represented by:
\begin{equation}
S_{F_{\rm th}F_{\rm th}}(\Omega)=4\hbar\left[\frac{1}{e^{\frac{\hbar\Omega}{k_BT}}-1}+\frac{1}{2}\right]\frac{{\rm Im}[\chi_m(\Omega)]}{|\chi_m(\Omega)|^2},
\end{equation}
which can be divide into the quantum and classical components:
\begin{equation}
\begin{split}
S_{f_{\rm cl}f_{\rm cl}}(\Omega)&=4\hbar\left[\frac{1}{e^{\frac{\hbar\Omega}{k_BT}}-1}\right]\frac{{\rm Im}[\chi_m(\Omega)]}{|\chi_m(\Omega)|^2}\approx4k_BTM\gamma_m,\\
S_{f_{\rm zp}f_{\rm zp}}(\Omega)&=2\hbar\frac{{\rm Im}[\chi_m(\Omega)]}{|\chi_m(\Omega)|^2}\approx2\hbar\Omega M\gamma_m.
\end{split}
\end{equation}
Typically, we have $k_BT\gg\hbar\Omega$ hence $S_{f_{\rm cl}f_{\rm cl}}(\Omega)\gg S_{f_{\rm zp}f_{\rm zp}}(\Omega)$, therefore the classical component dominates over the quantum component. This is why we call this prescription ``classical thermal noise prescription".

After reviewing the classical thermal noise prescription. Now we discuss its influence on our proposed protocol. Solving the Heisenberg equations of motion, the outgoing optical field can be solved as:
\begin{equation}
\begin{split}
\hat{b}_{+\theta}(\omega)&=\hat{b}_{+\theta q}(\omega)+\sqrt{\frac{M^3}{\hbar}}\Lambda\sin\theta_+\left[\frac{\omega^2_{\rm SN}\chi^{-1}_q(\omega)-M\delta\omega^4_{\rm SN}/4}{\chi^{-1}_{A}(\omega)\chi^{-1}_{B}(\omega)}\langle \hat x_+(\omega)\rangle+\frac{\chi^{-1}_m(\omega)\delta\omega^2_{\rm SN}/2}{\chi^{-1}_{A}(\omega)\chi^{-1}_{B}(\omega)}\langle \hat x_-(\omega)\rangle\right]\\
&\quad\qquad\qquad\qquad\qquad\qquad\qquad\qquad\qquad\qquad\qquad\qquad\qquad\qquad\qquad\qquad\qquad+\sqrt{\frac{M}{\hbar}}\frac{\Lambda\sin\theta_+}{\chi^{-1}_m(\omega)}[f_{\rm cl+}(\omega)+f_p(\omega)],\\
\hat{b}_{-\theta}(\omega)&=\hat{b}_{-\theta q}(\omega)+\sqrt{\frac{M^3}{\hbar}}\Lambda\sin\theta_-\left[\frac{\omega^2_{\rm SN}\chi^{-1}_q(\omega)-M\delta\omega^4_{\rm SN}/4}{\chi^{-1}_{A}(\omega)\chi^{-1}_{B}(\omega)}\langle \hat x_-(\omega)\rangle+\frac{\chi^{-1}_m(\omega)\delta\omega^2_{\rm SN}/2}{\chi^{-1}_{A}(\omega)\chi^{-1}_{B}(\omega)}\langle \hat x_+(\omega)\rangle\right]+\sqrt{\frac{M}{\hbar}}\frac{\Lambda\sin\theta_-}{\chi^{-1}_m(\omega)}f_{\rm cl-}(\omega).
\end{split}
\end{equation}
Since the thermal noise terms $f_{cl+}$ and $f_{cl-}$ are uncorrelated, thermal noise can not increase the correlation of the two outgoing fields $S_{y_{{\theta +}{\theta -}}}(\omega)$, but it can increase the diagonal spectrum of the single outgoing field $S_{y_{{\theta \pm}{\theta \pm}}}(\omega)$ as:
\begin{equation}
\begin{split}
&S_{y_{{\theta+}{\theta+}}}(\omega)=M^2\zeta\frac{\mathcal{F}_A(\omega)+\mathcal{F}_B(\omega)}{2|\chi^{-1}_m(\omega)|^2}+\frac{M\sin^2\theta\Lambda^2}{\hbar|\chi^{-1}_m(\omega)|^2}[4M\gamma_mk_BT+S_{f_{p}f_{p}}(\omega)],\\
&S_{y_{{\theta-}{\theta-}}}(\omega)=M^2\zeta\frac{\mathcal{F}_A(\omega)+\mathcal{F}_B(\omega)}{2|\chi^{-1}_m(\omega)|^2}+\frac{4M^2\sin^2\theta\Lambda^2\gamma_mk_BT}{\hbar|\chi^{-1}_m(\omega)|^2}.
\end{split}
\end{equation}
This simply means that the expectation value remains as the Eq.\,\ref{expectation_correlation_1} and the variance of the correlation indicator is modified as:
\begin{equation}
{\rm Var}[\hat{\bar{C}}_\theta(\omega)]=\mathcal{T}^2(1+\kappa)(\sin^2\theta\Lambda^4\zeta+4\sin^2\theta\Lambda^2\gamma_mk_BT/\hbar)^2,
\end{equation}
where we define:
\be
\kappa=\frac{M\sin^2\theta\Lambda^2}{\hbar|\chi^{-1}_m(\omega)|^2}\frac{S_{f_{p}f_{p}}(\omega)}{S_{y_{{\theta-}{\theta-}}}(\omega)}
\ee

In the classical thermal noise prescription, using the central limit theorem, the expectation value and variance of the statistical indicator $\hat{\chi}_N$ are:
\begin{equation}
\langle\hat{\chi}_N\rangle=\mathcal{T}\int_{\omega-\frac{1}{2\mathcal{T}}}^{\omega+\frac{1}{2\mathcal{T}}}\langle\hat{C}_\theta(\omega')\rangle d\omega'=2\mathcal{T}\delta \omega^2_{\rm SN}\Lambda^2\zeta\sin\theta,
\end{equation}
and 
\begin{equation}
{\rm Var}[\hat{\chi}_N]=\frac{\mathcal{T}^2(1+r)(\sin^2\theta\Lambda^4\zeta+4\sin^2\theta\Lambda^2(\gamma_mk_BT+S_{f_pf_p}/4M)/\hbar)^2}{\Gamma\mathcal{T}}.
\end{equation}
Now, we define the SNR as the ratio between the expectation value and the standard variance of the statistical indicator 
\begin{equation}
{\rm SNR}=\frac{\langle\hat{\chi}_N\rangle}{\sqrt{{\rm Var}[\hat{\chi}_N]}}.
\end{equation}
\begin{figure}
\centering
\includegraphics[width=0.49\textwidth]{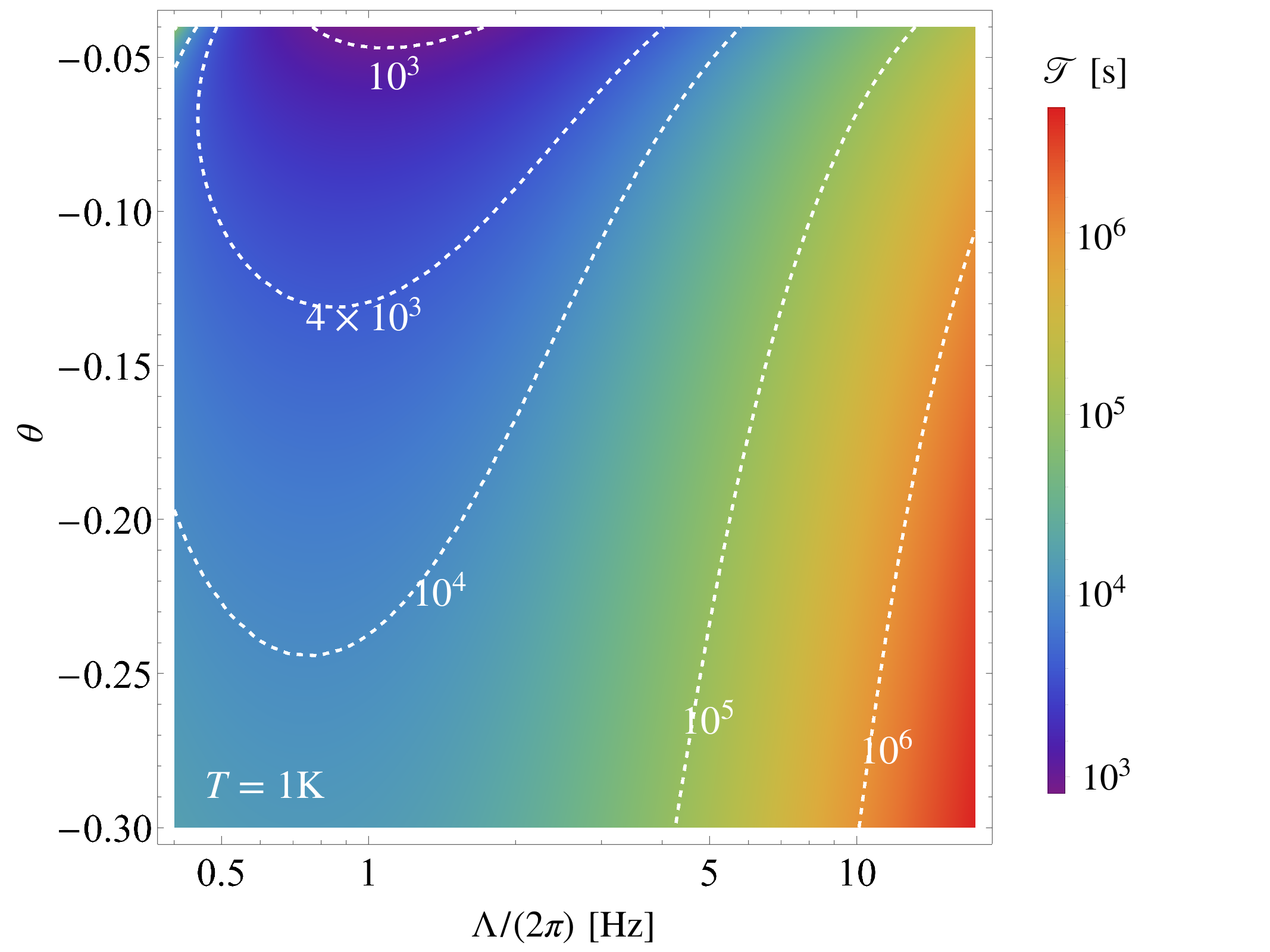}
\includegraphics[width=0.49\textwidth]{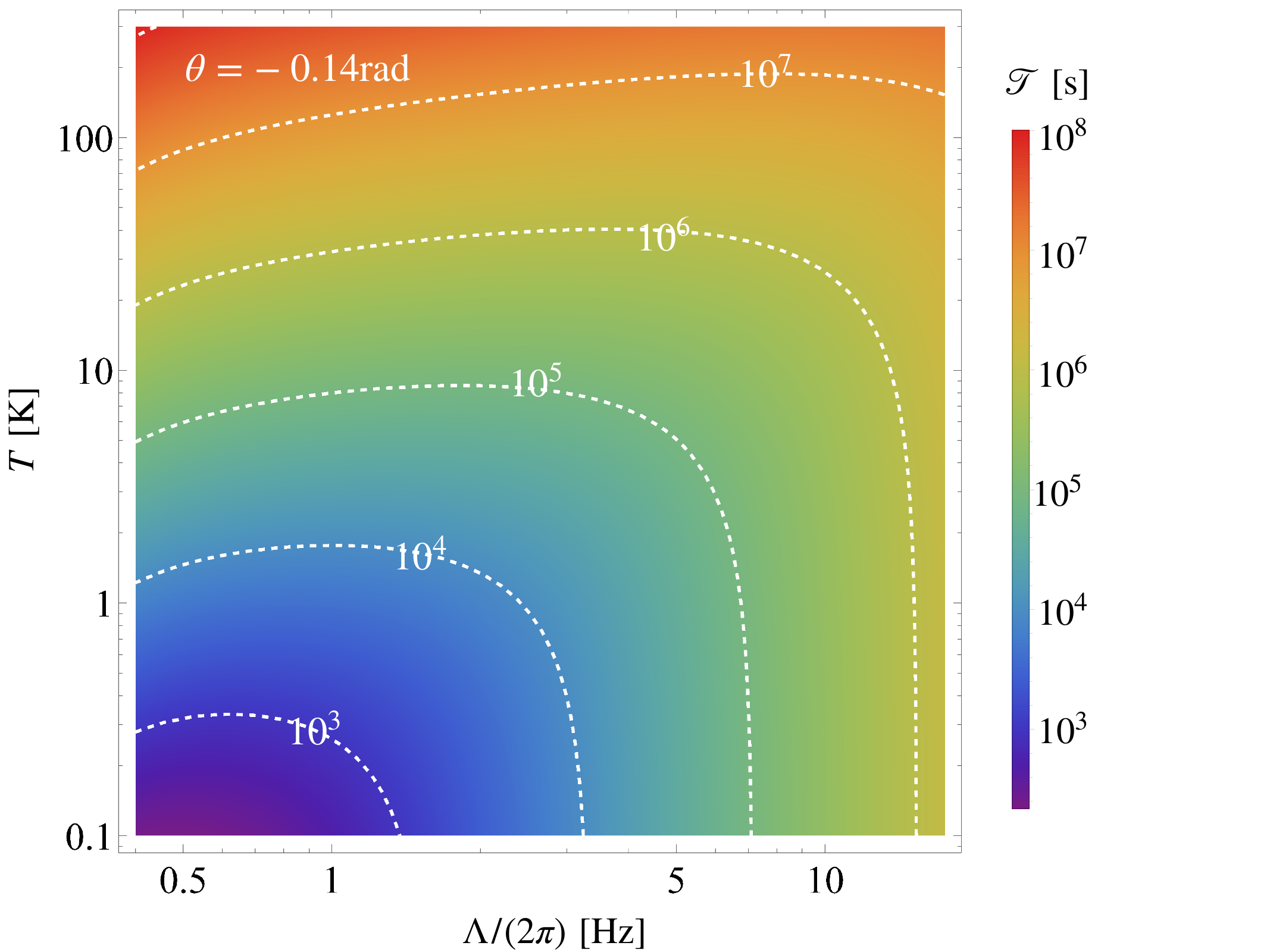}
\includegraphics[width=0.49\textwidth]{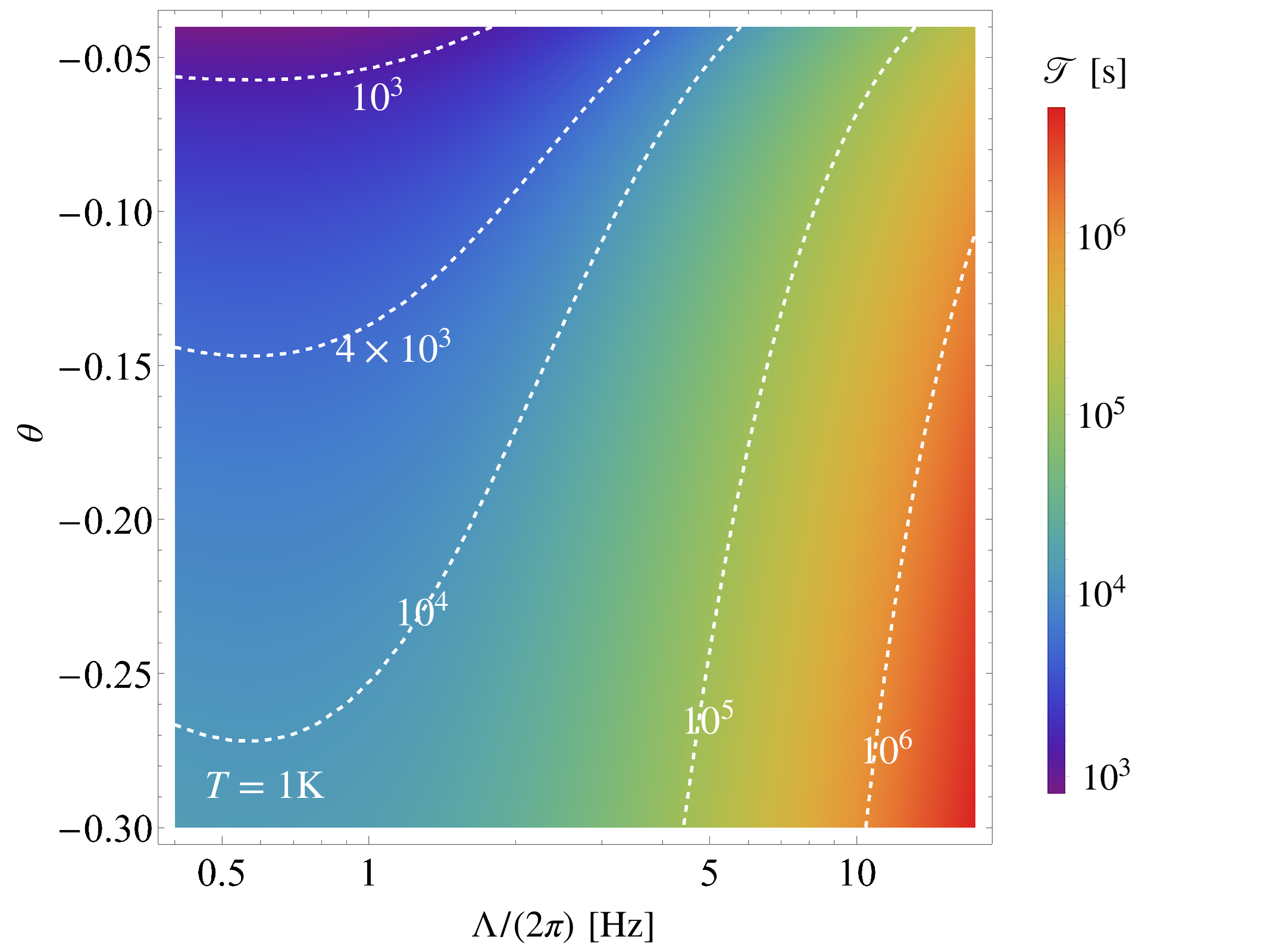}
\includegraphics[width=0.49\textwidth]{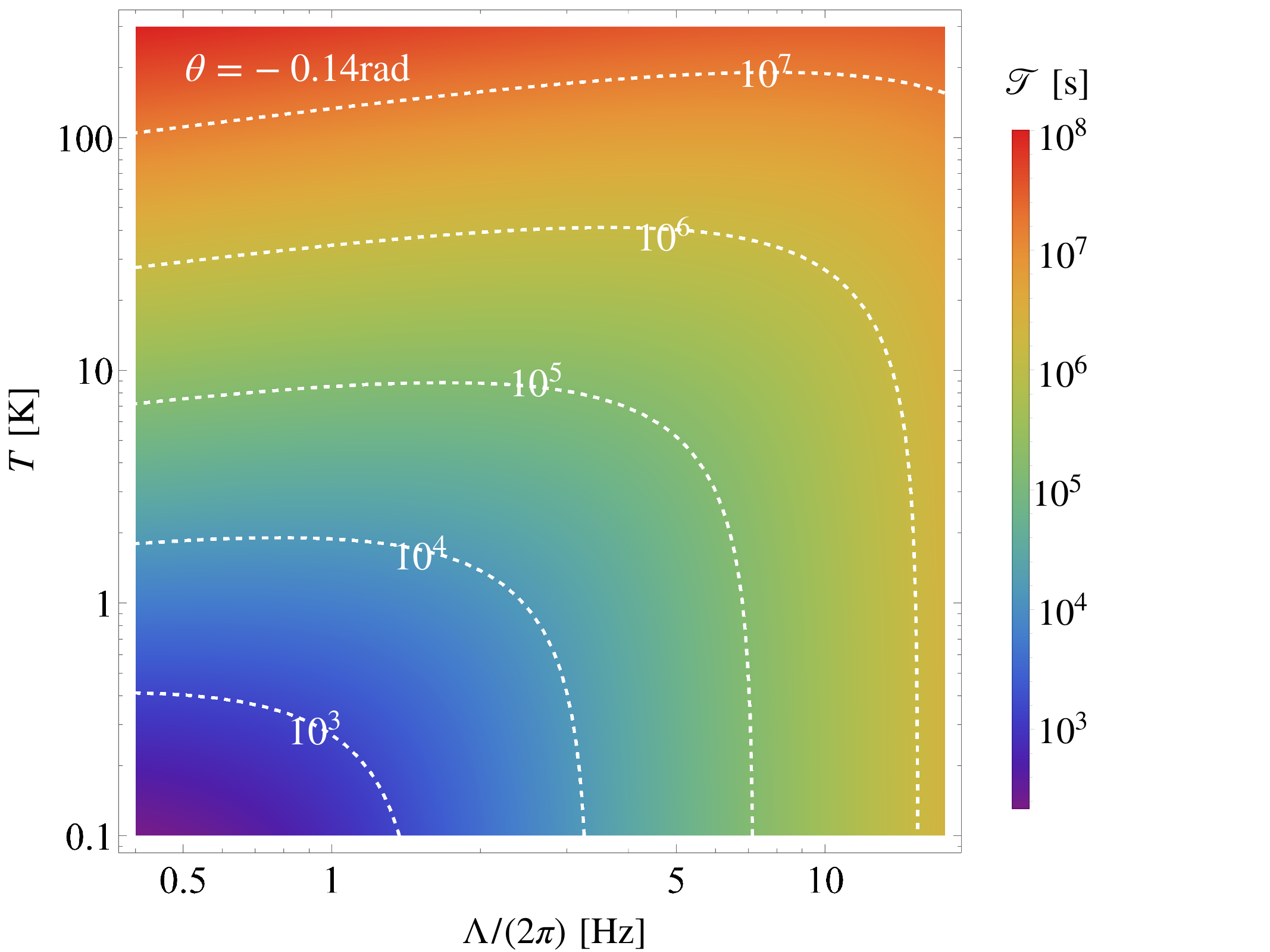}
\caption{Classical thermal noise prescription: dependence of measurement time $\mathcal{T}$ on the environmental temperature $T$, cooperativity $\Lambda$ and homodyne angle $\theta$um, where the signal-to-noise ratio is set to be ${\rm SNR}=1$. The optical input is set to be a 10\,dB phase squeezed vacuum. The Upper panel represents the result obtained using exact formula, while the lower panel is the one obtained using the approximated formula.}
\label{fig:time_parameter_dependence}
\end{figure}

Therefore, we can obtain the relationship between the measurement time and the SNR as
\begin{equation}
\mathcal{T}=(1+\kappa)\left[\frac{\sin^2\theta\Lambda^2\zeta+4\sin^2\theta\gamma_mk_BT/\hbar}{2\Gamma^{1/2} \delta \omega^2_{\rm SN}\zeta\sin\theta}\right]^2{\rm SNR}^2.
\end{equation}
The interferometer configuration offers a key advantage in measurement time scaling. While the classical radiation pressure due to the laser power noise affects only the common mode of the outgoing field, resulting in a measurement time proportional to $1+r$, the single-mirror scheme for testing semi-classical gravity would face a less favorable $(1+r)^2$ scaling when comparing noise spectra. This difference becomes particularly significant when classical radiation pressure noise dominates $(r\gg1)$, making our interferometric approach substantially more time-efficient in such regimes.

\subsection{The statistics of the correlation indicator: the influence of quantum thermal noise}

We now switch to the quantum thermal noise prescription, while we will not repeat the derivation of the laser power noise effect, which will be the same as the previous subsection. Different from the ensemble interpretation for density matrix, a mixed-state system density matrix is originated from tracing the environmental degrees of freedom from a pure system-environment joint state $|\Psi\rangle$:\begin{equation}
\rho_s={\rm Tr}_{\rm en}[|\Psi\rangle\langle\Psi|],
\end{equation}
and the expectation value of an operator $\hat O$ meets,
\begin{equation}
\langle\Psi|\hat{O}_s|\Psi\rangle={\rm Tr}[\rho_s\hat{O}_s].
\end{equation}
Therefore, the Hamiltonian in the Schr{\"o}dinger-Newton theory involving thermal environment should depend on the joint wavefunction $|\Psi\rangle$. 

In this case, 
the thermal noise operator $\hat{F}_{\rm th}$ is fully quantum and should not be decomposed into a classical and a quantum component as the above subsection:
\begin{equation}
S_{\hat F_{\rm th}\hat F_{\rm th}}(\Omega)=4\hbar\left[\frac{1}{e^{\frac{\hbar\Omega}{k_BT}}-1}+\frac{1}{2}\right]\frac{{\rm Im}[\chi_m(\Omega)]}{|\chi_m(\Omega)|^2}.
\end{equation}
In the interferometric configuration, the quantum thermal noise drives the motion of the test mass as follows,
\begin{equation}
\begin{split}
\dot{\hat{x}}_{q\pm}=\frac{\hat{p}_{q\pm}}{M},\quad\dot{\hat{p}}_{q\pm}&=-M(\omega_m^2+\omega^2_{\rm SN})\hat{x}_{q\pm}-\gamma_m\hat{p}_{q\pm}+\frac{M}{2}\delta\omega^2_{\rm SN}\hat{x}_{q\mp}+\hat{F}_{\rm th\pm}.
\end{split}
\end{equation}
The quantum part of the system $\hat{x}_{q\pm}$ and $\hat{b}_{\pm\theta q}$ can be solved as follows, 
\begin{equation}
\begin{split}
\hat{x}_{q\pm}(\omega)&=\frac{\sqrt{M\hbar}\Lambda\chi^{-1}_q(\omega)\hat{a}_{\pm1}(\omega)}{\chi^{-1}_A(\omega)\chi^{-1}_B(\omega)}-\frac{\sqrt{M\hbar}\Lambda M\delta\omega^2_{\rm SN}\hat{a}_{\mp1}}{2\chi^{-1}_A(\omega)\chi^{-1}_B(\omega)}+\frac{\chi^{-1}_q(\omega)\hat{f}_{{\rm th\pm}}}{\chi^{-1}_A(\omega)\chi^{-1}_B(\omega)}-\frac{M\delta\omega^2_{\rm SN}\hat{f}_{{\rm th\mp}}}{2\chi^{-1}_A(\omega)\chi^{-1}_B(\omega)}.
\end{split}
\end{equation}
and the quantum part of the outgoing field is,
\be
\begin{split}
\hat b_{\theta_\pm q}(\omega)&=\sin\theta\hat a_{\pm2}(\omega)+\cos\theta\hat a_{\pm1}(\omega)+\sqrt{M/\hbar}\Lambda\sin\theta\\&\times\left(\frac{\sqrt{M\hbar}\Lambda\chi^{-1}_q(\omega)\hat{a}_{\pm1}(\omega)}{\chi^{-1}_A(\omega)\chi^{-1}_B(\omega)}-\frac{\sqrt{M\hbar}\Lambda M\delta\omega^2_{\rm SN}\hat{a}_{\mp1}}{2\chi^{-1}_A(\omega)\chi^{-1}_B(\omega)}+\frac{\chi^{-1}_q(\omega)\hat{f}_{{\rm th\pm}}}{\chi^{-1}_A(\omega)\chi^{-1}_B(\omega)}-\frac{M\delta\omega^2_{\rm SN}\hat{f}_{{\rm th\mp}}}{2\chi^{-1}_A(\omega)\chi^{-1}_B(\omega)}\right).
\end{split}
\end{equation}
Interestingly, different from the classical thermal noise case, quantum thermal noise \emph{enhances} the crosstalk between common and differential modes of the test mass, hence strengthens correlations between outgoing fields $S_{y_{{\theta +}{\theta -}}}(\omega)$ as:
\begin{equation}
\begin{split}
S_{y_{\theta \pm}y_{\theta \mp}}(\omega)&=\frac{M^2\delta\omega^2_{\rm SN}\Lambda^2\left(\cosh2r\sin2\theta+2\cos(\theta-2\phi)\sin\theta\sinh2r\right)}{2|\chi^{-1}_m(\omega)|^2}\\
&-\frac{M^2\delta\omega^2_{\rm SN}\Lambda^2\left(2\sqrt{\xi}\sqrt{\sin^2\theta\zeta+\sin^2\theta\gamma_mk_BT/(\hbar\Lambda^2)}\right)}{2|\chi^{-1}_m(\omega)|^2},
\end{split}
\end{equation}
If we assume $\sin\theta\Lambda^2\gg(\sqrt{1+\cos^2\theta}+\cos^2\theta)\omega^2_{qA/B}$ and $\sin\theta\ll1$,  the spectrum of the single outgoing field $S_{y_{{\theta \pm}{\theta \pm}}}(\omega)$ becomes:
\begin{equation}
S_{y_{{\theta \pm}{\theta \pm}}}(\omega)=\frac{(\sin^2\theta\Lambda^4-2\omega^2_{\rm SN}|\beta|^2)\zeta+4\sin^2\theta\Lambda^2\gamma_mk_BT/\hbar}{|\chi^{-1}_m(\omega)|^2},
\end{equation}
where 
\begin{equation}
|\beta|^2\approx\sqrt{\frac{\sin^2\theta\Lambda^4\zeta+\sin^2\theta\Lambda^2\gamma_mk_BT/\hbar}{\xi}}.
\end{equation}

\begin{figure*}
\centering
\includegraphics[width=0.49\textwidth]{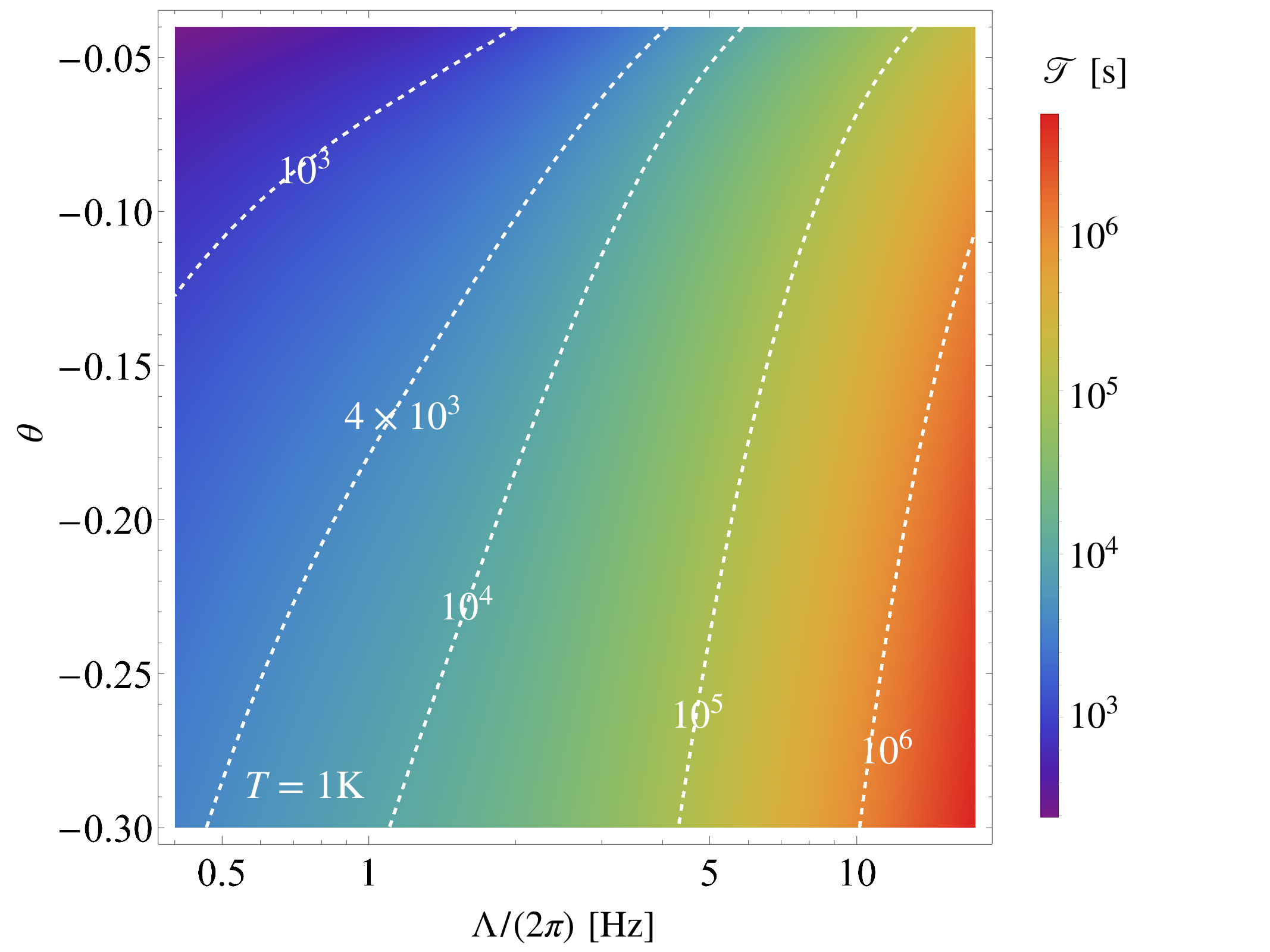}
\includegraphics[width=0.49\textwidth]{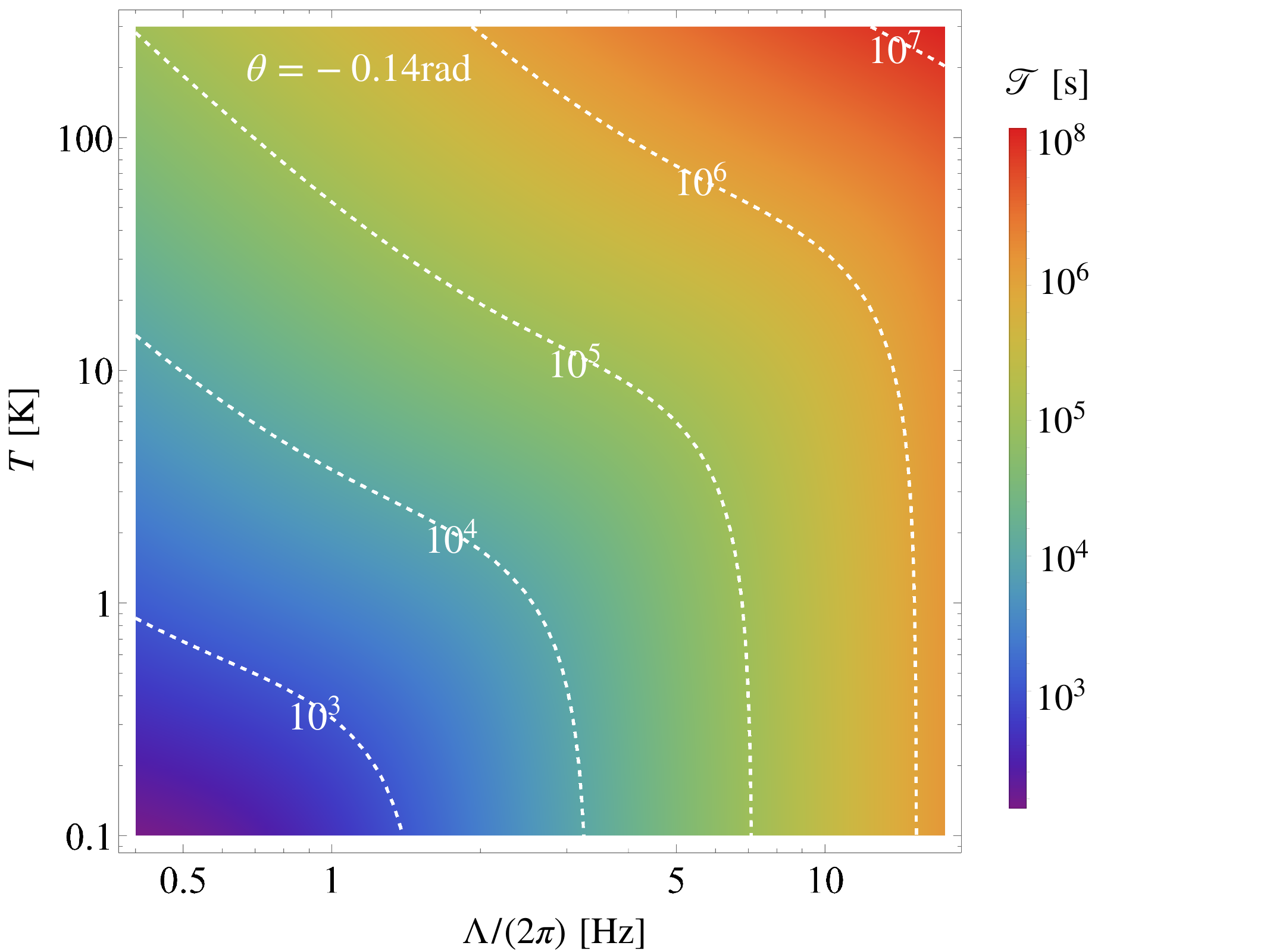}
\includegraphics[width=0.49\textwidth]{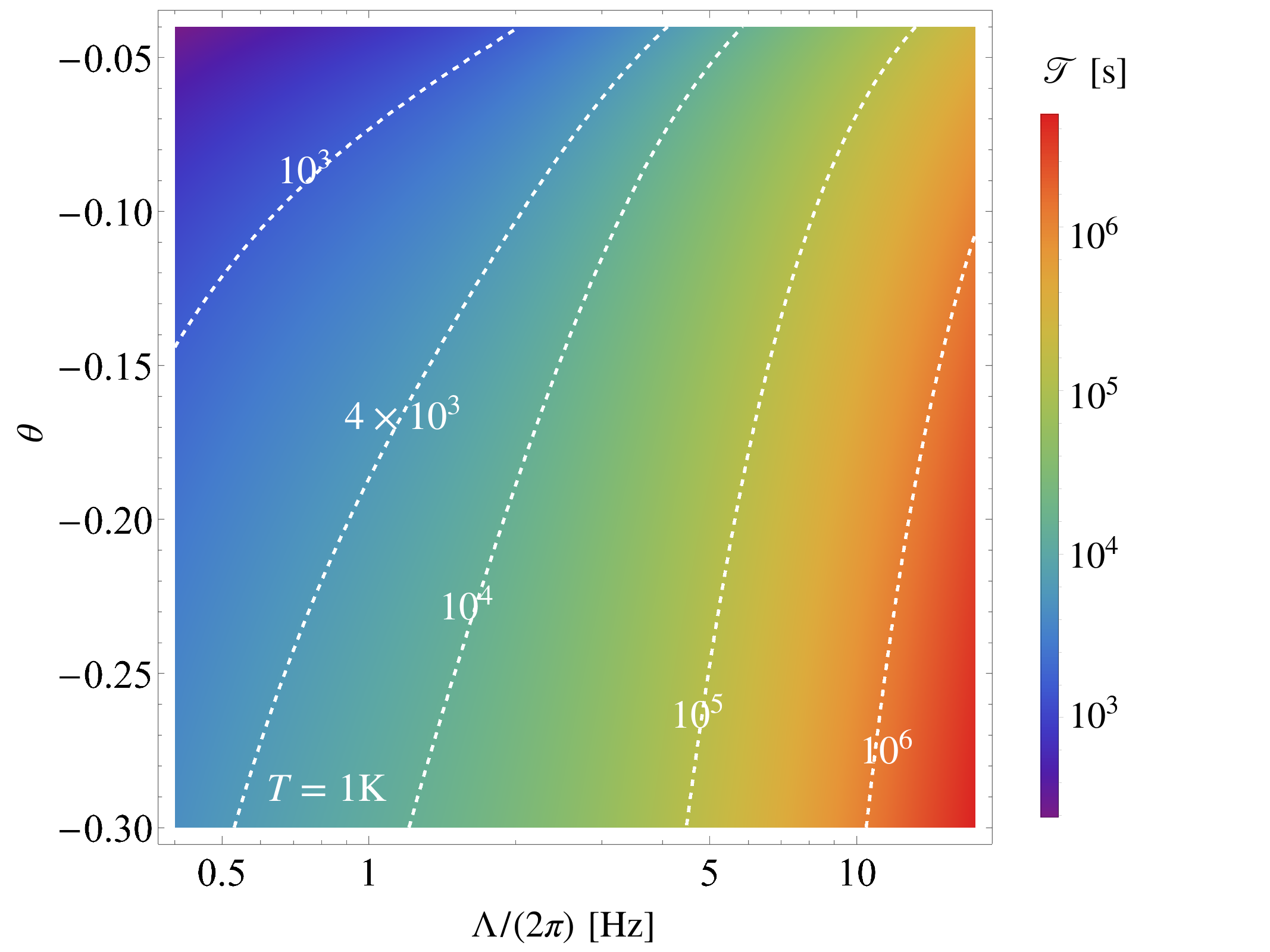}
\includegraphics[width=0.49\textwidth]{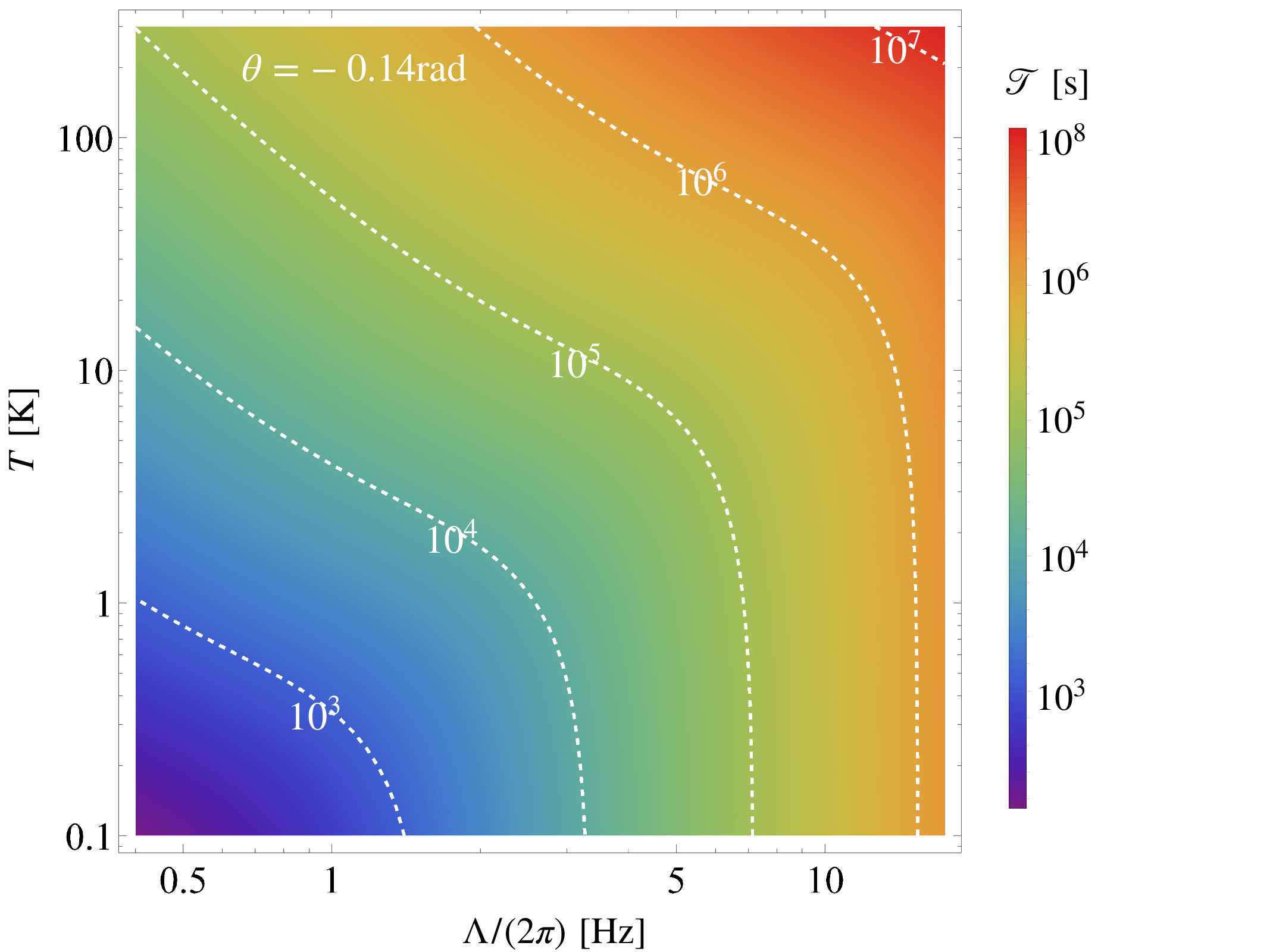}
\caption{Quantum thermal noise prescription: dependence of measurement time $\mathcal{T}$ on the environmental temperature $T$, cooperativity $\Lambda$ and homodyne angle $\theta$, where the signal-to-noise ratio is set to be ${\rm SNR}=1$. The optical input is set to be a 10\,dB phase squeezed vacuum. The Upper panel represents the result obtained using exact formula, while the lower panel is the one obtained using the approximated formula.}
\label{fig:time_parameter_dependence}
\end{figure*}

In the quantum thermal noise prescription, the expectation value and variance of the statistical indicator $\hat{\chi}_N$ are:

\begin{equation}
\langle\hat{\chi}_N\rangle=\mathcal{T}\delta\omega^2_{\rm SN}\Lambda^2\left(\sin\theta\zeta-\sqrt{\sin^2\theta\zeta(\zeta+\gamma_mk_BT/(\hbar\Lambda^2))}\right),
\end{equation}
\begin{equation}
{\rm Var}[\hat{\chi}_N]=\frac{\mathcal{T}^2(1+\kappa)[\zeta(\sin^2\theta\Lambda^4-2\omega^2_{\rm SN}|\beta|^2)+4\sin^2\theta\Lambda^2\gamma_mk_BT/\hbar]^2}{\Gamma \mathcal{T}}.
\end{equation}
Similarly, we can obtain the relationship between the measurement time and the SNR,
\begin{equation}
\mathcal{T}=(1+\kappa)\left[\frac{\zeta(\sin^2\theta\Lambda^4-2\omega^2_{\rm SN}|\beta|^2)+4\sin^2\theta\Lambda^2\gamma_mk_BT/\hbar}{\Gamma^{1/2}\delta\omega^2_{\rm SN}\Lambda^2\left(\sin\theta\zeta-|\sin\theta|\sqrt{\zeta(\zeta+\gamma_mk_BT/(\hbar\Lambda^2))}\right)}\right]^2{\rm SNR}^2.
\end{equation} 

\section{Device imperfections as a source of false-positives}

The preceding analysis assumes idealized interferometer conditions, but practical implementations must account for device imperfections that break the arm symmetry. Such symmetry breaking introduces spurious correlations between the common and differential optical modes, which can both obscure the targeted SN signatures and generate false positives. This section examines how these imperfection-induced effects impact the experimental feasibility of our protocol, providing quantitative requirements for the allowable levels of imperfections in the interferometric setup.

\subsection{Effect of test-mass parameter-mismatches: $M$, $Q$, and $\omega_m$}
In the real experiment, the mechanical parameters of the two test masses, such as mass $M$, mechanical quality factor $Q$, and resonant frequency $\omega_m$, will be independently deviated from the ideal values. The mismatches will break the system's symmetry and introduce spurious correlations, providing the false-positives to the SN-induced correlation. Taking these mismatches into account and expanding to the leading order of the relative mismatch, we can modify the equation of the common ($\hat{x}_{+}$) and differential ($\hat{x}_{-}$) mechanical modes as follows:

\begin{equation}
\begin{split}
\hat{x}_{+}(\omega)&=\chi_m(\omega)\left[1+(iM\gamma_m\omega\bar{\epsilon}_{Q}-2M\omega_m^2\bar{\epsilon}_{\omega_m})\chi_m(\omega)-\bar{\epsilon}_M\right](\Lambda\sqrt{M\hbar}\hat{a}_{1+}+f_{\rm th+})\\
&+\frac{\chi_m(\omega)}{2}\left[(iM\gamma_m\omega\delta\epsilon_{Q}-2M\omega_m^2\delta\epsilon_{\omega_m})\chi_m(\omega)-\delta\epsilon_M\right](\Lambda\sqrt{M\hbar}\hat{a}_{1-}+f_{\rm th-}),\\
\hat{x}_{-}(\omega)&=\chi_m(\omega)\left[1+(iM\gamma_m\omega\bar{\epsilon}_{Q}-2M\omega_m^2\bar{\epsilon}_{\omega_m})\chi_m(\omega)-\bar{\epsilon}_M\right](\Lambda\sqrt{M\hbar}\hat{a}_{1-}+f_{\rm th-})\\
&+\frac{\chi_m(\omega)}{2}\left[(iM\gamma_m\omega\delta\epsilon_{Q}-2M\omega_m^2\delta\epsilon_{\omega_m})\chi_m(\omega)-\delta\epsilon_M\right](\Lambda\sqrt{M\hbar}\hat{a}_{1+}+f_{\rm th+}).
\end{split}
\end{equation}
where $\bar{\epsilon}_{{Q}/{\omega_m}/M}=(\epsilon^A_{{Q}/{\omega_m}/M}+\epsilon^B_{{Q}/{\omega_m}/M})/2$ represents the averaged mismatch, and $\delta\epsilon_{{Q}/{\omega_m}/M}=(\epsilon^A_{{\rm Q}/{\omega_m}/M}-\epsilon^B_{{Q}/{\omega_m}/M})$ denotes the difference of the two mirrors' mismatches for the respective parameters. The $\epsilon^{A/B}_{{Q}/{\omega_m}/M}$ are defined as the relative mismatches between the actual parameters of test masses A and B and the ideal values:
\begin{equation}
\epsilon^{A/B}_{{\rm Q}}=\frac{Q_{A/B}-Q}{Q},\quad\epsilon^{A/B}_{{\omega_m}}=\frac{\omega_{mA/B}-\omega_m}{\omega_m},\quad\epsilon^{A/B}_M=\frac{M_{A/B}-M}{M}.
\end{equation}

The mismatch in $Q$ affects the power spectrum of the thermal force noise, yielding:
\begin{equation}
\begin{split}
S_{f_{\rm th+}f_{\rm th-}}(\omega)&=-2\delta\epsilon_{Q}M\gamma_mk_BT,\\
S_{f_{\rm th\pm} f_{\rm th\pm}}(\omega)&=(1+\bar{\epsilon}_{Q})4M\gamma_mk_BT.
\end{split}
\end{equation}

When the system dynamics is thermal-noise-limited, the correlation $S^{\rm mis}_{y_{\theta\pm}y_{\theta\mp}}(\omega)$ between the common and differential modes of the outgoing field, induced by these mismatches, can be approximated by:
\begin{equation}
\begin{split}
S^{\rm mis}_{y_{\theta\pm}y_{\theta\mp}}(\omega)\approx&4\sin^2\theta|\chi_m(\omega)|^2\Lambda^2{\rm Re}\left[(iM\gamma_m\omega\delta\epsilon_{Q}-2M\omega_m^2\delta\epsilon_{\omega_m})\chi_m(\omega)-\delta\epsilon_M\right]\frac{M^2\gamma_mk_BT}{\hbar}\\
&-2\sin^2\theta|\chi_m(\omega)|^2\Lambda^2\delta\epsilon_{Q}\frac{M^2\gamma_mk_BT}{\hbar},
\\\approx&4\sin^2\theta\Lambda^2|\chi_m(\omega)|^2\frac{M^2\gamma_mk_BT}{\hbar}\left[\frac{2\omega^2_m(\omega^2-\omega^2_m)}{(\omega^2-\omega^2_m)^2+\gamma_m^2\omega^2}\delta\epsilon_{\omega_m}-\left(\frac{1}{2}+\frac{\gamma_m^2\omega^2}{(\omega^2-\omega^2_m)^2+\gamma_m^2\omega^2}\right)\delta\epsilon_{Q}-\delta\epsilon_M\right].
\end{split}
\end{equation}
To ensure that these mismatch-induced correlations do not obscure the target correlations induced by SN gravity, denoted $S_{y_{\theta\pm}y_{\theta\mp}}(\omega)$), the condition $S^{\rm mis}_{y_{\theta\pm}y_{\theta\mp}}(\omega) < S_{y_{\theta\pm}y_{\theta\mp}}(\omega)$ must be satisfied. This imposes the following constraints on the test mass parameter mismatches:
\begin{equation}
\delta\epsilon_{\omega_m}<\frac{\hbar\delta\omega^2_{\rm SN}\zeta[(\omega^2-\omega^2_m)^2+\gamma_m^2\omega^2]}{4\sin\theta\omega_m^2(\omega^2-\omega_m^2)\gamma_mk_BT},\quad\delta\epsilon_{Q}<\frac{\hbar\delta\omega^2_{\rm SN}[(\omega^2-\omega^2_m)^2+\gamma_m^2\omega^2]\zeta}{\sin\theta\gamma_mk_BT[(\omega^2-\omega^2_m)^2+3\gamma_m^2\omega^2]},\quad\delta\epsilon_M<\frac{\hbar\delta\omega^2_{\rm SN}\zeta}{2\sin\theta\gamma_mk_BT}.
\end{equation}
The high quality factor $Q$ of the test masses renders their dynamics particularly sensitive near the resonant frequency $\omega_m$. Consequently, symmetry breaking due to test mass parameter mismatches can lead to significant correlations in this frequency region. Therefore, the aggregated data analysis procedure should exclude the data within the frequency range around $\omega_m$, while still keeping the SN signature. Specifically, when the observational bandwidth is set to be $\Gamma \gg \omega_m$, data within the interval $[\omega_m/2, 3\omega_m/2]$ can be discarded when extracting SN-induced correlations. Under this data exclusion strategy, the constraints on test mass mismatches can be simplified as:
\begin{equation}
\delta\epsilon_{\omega_m}<\frac{\hbar\delta\omega^2_{\rm SN}\zeta}{4\sin\theta\gamma_mk_BT},\quad\delta\epsilon_{Q}<\frac{\hbar\delta\omega^2_{\rm SN}\zeta}{\sin\theta\gamma_mk_BT},\quad\delta\epsilon_M<\frac{\hbar\delta\omega^2_{\rm SN}\zeta}{2\sin\theta\gamma_mk_BT}.
\end{equation}
For illustrative purposes, with $T=1\,$K and $10\,$dB of input field squeezing, these constraints translate to $\delta\epsilon_{\omega_m} < 0.84 \times 10^{-2}$, $\delta\epsilon_{Q} < 3.35 \times 10^{-2}$, and $\delta\epsilon_M < 1.68 \times 10^{-2}$.

\subsection{Mismatch of the optical cavity parameters: optical cavity bandwidth $\gamma$}

A mismatch in the cavity bandwidths between the two arms (i.e., $\gamma_A \neq \gamma_B$, leading to deviations from an ideal $\gamma$) can also induce correlations in the outgoing fields. Incorporating such a mismatch, the equations of motion for the common and differential modes (Eq.\,\ref{eq:mismatch_cavity}) become:
\begin{equation}\label{eq:mismatch_cavity}
\dot{\hat{x}}_{\pm}=\frac{\hat{p}_{\pm}}{M},\quad
\dot{\hat{p}}_{\pm}=-M\omega_m^2\hat{x}_{\pm}-\gamma_m\hat{p}_{\pm}+\sqrt{M\hbar}\Lambda\left(1-\frac{\bar{\epsilon}_{\gamma}}{2}\right)\hat{a}_{1\pm}-\frac{\delta\epsilon_{\gamma}}{4}\sqrt{M\hbar}\Lambda\hat{a}_{1\mp}+f_{\rm th\pm},
\end{equation}
where $\bar{\epsilon}_{\gamma}=(\epsilon_{\gamma_A}+\epsilon_{\gamma_B})/2$ is the average mismatch and $\delta\epsilon_{\gamma}=(\epsilon_{\gamma_A}-\epsilon_{\gamma_B})$ is the bandwidth difference of the two arm cavities. The relative mismatch of cavity A and B $\epsilon_{\gamma_{A/B}}$ to the ideal value are defined by:
\begin{equation}
\epsilon_{\gamma_{A/B}}=\frac{\gamma_{A/B}-\gamma}{\gamma},
\end{equation}
with $\gamma_{A/B}$ being the actual bandwidth for the cavities in arms A and B, $\gamma$ is the theoretically recommended cavity bandwidth. As evident from Eq.\,(\ref{eq:mismatch_cavity}), the differential mismatch $\delta\epsilon_{\gamma}$ introduces a coupling term, thereby generating correlation between the common and differential motion of the test masses. This, in turn, modifies the outgoing field quadratures $\hat{b}_{\pm\theta}$ as:
\begin{equation}
\begin{split}
\hat{b}_{\pm\theta}&=\cos\theta\hat{a}_{1\pm}+\sin\theta\hat{a}_{2\pm}+\left(1-\frac{\bar{\epsilon}_{\gamma}}{2}\right)\sin\theta\sqrt{\frac{M}{\hbar}}\Lambda\chi_m(\omega)f_{\rm th\pm}+\sin\theta M\Lambda^2(1-\bar{\epsilon}_{\gamma})\chi_m(\omega)\hat{a}_{1\pm}\\
&-\frac{\delta\epsilon_{\gamma}}{4}\sqrt{\frac{M}{\hbar}}\sin\theta\Lambda\chi_m(\omega)f_{\rm th\mp}-\frac{\delta\epsilon_{\gamma}}{2}\sin\theta M\Lambda^2\chi_m(\omega)\hat{a}_{1\mp}.
\end{split}
\end{equation}
The resulting error correlation $S^{\rm de}_{y_{\theta\pm}y_{\theta\mp}}(\omega)$ due to the mismatch $\delta\epsilon_{\gamma}$ is represented as:
\begin{equation}
S^{\rm de}_{y_{\theta\pm}y_{\theta\mp}}(\omega)=-\delta\epsilon_{\gamma}M^2\Lambda^2|\chi_m(\omega)|^2\gamma_mk_BT/\hbar-\delta\epsilon_{\gamma}M^2\Lambda^4|\chi_m(\omega)|^2/2.
\end{equation}

Using a similar approach as the above subsection and assuming a thermal-noise-limited dynamics, the requirement to distinguish SN-induced correlations implies that the mismatch $\delta\epsilon_{\gamma}$ must satisfy:
\begin{equation}
\delta\epsilon_{\gamma}<\frac{2\hbar\delta\omega^2_{\rm SN}\zeta}{\sin\theta\gamma_mk_BT}.
\end{equation}
For $T=1\,$K and $10\,$dB input phase-squeezed vacuum, this constraint is evaluated as $\delta\epsilon_{\gamma} < 6.70 \times 10^{-2}$.

\subsection{Beam splitter imperfections}

We now consider the impact of beam splitter (BS) imperfections on the correlation of the outgoing fields. Assuming the thermal-noise-limited case, the influence brought by the modification of the ingoing field $\hat{a}_{\rm in}$ due to the BS imperfections is typically negligible. Our focus is thus on how BS imperfections affect the recombination of the output fields $\hat{b}_{A\theta}$ and $\hat{b}_{B\theta}$ from the two optical cavities into the common $\hat{b}_{\theta+}$ and differential $\hat{b}_{\theta-}$ output fields (see Fig.\,\ref{fig:the_BS}).

\begin{figure}[h!]
\centering
\includegraphics[width=0.8\textwidth]{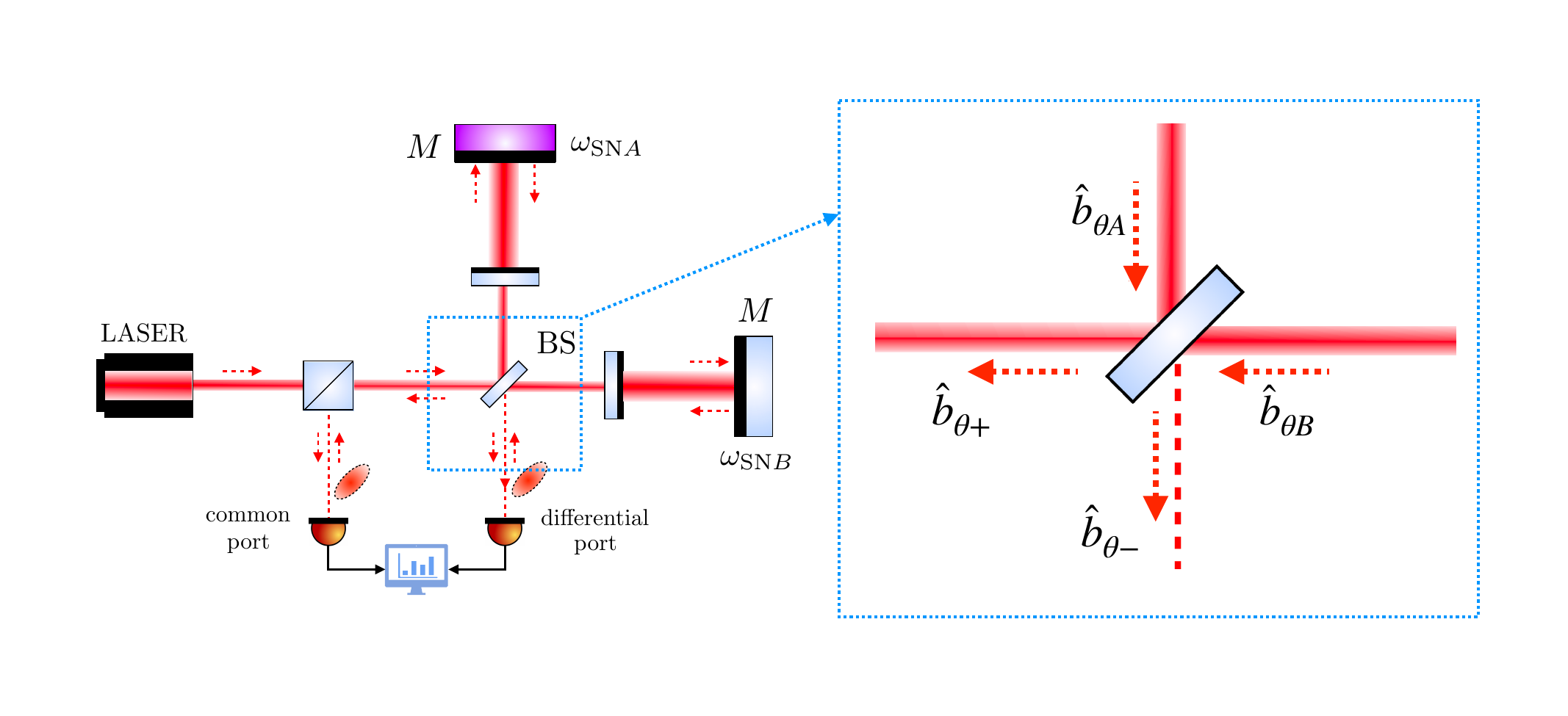}
\caption{Schematic illustrating the analysis of beam splitter imperfections, including losses ($\epsilon_{\rm BS}^{c/d}$) and splitting ratio errors ($\Delta_{c/d}$).}
\label{fig:the_BS}
\end{figure}

Taking into account the optical losses and the mismatch from the ideal 50:50 BS ratio, the general input-output relations for the BS can be expressed as:
\begin{equation}
\begin{split}
\hat{b}_{\theta+}&=\sqrt{1-\epsilon_{\rm BS}^c}\left[\sqrt{\frac{1}{2}+\Delta_c}\hat{b}_{\theta A}+\sqrt{\frac{1}{2}-\Delta_c}\hat{b}_{\theta B}\right]+\sqrt{\epsilon_{\rm BS}^c}\hat{n}_c,\\
\hat{b}_{\theta-}&=\sqrt{1-\epsilon_{\rm BS}^d}\left[\sqrt{\frac{1}{2}+\Delta_d}\hat{b}_{\theta A}-\sqrt{\frac{1}{2}-\Delta_d}\hat{b}_{\theta B}\right]+\sqrt{\epsilon_{\rm BS}^d}\hat{n}_d,
\end{split}
\end{equation}
where $\epsilon_{\rm BS}^{c/d}$ represents the effective optical loss and $\Delta_{c/d}$ denotes the deviation from a perfect 50:50 BS ratio for the common and differential mode paths through the BS. Note that the noise $\hat n_{c/d}$ associated with the optical loss has no common-differential correlations, hence will be neglected in the following analysis. For small imperfections, i.e., $\epsilon_{\rm BS}^{c/d} \ll 1$ and $\Delta_{c/d} \ll 1$, these in-out relations can be simplified as:
\begin{equation}
\begin{split}
\hat{b}_{\theta+}&=\left[1-\frac{\epsilon^{c}_{\rm BS}}{2}\right]\frac{1}{\sqrt{2}}(\hat{b}_{\theta A}+\hat{b}_{\theta B})+\Delta_c\frac{1}{\sqrt{2}}(\hat{b}_{\theta A}-\hat{b}_{\theta B}),\\
\hat{b}_{\theta-}&=\left[1-\frac{\epsilon^{d}_{\rm BS}}{2}\right]\frac{1}{\sqrt{2}}(\hat{b}_{\theta A}-\hat{b}_{\theta B})+\Delta_d\frac{1}{\sqrt{2}}(\hat{b}_{\theta A}+\hat{b}_{\theta B}).
\end{split}
\end{equation}
The correlation $S^{\rm mis}_{y_{\theta\pm}y_{\theta\mp}}(\omega)$ arising from these BS imperfections is then found to be:
\begin{equation}
\begin{split}
S^{\rm mis}_{y_{\theta\pm}y_{\theta\mp}}(\omega)&=\left[1-\frac{\epsilon^{c}_{\rm BS}}{2}\right]\frac{\Delta_d}{2}(S_{y_{\theta+}y_{\theta+}}(\omega)+S_{y_{\theta-}y_{\theta-}}(\omega))+\left[1-\frac{\epsilon^{d}_{\rm BS}}{2}\right]\frac{\Delta_c}{2}(S_{y_{\theta+}y_{\theta+}}(\omega)+S_{y_{\theta-}y_{\theta-}}(\omega))\\&=\left[\frac{\Delta_c+\Delta_d}{2}-\frac{\epsilon^{d}_{\rm BS}\Delta_c+\epsilon^{c}_{\rm BS}\Delta_d}{4}\right](S_{y_{\theta+}y_{\theta+}}(\omega)+S_{y_{\theta-}y_{\theta-}}(\omega)),
\end{split}
\end{equation}
where $S_{y_{\theta\pm}y_{\theta\pm}}(\omega)$ is the diagonal power spectrum of the outgoing field in the ideal case. In the thermal-noise-limited case, it is approximated by:
\begin{equation}
S_{y_{\theta\pm}y_{\theta\pm}}(\omega)\approx4\sin^2\theta|\chi_m(\omega)|^2M^2\Lambda^2\gamma_mk_BT/\hbar.
\end{equation}
Note that the term $\propto \epsilon^{d/c}_{\rm BS}\Delta_{c/d}$ is a second-order small quantity and can be ignored.  Finally, we obtain:
\begin{equation}
S^{\rm mis}_{y_{\theta\pm}y_{\theta\mp}}(\omega) \approx 8\sin^2\theta\bar{\Delta}_{\rm BS}|\chi_m(\omega)|^2M^2\Lambda^2\gamma_mk_BT/\hbar,
\end{equation}
where $\bar{\Delta}_{\rm BS}=(\Delta_{c}+\Delta_{d})/2$ is the averaged BS mismatch. To ensure that these BS-mismatch-induced correlations do not disturb the targeted SN effects, $\bar{\Delta}_{\rm BS}$ must satisfy the constraint:
\begin{equation}
\bar{\Delta}_{\rm BS}<\frac{\hbar\delta\omega^2_{\rm SN}\zeta}{4\sin\theta\gamma_mk_BT}.
\end{equation}
With $T=1\,$K and $10\,$dB input squeezing, this requires $\bar{\Delta}_{\rm BS} < 0.84 \times 10^{-2}$.

\section{Circumventing the seismic noise and  the complete rejection of common-mode noise}

$\bullet$ \textbf{Seismic Noise:} Part of the seismic noise can be bypassed by special designs of the interferometer configuration. For example, in Fig.\,\ref{fig:folded_configuration}, an exemplary design is presented in which the two arms are folded so that the two movable test mass mirrors are linearly aligned.

\begin{figure*}
\centering
\includegraphics[width=0.5\textwidth]{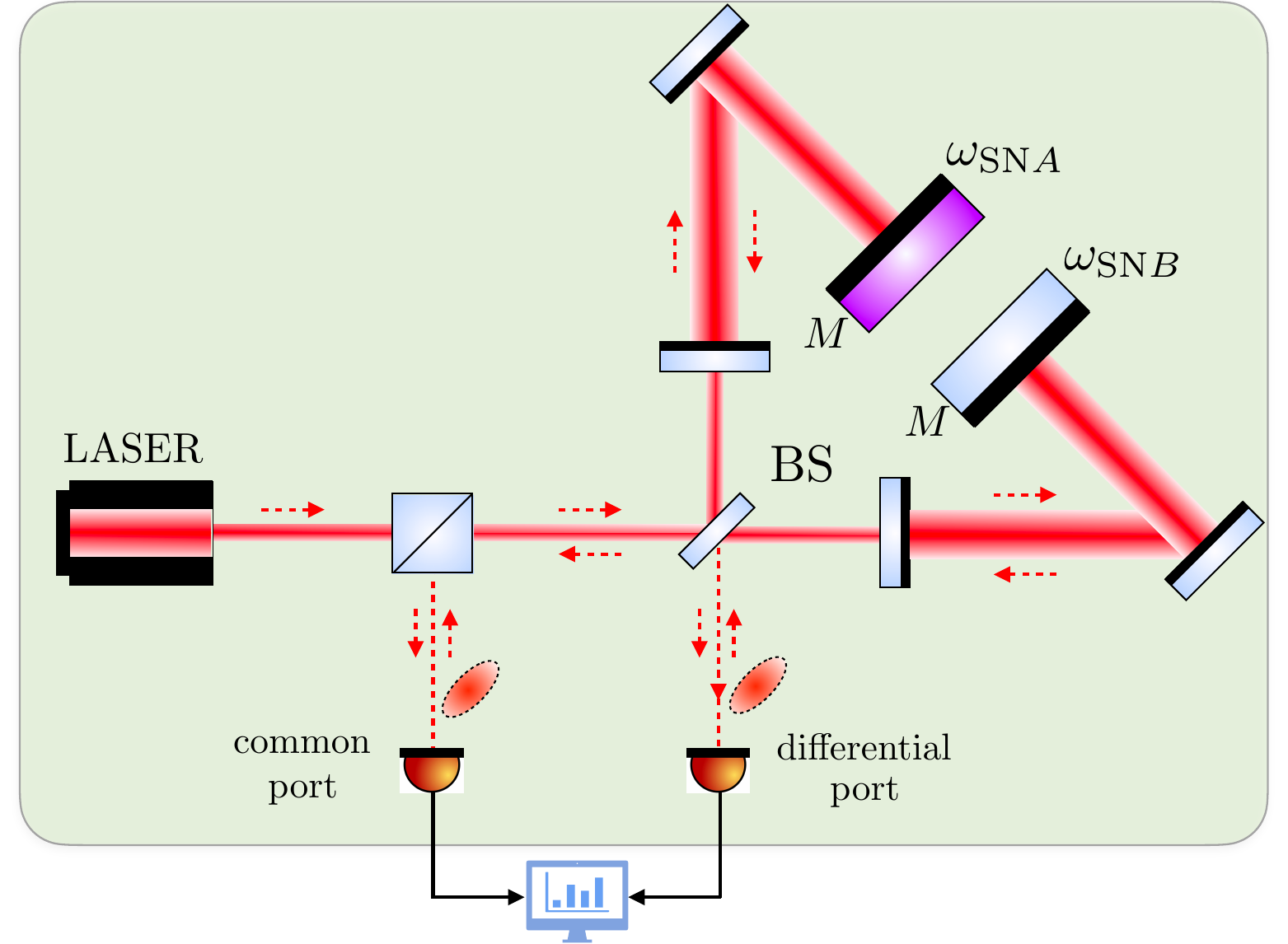}
\caption{Folded interferometric configuration: two test mass mirrors are linearly aligned and the arm cavities are folded. The background is the experimental platform that will have displacement noises due to its connection with the ground motion.}
\label{fig:folded_configuration}
\end{figure*}

Suppose the seismic ground motion shakes the experimental platform, since the typical size of the platform is much smaller than the wavelength of the typical earthquake wave $\lambda_{\rm seismic}\sim v_{\rm seismic}/f\sim 5\times 10^5$\,m, the seismic motion will exert a common displacement noise to all the apparatus on the experimental platform, in particular the two test mass mirrors. If the two arms are not folded, as in the usual Michelson interferometer, this motion will create asymmetry between the two arms so that there will be cross-talk noise that contaminate the Schr{\"o}dinger-Newton signature. In contrast, because the two mirrors are linearly aligned in the folded interferometric configuration, the common noise motion can not induce the cross-talk between the common and differential mechanical/optical modes. Fig.\,\ref{fig:folded_configuration} is merely an example of the upgrading of the interferometer configuration; a more detailed configurational design is beyond the scope of this work, can be further explored considering the detailed experimental requirements.\\

$\bullet$ \textbf{Common-mode Noise:} For a complete rejection of the common-mode noise so that $\kappa=0$, a more complicated design is required and more sophisticated interferometer balancing condition is also required. In this Supplementary Material, we provide a possible design shown in Fig.\,\ref{fig:common_noise_rejection}, where the configuration consists of three interferometric structures. The two interferometers A/B shares the carrier laser field that is split from the the beams-splitter of the third interferometric structure. The interferometer A/B are balanced interferometer on their own. However, the test mass in A/B, though share the same mass $M$,  are made with different materials (silicon and osmium, for instance) hence are characterized by different SN frequencies. The symmetry-breaking happens between the interferometer A/B. The fourth beam splitter recombines the differential output fields $(\hat b_-^A,\hat b_-^B)$ from interferometer A and B, generating ``common" and ``differential" output fields $(\hat b_-^{A+B},\hat b_-^{A-B})$. The cross-correlation of $\hat b_-^{A+B}$ and $\hat b_-^{A-B}$ will carry the binary SN-induced signature, while the common-mode noise is almost completely rejected if the interferometers are fine-tuned. The mathematical details for analyzing this configuration is almost exactly the same as the main text and the previous section of the Supplementary Material, which will not be repeated here.

However, we want to emphasize that this configuration involves more complex interferometric structures and thus requires more delicate fine-tuning. In this sense, the single-interferometer setup presented in the main text is technically less demanding. Additionally, recent advances in laser noise suppression technology have significantly improved, which could further reduce the $\kappa$ factor for the single-interferometer case. As such, it remains unclear which configuration is superior. We believe only a thorough experimental design study—best conducted by the experimentalists with greater expertise—can provide a definitive answer to this question.

\begin{figure*}
\centering
\includegraphics[width=0.6\textwidth]{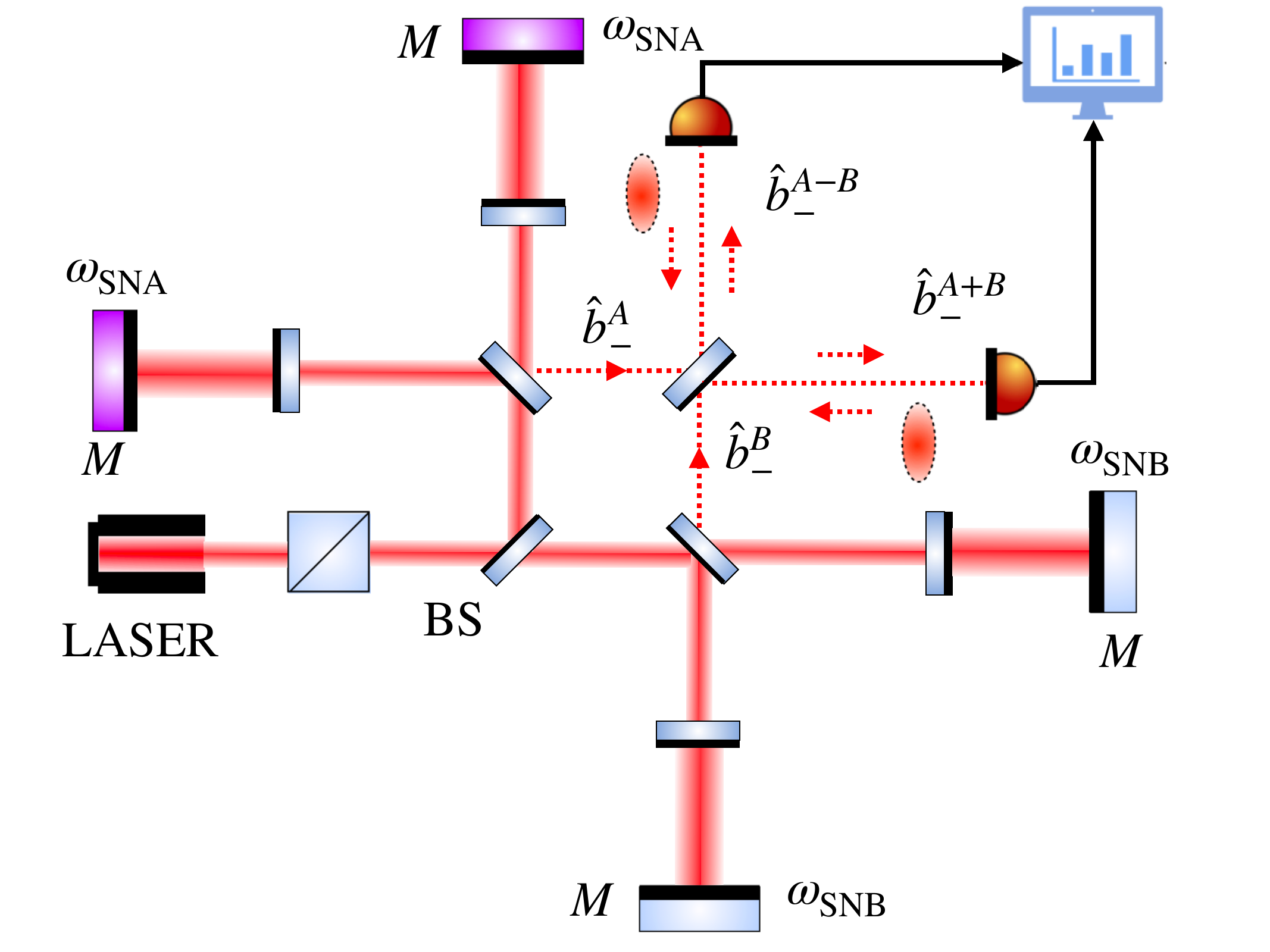}
\caption{Double-interferometer configuration: A possible configuration that can completely reject the common-mode noise, at the price of more sophisticated and complicated fine-tuning of the interferometer.}
\label{fig:common_noise_rejection}
\end{figure*}

\end{widetext}
\end{document}